\documentclass[a4paper,fleqn,twocolumn,margin=1in]{aastex62}
\usepackage{float}
\usepackage{graphicx}
\usepackage{ae,aecompl} 
\usepackage{enumitem} 
\usepackage{amssymb} 
\usepackage{amsmath} 
\usepackage[lofdepth,lotdepth,caption=false]{subfig}

\def\muhz{\mu \mathrm{Hz}}

\def\pmax{A_{\mathrm{max}}}
\def\numax{\nu_{\mathrm{max}}}
\def\teff{T_{\mathrm{eff}}}
\def\pg{A_{\mathrm{excess}}}
\def\dnu{\Delta \nu}
\def\lsun{L_{\odot}}
\def\teffsun{T_{\mathrm{eff,\ }\odot}}
\def\K{K_{\mathrm{s}}}

\def\msun{M_{\odot}}
\def\ben{\begin{longenum}}
\def\een{\end{longenum}}
\def\gax{\mathrel{\raise.3ex\hbox{$>$}\mkern-14mu\lower0.6ex\hbox{$\sim$}}}
\def\lax{\mathrel{\raise.3ex\hbox{$<$}\mkern-14mu\lower0.6ex\hbox{$\sim$}}}
\def\gtorder{\mathrel{\raise.3ex\hbox{$>$}\mkern-14mu
             \lower0.6ex\hbox{$\sim$}}}
\def\ltorder{\mathrel{\raise.3ex\hbox{$<$}\mkern-14mu
    \lower0.6ex\hbox{$\sim$}}}
\def\sigmameso{\sigma_{\mathrm{meso}}}
\def\taumeso{\tau_{\mathrm{meso}}}
\def\sigmagran{\sigma_{\mathrm{gran}}}
\def\taugran{\tau_{\mathrm{gran}}}
\def\thetameso{\theta_{\mathrm{meso}}}
\def\thetagran{\theta_{\mathrm{gran}}}
\def\thetaexcess{\theta_{\mathrm{excess}}}
\def\thetadnu{\theta_{\dnu}}

\begin{document}
\title{The Bayesian Asteroseismology data Modeling pipeline and its application to {\it K2} data}
\author{Joel C. Zinn}
\affiliation{School of Physics, University of New South Wales, Barker
  Street, Sydney, NSW 2052, Australia}
\author{Dennis Stello}
\affiliation{School of Physics, University of New South Wales, Barker
  Street, Sydney, NSW 2052, Australia}
\affiliation{Sydney Institute for Astronomy (SIfA), School of Physics,
  University of Sydney, NSW 2006, Australia}
\affiliation{Stellar Astrophysics Centre, Department of Physics and
Astronomy, Aarhus University, Ny Munkegade 120, DK-8000
Aarhus C, Denmark}
\affiliation{ARC Centre of Excellence for All Sky Astrophysics in 3 Dimensions (ASTRO 3D)}
\author{Daniel Huber}
\affiliation{ Institute for Astronomy, University of Hawai`i, 2680
  Woodlawn Drive, Honolulu, HI 96822, USA}
\affiliation{Sydney Institute for Astronomy (SIfA), School of Physics,
  University of Sydney, NSW 2006, Australia}
\affiliation{Stellar Astrophysics Centre, Department of Physics and
Astronomy, Aarhus University, Ny Munkegade 120, DK-8000
Aarhus C, Denmark}
\affiliation{SETI Institute, 189 Bernardo Avenue, Mountain View, CA
  94043, USA}
\author{Sanjib Sharma}
\affiliation{Sydney Institute for Astronomy (SIfA), School of Physics,
University of Sydney, NSW 2006, Australia}
\affiliation{ARC Centre of Excellence for All Sky Astrophysics in 3 Dimensions (ASTRO 3D)}
\correspondingauthor{Joel C. Zinn}
\email{j.zinn@unsw.edu.au}

\begin{abstract}

We present the Bayesian Asteroseismology data Modeling (BAM) pipeline,
an automated asteroseismology pipeline that returns global
oscillation parameters and granulation parameters from the analysis of
photometric time-series. BAM also determines if a star is likely to
be a solar-like oscillator. We have designed BAM to specially process
{\it K2} light curves, which suffer from unique noise signatures that
can confuse asteroseismic analysis, though it may be used on any
photometric time
series --- including those from {\it Kepler} and TESS. We demonstrate
the BAM oscillation parameters are consistent within $\sim
1.53\% (\mathrm{random}) \pm 0.2\% (\mathrm{systematic})$ and $1.51\% (\mathrm{random}) \pm
0.6\% (\mathrm{systematic})$ for $\numax$ and $\dnu$ with benchmark results for typical {\it K2}
red giant stars in the {\it K2} Galactic Archaeology Program's (GAP)
Campaign 1 sample. Application of
BAM to $13016$ {\it K2} Campaign 1 targets not in the GAP sample yields
$104$ red giant solar-like oscillators. Based on the number of
serendipitous giants we find, we estimate an upper limit on the average purity in dwarf selection among C1
proposals is $\approx 99\%$, which could be lower when considering
incompleteness in BAM detection efficiency, and proper motion cuts specific to C1 Guest Observer proposals.
\end{abstract}

\section{introduction}
\label{sec:introduction}
Solar-like oscillators are stars that support standing acoustic waves
excited by surface convection, and whose global frequency characteristics are determined by
the stellar density and surface gravity \citep[e.g.,][]{ulrich1986,
  brown+1991, kjeldsen&bedding1995}. The frequencies may be measured in radial velocity variations or in
photometric variability. Detecting mode frequencies in solar-like
oscillators yields precise determinations of fundamental stellar
parameters like mass and radius.
However, only about a dozen stars had been observed to exhibit solar-like
oscillations prior to the results from the space-based CoRoT
\citep{baglin+2006} and {\it Kepler} \citep{borucki+2008}
missions. With improved photometric precision compared to ground-based
observations, and continuous monitoring of many stars simultaneously
for up to four years with {\it
  Kepler}, solar-like oscillations have been photometrically detected in thousands of
stars --- mostly red giants \citep[e.g.,][]{de_ridder+2009,hekker+2009,bedding+2010,mosser+2010,stello+2013,yu+2018}. In light of these large asteroseismic data sets, several pipelines have been
developed in order to automatically extract asteroseismic parameters
(e.g., OCT [\citealt{hekker+2010}], CAN
[\citealt{kallinger+2010,kallinger+2014,kallinger+2016}], COR
[\citealt{mosser&appourchaux2009}], A2Z [\citealt{mathur+2010}]).

Among these pipelines is SYD \citep{huber+2009}, much of whose success can be attributed to taking advantage of known scaling
relations among stellar granulation, the frequency of maximum power ($\numax$), and the overtone frequency
separation ($\dnu$) \citep{kjeldsen&bedding2011} to provide accurate initial guesses for fitting
parameters. A
significant shortcoming of SYD (and other similar pipelines) is that
it does not assess if a given star shows excess power from
oscillations in a statistically robust way, hence requiring
post-processing and often visual verification. This introduces
significant unknown, and subjective, detection bias, which hampers
population analyses of the seismic sample. Ensuring reproducible
selection functions is particularly
important for applications aimed to perform Galactic archaeology
studies \cite{stello+2017}. 

In this paper we introduce a new pipeline, the Bayesian
Asteroseismology data Modelling Pipeline (BAM), which builds on the SYD pipeline with an eye toward automatic, robust classification of light curves. BAM formalizes relations among granulation, $\numax$, and $\dnu$ through a Bayesian
framework in which these relations are implemented as priors. It is this Bayesian framework that then allows for a self-consistent, statistical separation of oscillators from non-oscillators.

BAM was also developed with the particular challenges involved in extracting
asteroseismic parameters from the re-purposed {\it Kepler} mission,
{\it K2}, in mind. Following the failure of two of its reaction wheels, the
{\it Kepler} satellite was re-aligned to point in the ecliptic
plane. As opposed to {\it Kepler}'s single field of view in Cygnus,
the {\it K2} pointing pattern covers the ecliptic plane with a
footprint of about 100 square degrees, which is repositioned every
$\sim 80$ days by typically $\sim 90$ degrees along the ecliptic. However,
periodic small-angle pointing corrections are performed every six
hours by firing the spacecraft thrusters, which introduce instrumental signatures in {\it K2} light
curves. These features unfortunately correspond to typical frequencies
of red giant oscillations, and can mimic true asteroseismic
oscillations near $\sim 47 \muhz$ (the 6 hour thruster firing
frequency period). Because this instrumental feature overlaps in
frequency with where a typical red clump star shows maximum oscillation
power, it can hinder recovering red clump stars, which comprise the largest
population of red giants in the Galaxy. BAM's Bayesian framework uses information like the amplitude of the power
excess and the shape of the rest of the power spectrum to distinguish
between {\it K2} thruster firing noise and genuine oscillations. In addition to this
instrumental feature, the {\it K2} white noise level is typically
larger than the white noise of the original {\it Kepler} mission by a
factor of about two, depending on how the data are processed. (However,
several {\it K2} light curve processing pipelines have reported
near-{\it Kepler} white noise levels [\citealt{vanderburg&johnson2014,lund+2015,armstrong+2015,aigrain_parviainen&pope2016,luger+2016}].)

In addition to describing how BAM works in this paper, we apply it to
extract global oscillation parameters for red giants observed
serendipitously by {\it K2} through Guest Observer (GO) programs targeting dwarf
stars during Campaign 1. This new sample of giants therefore adds to
the already known red giant sample from \cite{stello+2017}.

\section{data}
\label{sec:data}
In this paper, we work with two sets of {\it K2} light curves: 1) the
Campaign 1 (C1) target sample from the {\it K2} Galactic
Archaeology Program
\citep[GAP;][]{stello+2015,stello+2017}\footnote{\url{http://www.physics.usyd.edu.au/k2gap/}\\\url{https://archive.stsci.edu/prepds/k2gap/}},
which comprises 8630 stars, and 2) all non-GAP C1 targets, of $13016$ in
total.\footnote{We exclude the Trans-Neptunian object, EPIC 200001049.}
Results from BAM for the former sample have been published in
\cite{stello+2017}. We review some of those results here, and extend the
application of BAM to the latter sample in order to identify serendipitous red giants.

All our C1 light curves have been generated by
\cite{vanderburg&johnson2014} (VJ), who perform aperture photometry on
     {\it K2} images and remove trends associated with centroid errors
     caused by the spacecraft's unstable pointing. We will show below
     that this preprocessing does not completely remove the thruster-induced
     instrumental features from the data, and therefore requires
     additional processing in BAM.

We begin by first removing trends on
time-scales much longer than solar-like oscillation time-scales for
the stars we are interested in. For each light curve, we perform high-pass filtering by
dividing the VJ light curve by a 4-day wide boxcar-smoothed version of the light
curve, thus imposing a high-pass
cutoff frequency of $\sim 3\muhz$; frequencies below this limit
are not considered in any of our analysis.\footnote{We do, however, identify
red giants with solar-like oscillations at frequencies $\sim 3\muhz$,
but the measured frequencies are upper limits and are not assigned errors.} Next, we fill in small gaps in the light curve
of up to three consecutive points with linear interpolation, and remove $4\sigma$ outliers. This procedure
results in a smoother power spectrum and less contamination from the
spectral window, without biasing global oscillation parameters
\citep{stello+2015}. We will see, however, that for some stars, additional measures are required to account for spectral window effects. We then calculate a power spectrum of the resulting light curve with a Lomb-Scargle periodogram \citep{scargle1982}.
\begin{figure}
\includegraphics[width=0.5\textwidth]{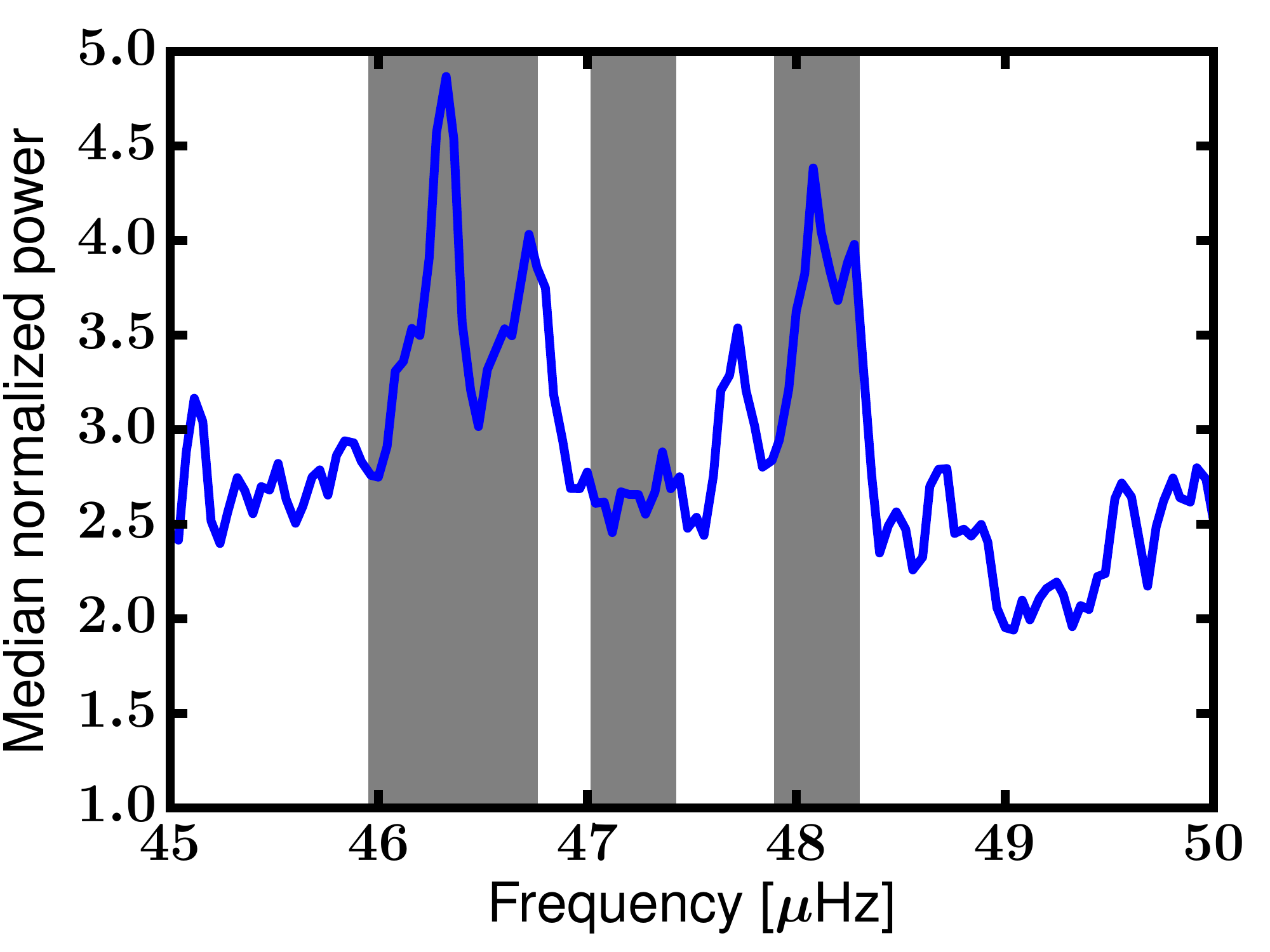}
\caption{The median spectrum for all C1 objects. We identify two
  regions particularly affected by {\it K2} noise in VJ light curves: $46.3\muhz
  \pm 0.4\muhz$ (left) and $48.1 \muhz \pm 0.2\muhz$ (right). The
  middle grey shaded region ($47.22 \muhz \pm 0.2\muhz$) corresponds
  to the nominal thruster firing frequency of the spacecraft. These
  regions are treated specially in BAM, as described in the text.}
\label{fig:median_noise}
\end{figure}
\begin{figure}
\centering
\includegraphics[width=0.5\textwidth]{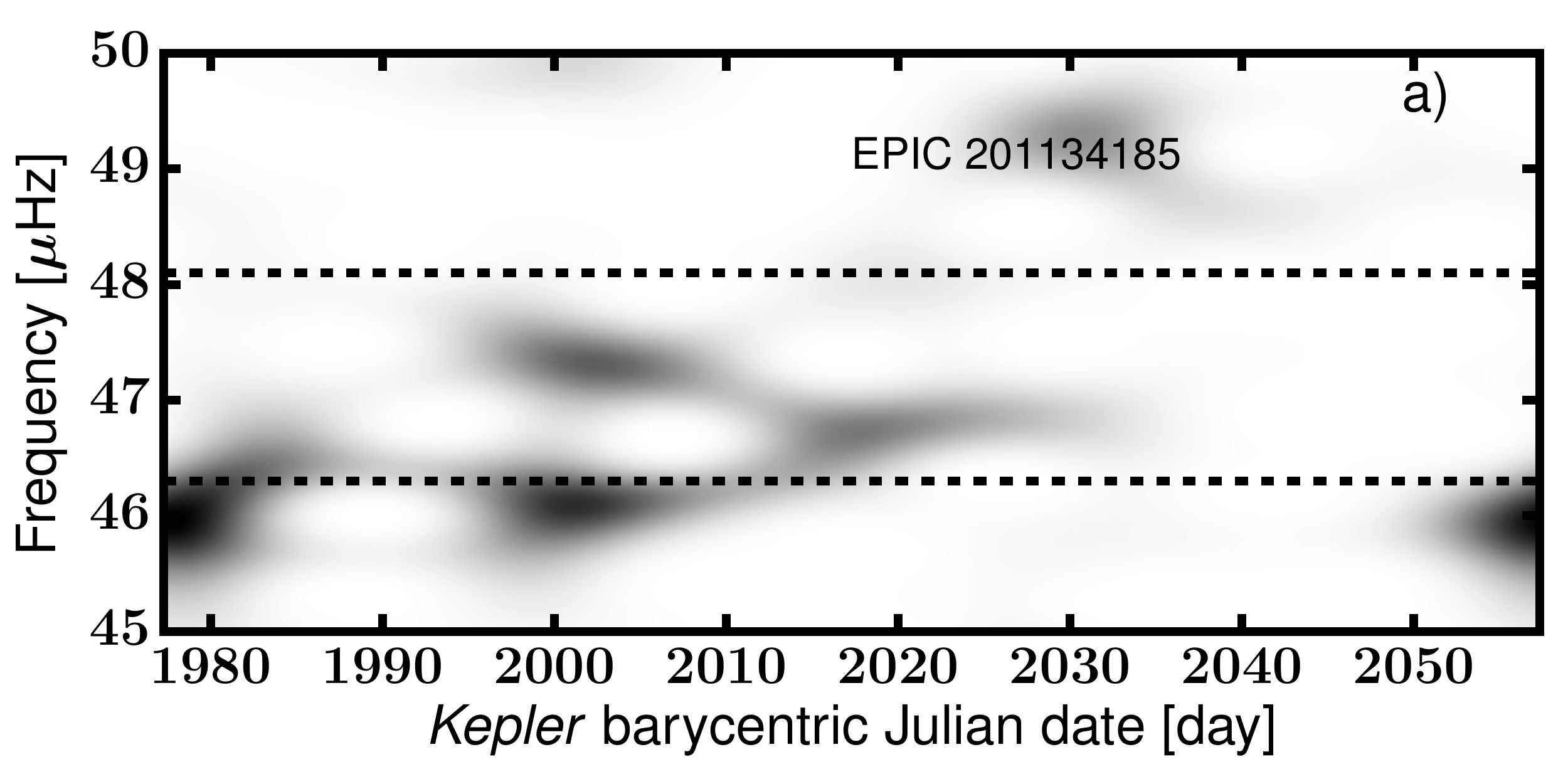}
\\
\includegraphics[width=0.5\textwidth]{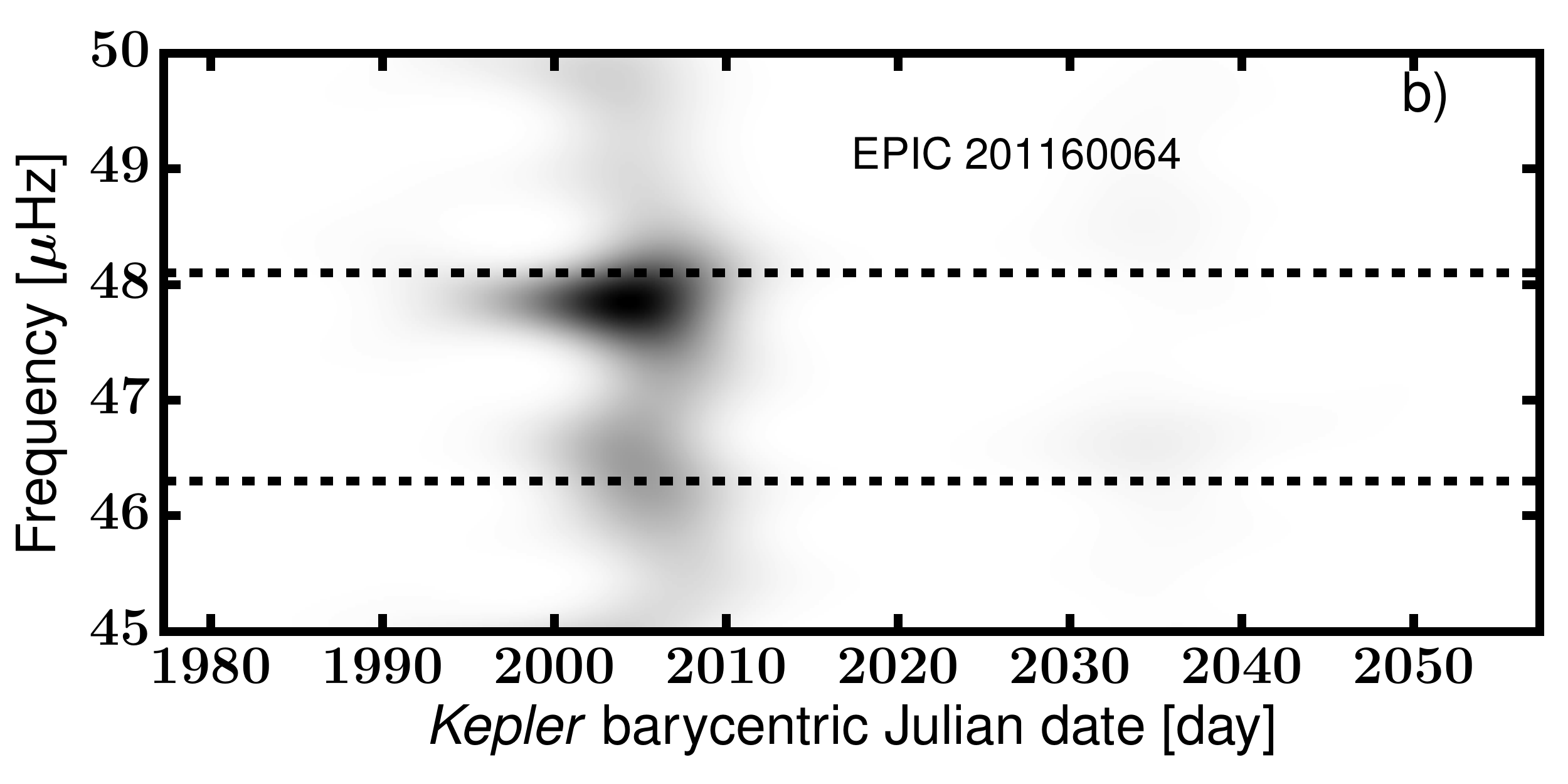}
\caption{Two examples of a wavelet analysis of the
  same frequency range around the nominal thruster firing frequency
  as shown in Figure~\protect\ref{fig:median_noise}, for EPIC 201134185 (top) and EPIC 201160064
(bottom). Clearly the $48.1 \muhz$ and $46.3 \muhz$
  instrumental features seen in the median spectrum
  (Fig.~\ref{fig:median_noise}) are not necessarily both present in
  every light curve at the same level, and do not necessarily persist over the entire time baseline.}
\label{fig:noise_wavelet}
\end{figure}

Despite the efforts to remove systematic errors, the VJ light curves still exhibit non-negligible contamination at frequencies of 48.1 $\muhz$ and 46.3
$\muhz$ due to thruster firings. Generally, we do not find excess power at the
nominal thruster firing frequency of 47.22
$\muhz$. Figure~\ref{fig:median_noise} shows a median power spectrum across all GAP C1 spectra  (8630
spectra in total) in
a region around the thruster firing frequency. To calculate this spectrum, we normalized each
spectrum to the white noise level, defined to be the median
power density in a range from 250 $\muhz$ to the Nyquist frequency of
283 $\muhz$.

In order to investigate whether the thruster firing noise
features showed temporal variation over the course of the campaign, we
computed a wavelet periodogram using the \texttt{astroML} library \citep{astroML}. The
chosen wavelet has the form
\[
w(t, t_0, f_0, Q) \propto e^{-[f_0 (t - t_0) / Q]^2}e^{2 \pi i f_0 (t - t_0)},
\]
where $t_0$ and $f_0$ are the time and frequency of the 2D wavelet
transform, $t$ is the time coordinate for the entire baseline
considered, and $Q$ is a factor determining the time resolution of the
wavelet transform : $Q \rightarrow \infty$ recovers a Fourier
transform and $Q \rightarrow 0$ yields a wavelet periodogram with
infinite temporal resolution. We set $Q = 30$ for analyzing the noise
feature of interest, which allows for resolving features in time of approximately $1/10$ the baseline of C1, i.e., 8 days.

Two representative wavelet
periodograms for C1 are shown in Figure~\ref{fig:noise_wavelet}. We
find that there are definite temporal structures in the frequency
domain of the {\it K2} thruster firing noise. We note that
C1 light curves reduced by \cite{angus_foreman-mackey_johnson2016}
also exhibit qualitatively similar features.

Given these noise features are present in most of the VJ light curves, we remove the affected regions of the power spectra in
Fourier space by replacing each frequency bin in $0.2\muhz$-wide
regions on either side of $47.2\muhz$ and $48.1\muhz$, and a
$0.4\muhz$-wide region on either side of $46.3\muhz$. We replace the
power density in this region with power drawn from a chi-square distribution scaled to a linear interpolation
between the median power in regions $5 \muhz$ on either side of the
affected regions.
\section{methods}
\label{sec:methods}
After the pre-processing of the power spectrum with power, $A_o(\nu_j)$, at discrete
frequencies, $\nu_j$, which constitutes our data, $D$, we then fit a smooth background component
to the power spectrum, whose sets of parameters, $\thetameso$ and
$\thetagran$, are used as guesses for a
subsequent stage of determining the global asteroseismic parameter
$\numax$ and the other parameters describing the oscillation excess,
$\thetaexcess$, which is finally used to guide fitting the global
asteroseismic parameters related to $\dnu$, $\thetadnu$.

We discuss each step in turn below.
\subsection{Granulation calculation}
\label{sec:gran}
BAM first fits a two-component Harvey-like model that \cite{kallinger+2014} find best describes the smooth background component of {\it Kepler} red giant power spectra:
\begin{align}
  A(\nu_j) &= \left[ W_N(\nu_j) \mathrm{sinc}\left(\frac{\pi}{2}\frac{\
      \nu_j}{\nu_{\mathrm{Nyq}}}\right)\right]^2\sum_{i=1,2} \frac{\sigma_i^2 \tau_i}{1 + (\pi \nu_j \tau_i)^4} + \mathrm{WN} \\
&= A_{\mathrm{meso}}(\nu_j) + A_{\mathrm{gran}}(\nu_j) + \mathrm{WN},
\label{eq:numax}
\end{align}
where WN represents a white noise term, which will
dominate red giant power spectra at high frequencies; $\sigma_i$ are amplitudes of each
so-called Harvey components; and $\tau_i$ are their characteristic time-scales. $A_{\mathrm{meso}}(\nu_j)$ and $A_{\mathrm{gran}}(\nu_j)$ are defined here to be the two Harvey components of the granulation background. The sinc pre-factor with dependence on the Nyquist frequency,
$\nu_{\mathrm{Nyq}}$, arises due to {\it K2}'s finite exposure times,
and $W_N(\nu_j)$ is the spectral window function (see \citealt{kallinger+2014} for more details).

Of the two Harvey-like components, the component at higher
frequency is attributed to granulation, whereby the integrated
light from the stellar disk varies due to convective cell brightness
variations. The lower-frequency component is attributed to
meso-granulation, which is likely due to the variation in
convective cell brightness for cells with sizes around $5-10$ times
that of granular cells (for a review of convection on the stellar
surface, see \citealt{nordlund_stein_asplund2009}). For bookkeeping
purposes, we require that the
second component always be identified with the granulation background
for which $\tau_{\mathrm{meso}} > \tau_{\mathrm{gran}}$ and $\sigma_{\mathrm{gran}}^2 \tau_{\mathrm{gran}} < \sigma_{\mathrm{meso}}^2 \tau_{\mathrm{meso}}$.

We achieve a robust fit to the granulation background by taking
advantage of scaling relations between $\numax$ and the granulation
parameters ($\sigma$ and $\tau$) noted by previous work
\citep[e.g.,][]{kjeldsen&bedding2011,kallinger+2010}. These relations naturally translate
into priors in a Bayesian
framework. We construct priors on the
granulation parameters as detailed in
Table~\ref{tab:numax_priors}. The final prior for a set of trial
parameters is the product of the individual priors according to:
\begin{equation}
  \begin{split}
    P(\thetameso&=\{\sigmameso, \taumeso\}, \thetagran=\{\sigmagran,  \taugran\}, 
    \thetaexcess)\\
    &= P(\sigmameso | \taumeso, \sigmagran, \taugran| \thetaexcess) \\
    &P(\taumeso, \taugran | \sigmagran, \thetaexcess) P(\sigmagran |
    \thetaexcess)P(\thetaexcess)\\
    &= P(\sigmameso|\numax) P(\taumeso|\numax) P(\taugran|\numax) P(\frac{\tau_{\mathrm{meso}}}{\tau_{\mathrm{gran}}})\\
    &P(\sigmagran|\numax),
  \end{split}
\end{equation}
for which we introduce the notation $\thetaexcess$ to indicate parameters describing the solar-like oscillations (as distinguished from the granulation parameters), and whose parameters (other than $\numax$) are defined later. The granulation priors are conditional upon $\numax$, and, in this sense, $\numax$ is a latent variable that defines the relationships among all the granulation parameters.

Subsequently, we define a posterior probability given by
\begin{equation}
  \begin{split}
    P(&\thetameso, \thetagran | D=\{(\nu_j, A_{o}(\nu_j)), j=0,1,2,...\}, \thetaexcess) \\
    &\propto P(\thetameso,
    \thetagran | \thetaexcess) \prod_{j} \left[\frac{1}{A(\nu_j)} \exp\left(-\frac{A_o(\nu_j)}{A(\nu_j)}\right)\right].
\end{split}
\label{eq:logl_harvey}
\end{equation}
Here, $A_o(\nu_j)$ is the observed spectral density and $A(\nu_j)$ is the model given by Equation~\ref{eq:numax}. Note that the above
expression assumes $\chi^2$ statistics and not Gaussian statistics to describe
$A_o(\nu_j)/\langle A(\nu_j) \rangle \sim \chi^2(2)$, where the observed spectrum is critically-sampled and the observed
spectrum is modeled by $A(\nu_j)$. 

Given a Bayesian model for the data, we explore the parameter space with Monte
Carlo Markov Chains (MCMC), as implemented in \texttt{emcee}
\citep{foreman-mackey+2013}, and report best-fitting parameters as the
median of their marginalized posterior distributions, and the
uncertainty as the average of the range around the median encompassing
$64\%$ of the distribution. Of course, the prior factor, $P(\thetameso, \thetagran | \thetaexcess)$ depends on
$\numax$ (see Table~\ref{tab:numax_priors}). We simultaneously fit for
$\numax$ and the background parameters, with a guess for $\numax$ calculated from a smoothed
version of the spectrum, as in the SYD pipeline
\citep{huber+2009}. Note that in this step, the region of power
excess is not explicitly modeled, and so $\numax$ is implemented effectively as a dummy variable for this granulation model fitting stage of the process. The resulting best-fitting parameters are then used as initial guesses
for a more complicated model that adds an additional component to
describe the oscillation excess power, which we describe next.

Ultimately, BAM allows the user to choose which of the priors listed in Table~\ref{tab:numax_priors} are to be used. The results presented in this paper do not use the first four priors of Table~\ref{tab:numax_priors} for this granulation background fitting step, though they are used for the subsequent fitting step that determines $\numax$ and $\pmax$, as described in the next section. The extent to which the priors in Table~\ref{tab:numax_priors} are applied does not significantly affect the resulting $\numax$ value. 

\subsection{$\numax$ and $\pmax$ calculation}
\label{sec:numax}
In the subsequent step, we add another component to the
model such that
\begin{equation}
A_{\mathrm{tot}}(\nu_j) = A_{\mathrm{meso}}(\nu_j) + A_{\mathrm{gran}}(\nu_j) + \pg(\nu_j) + \mathrm{WN},
\label{eq:tot}
\end{equation}
where $\pg$ represents the power excess from solar-like
oscillations, and $A_{\mathrm{meso}}(\nu_j)$, $A_{\mathrm{gran}}(\nu_j)$, and WN are defined in
Equation~\ref{eq:numax}. We model $\pg$ as a Gaussian profile
\begin{equation}
\pg = A_{\mathrm{max}} \left[ W_N(\nu_j) \mathrm{sinc}\left(\frac{\pi}{2}\frac{\nu_j}{\nu_{\mathrm{Nyq}}}\right)\right]^2 \exp\left[-\frac{(\nu_j - \numax)^2}{2 b^2}\right].
\label{eq:excess}
\end{equation}

\begin{table}
  \begin{tabular}{ccc}
    \hline \hline
    Parameter & Prior Distribution & Use \\ \hline
    $\ln \sigmagran$ & $\mathcal{N}(-0.609 \ln \numax + 8.70, 0.165)$
    & Eq.~\ref{eq:logl_harvey} \& Eq.~\ref{eq:logl_numax}\\
    $\ln \taugran$ & $\mathcal{N}(-0.992 \ln \numax - 1.09, 0.0870)$ &
    Eq.~\ref{eq:logl_harvey} \& Eq.~\ref{eq:logl_numax}\\
    $\ln \sigmameso$ & $\mathcal{N}(-0.609 \ln \numax + 8.70, 0.165)$
    & Eq.~\ref{eq:logl_harvey} \& Eq.~\ref{eq:logl_numax}\\
    $\ln \taumeso$ & $\mathcal{N}(-0.970 \ln \numax + 0.00412, 0.970)$ &
    Eq.~\ref{eq:logl_harvey} \& Eq.~\ref{eq:logl_numax}\\
    $\ln \frac{\taumeso}{\taugran}$ & $\mathcal{N}(1.386, 0.316)$ &
    Eq.~\ref{eq:logl_harvey} \& Eq.~\ref{eq:logl_numax}\\
    $\ln b$ & $\mathcal{N}(1.05 \ln \numax -1.91, 0.198)$ &
    Eq.~\ref{eq:logl_numax}\\
    $\ln A_{\mathrm{max}} + \ln b$ & $\mathcal{N}(-1.32 \ln \numax + 14.5, 1.22)$ & Eq.~\ref{eq:logl_numax}\\
  \end{tabular}
  \caption{Priors used for the full power spectrum fit,
    Equation~\ref{eq:logl_numax}, adapted from
    \protect\cite{kallinger+2010}. The notation $\mathcal{N}(a, b)$
    indicates a Gaussian distribution with mean $a$ and standard
    deviation $b$. Whether or not a given prior
    enters into Equation~\ref{eq:logl_harvey} or Equation~\ref{eq:logl_numax} is
    indicated in the final column.}
  \label{tab:numax_priors}
\end{table}

Our prior is now:
\begin{equation}
  \begin{split}
    P(&\thetameso, \thetagran, \thetaexcess=\{A_{\mathrm{max}},
    \numax, b\})  = \\
    &P(\sigmameso | \taumeso, \taugran, \sigmagran, \thetaexcess)\\
  &P(\taumeso, \taugran | \sigmagran, \thetaexcess) P(\sigmagran | \thetaexcess) P(\thetaexcess)\\
  &= P(\sigmameso|\numax) P(\taumeso|\numax) P(\taugran|\numax) P(\frac{\tau_{\mathrm{meso}}}{\tau_{\mathrm{gran}}})\\
  &P(\sigmagran|\numax) P(b, A_{\mathrm{max}}, \numax)\\
  &=P(\sigmameso|\numax) P(\taumeso|\numax) P(\taugran|\numax)
  P(\frac{\tau_{\mathrm{meso}}}{\tau_{\mathrm{gran}}\
  })\\
  &P(\sigmagran|\numax) P(b | \numax) P(A_{\mathrm{max}}, b | \numax).
  \end{split}
\label{eq:prior_numax}
\end{equation}

We construct a posterior probability given by:

\begin{equation}
  \begin{split}
P(\thetameso, \thetagran, \thetaexcess | D) &\propto P(\thetameso, \thetagran,
\thetaexcess) \\
&\prod_{j} \left[\frac{1}{A_{\mathrm{tot}}(\nu_j)} \exp \left(-\frac{A_o(\nu_j)}{A_{\mathrm{tot}}(\nu_j)}\right) \right].
\end{split}
\label{eq:logl_numax}
\end{equation}

In this case, the total prior is a product over all priors listed in
Table~\ref{tab:numax_priors}. By first fitting the parameters of the
granulation as described in \S\ref{sec:gran} and subsequently using
these as priors for the fit involving both the granulation model and the
Gaussian excess, we reduce the burn-in time and the
chance of getting stuck at local maxima. It will also make more convenient our oscillator selection process, described in \S\ref{sec:selection}.

\begin{figure*}
\centering
\includegraphics[width=0.5\textwidth]{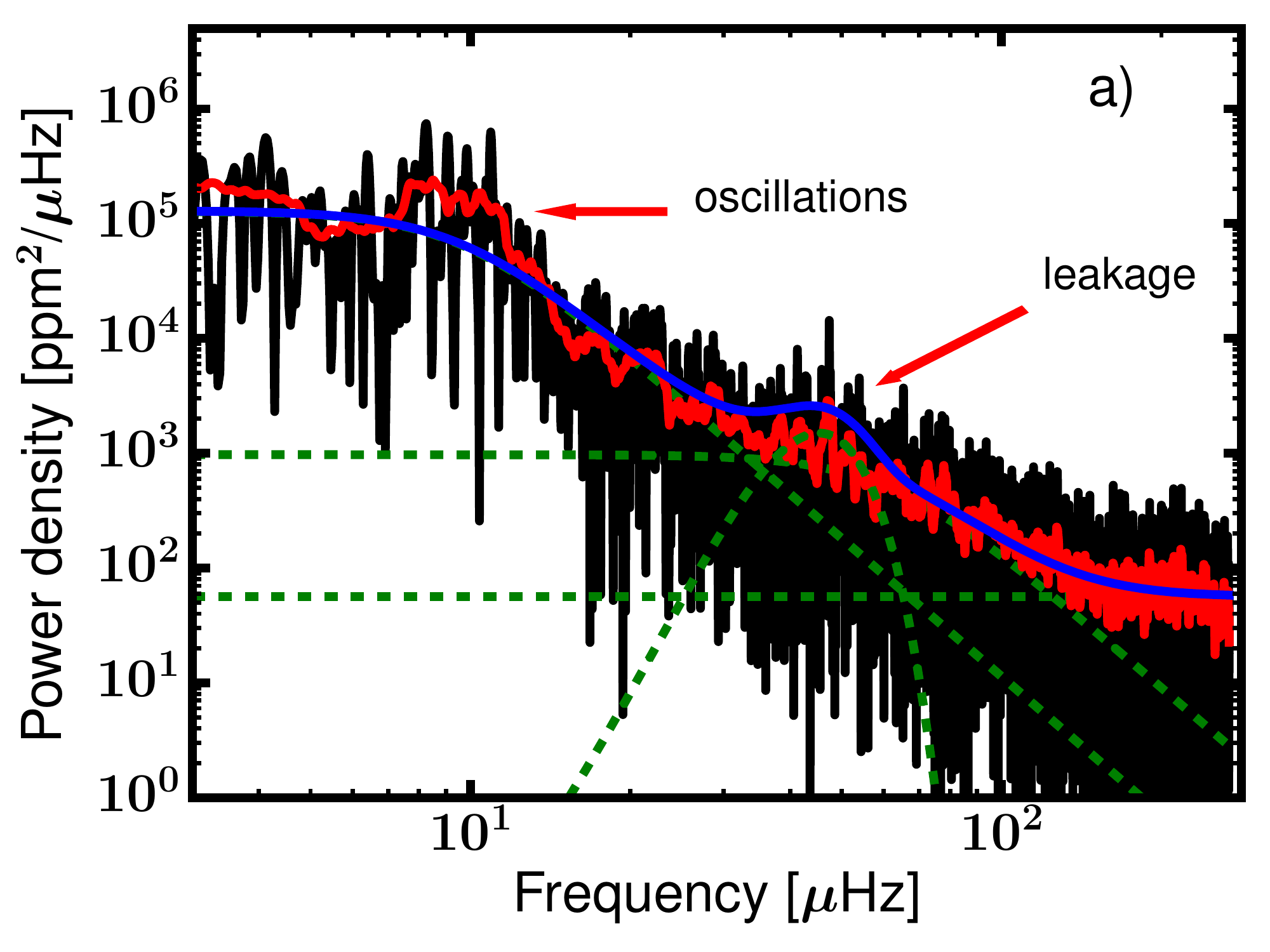}
\includegraphics[width=0.5\textwidth]{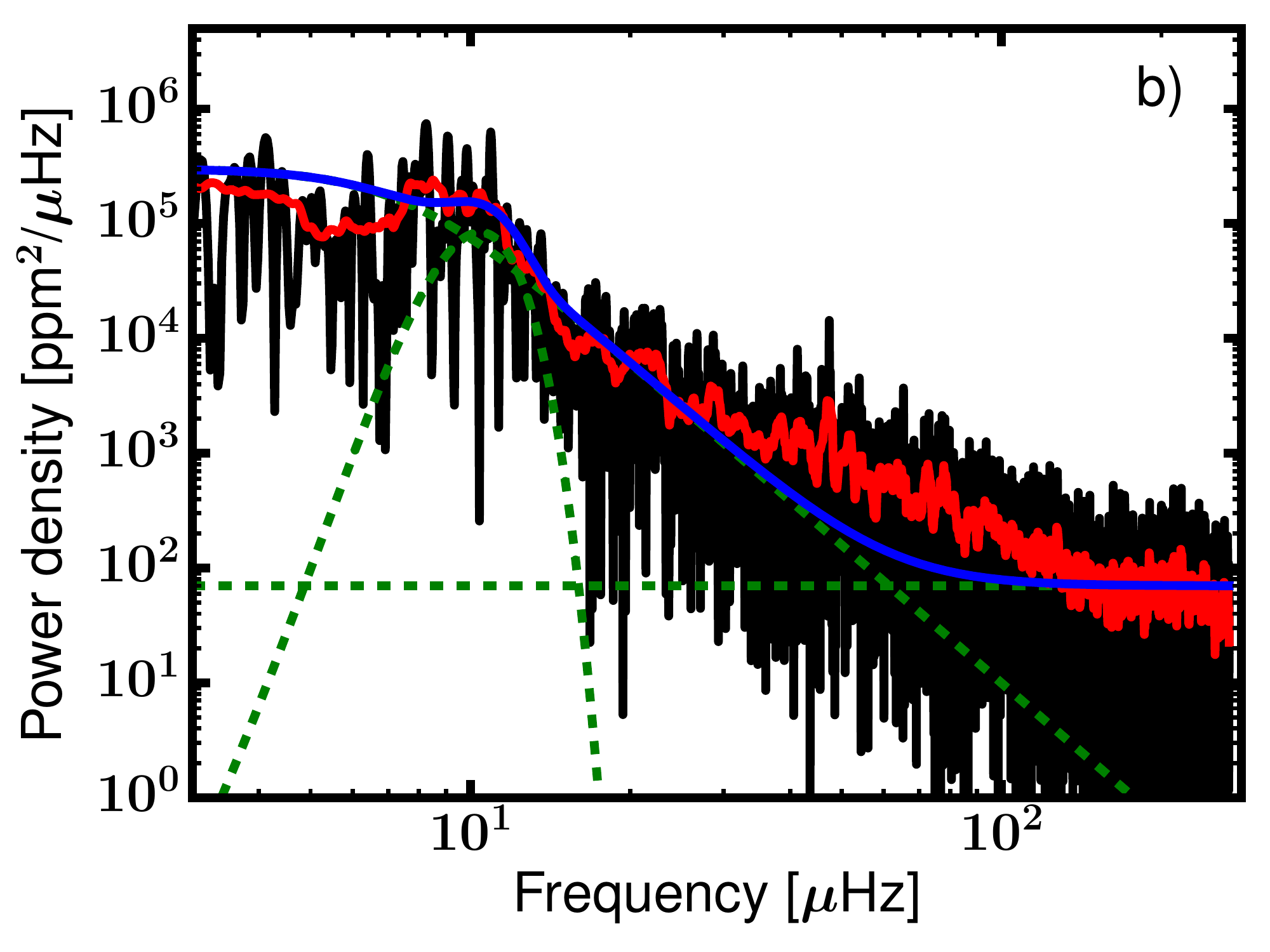}
\caption{Raw (black) and smoothed (red) power spectrum of EPIC
  201186616, and model fits without (a) and with (b) spectral window
  corrections (blue). Each component of the models is shown in green
  dashed curves (white noise, Gaussian excess, and Harvey components). The
  meso-granulation component does not contribute significantly to the fit upon
  spectral window correction, and so is not shown in (b).}
\label{fig:numax_ex}
\end{figure*}

\subsection{Low frequency oscillators}
We find that objects oscillating at frequencies $\numax \lesssim
15\muhz$ exhibit significant spectral leakage at frequencies $30 \muhz \lesssim \nu \lesssim 100 \muhz$, often confusing the pipeline to fit a $\numax$ at the location of the leakage, as shown in Figure~\ref{fig:numax_ex}a.
We correct for this leakage at each step in our MCMC chains: for each trial model granulation spectrum (Eq.~\ref{eq:numax}), we compute an amplitude spectrum, with each
frequency in the spectrum being assigned a random spectral phase. This amplitude spectrum is
then convolved with the spectral window, and squared to yield a power
spectrum (see \citealt{murphy+2013a} for a worked example of how to contend with the spectral window in the context of asteroseismology, specifically). A lightly smoothed version of this convolved granulation power spectrum is added to the power excess term to create a model of the power spectrum that takes into account spectral leakage. This model is then fitted to the observed power spectrum
within the Bayesian framework. Note that the trial power excess term is not convolved with the window function, as it turns out it adds minimally to the spectral leakage compared to the granulation background, and it can lead to unstable fits in which the entire spectrum is modeled as a Gaussian excess plus its resulting spectral leakage. We find that this procedure
results in correct $\numax$ identifications for $\numax \lesssim
15$. Correcting for spectral leakage results in a statistically
significant difference in fitted
granulation parameters for low frequency oscillators
(Fig.~\ref{fig:numax_ex}b; note difference in shape of blue curve
in regions dominated by granulation).

A caveat for these stars is that the lowest
$\numax$ ($ \numax \lesssim 4 \muhz$) values likely represent upper limits for $\numax$ because the {\it K2} resolution prevents an unambiguous
determination of $\numax$. Indeed, at frequencies near $\sim 3\muhz$,
there may only be three modes visible \citep[e.g.,][]{stello+2014}, which limits the precision
with which a central $\numax$ may be defined using the Gaussian
to model oscillation excess (Eq.~\ref{eq:excess}). 

\subsection{$\dnu$ calculation}
\label{sec:dnu}
  \begin{table}
    \begin{tabular}{cc}
      \hline \hline
      Parameter & Prior Distribution \\ \hline
      $\delta_{01}$ & $\mathcal{N}(-0.025 \Delta \nu^{a}, 0.1\dnu^{b})$ \\
      $\delta_{02}$ & $\mathcal{N}(0.121 \Delta \nu^{a} + 0.047^{a}, 0.1\dnu^{b})$ \\
      $A_0$ & $\mathcal{N}(1.0^{c}, 0.15^{b})$ \\
      $A_1$ & $\mathcal{N}(0.5^{c}, 0.15^{b})$ \\
      $A_2$ & $\mathcal{N}(0.8^{c}, 0.15^{b})$ \\
      $\mathrm{FWHM}_0$ & $\mathcal{U}(0.035\dnu_{\mathrm{guess}}^{b}, 0.45\dnu_{\mathrm{guess}}^{b})$ \\
      $\mathrm{FWHM}_1$ & $\mathcal{U}(0.035\dnu_{\mathrm{guess}}^{b}, 0.9\dnu_{\mathrm{guess}}^b)$ \\
      $\mathrm{FWHM}_2$ & $\mathcal{U}(0.035\dnu_{\mathrm{guess}}^{b}, 0.45\dnu_{\mathrm{guess}}^b)$ \\
      $\Delta \nu$ & $\mathcal{N}(\dnu_{\mathrm{guess}}^{b}, 0.15^{b}\dnu_{\mathrm{guess}})$ \\
      $C$ & $\mathcal{U}(0.001^b, 0.1^b)$ \\
    \end{tabular}
    \caption{Priors used for the fit to $\dnu$,
      Equation~\protect\ref{eq:dnu}. The notation $\mathcal{U}(a, b)$
      indicates a uniform distribution between $a$ and $b$. Priors adapted from
      \protect\cite{huber+2010}$^{a}$; Current work$^{b}$;
      and
      \protect\cite{stello+2016}$^{c}$. $\dnu_{\mathrm{guess}}$ is the
      expected $\dnu$ given a $\numax$, from
      \protect\cite{stello+2009}: $\dnu_{\mathrm{guess}} \equiv 0.263 \numax^{0.772}$.}
    \label{tab:dnu_priors}
  \end{table}

\begin{figure}
\centering
    \includegraphics[width=0.5\textwidth]{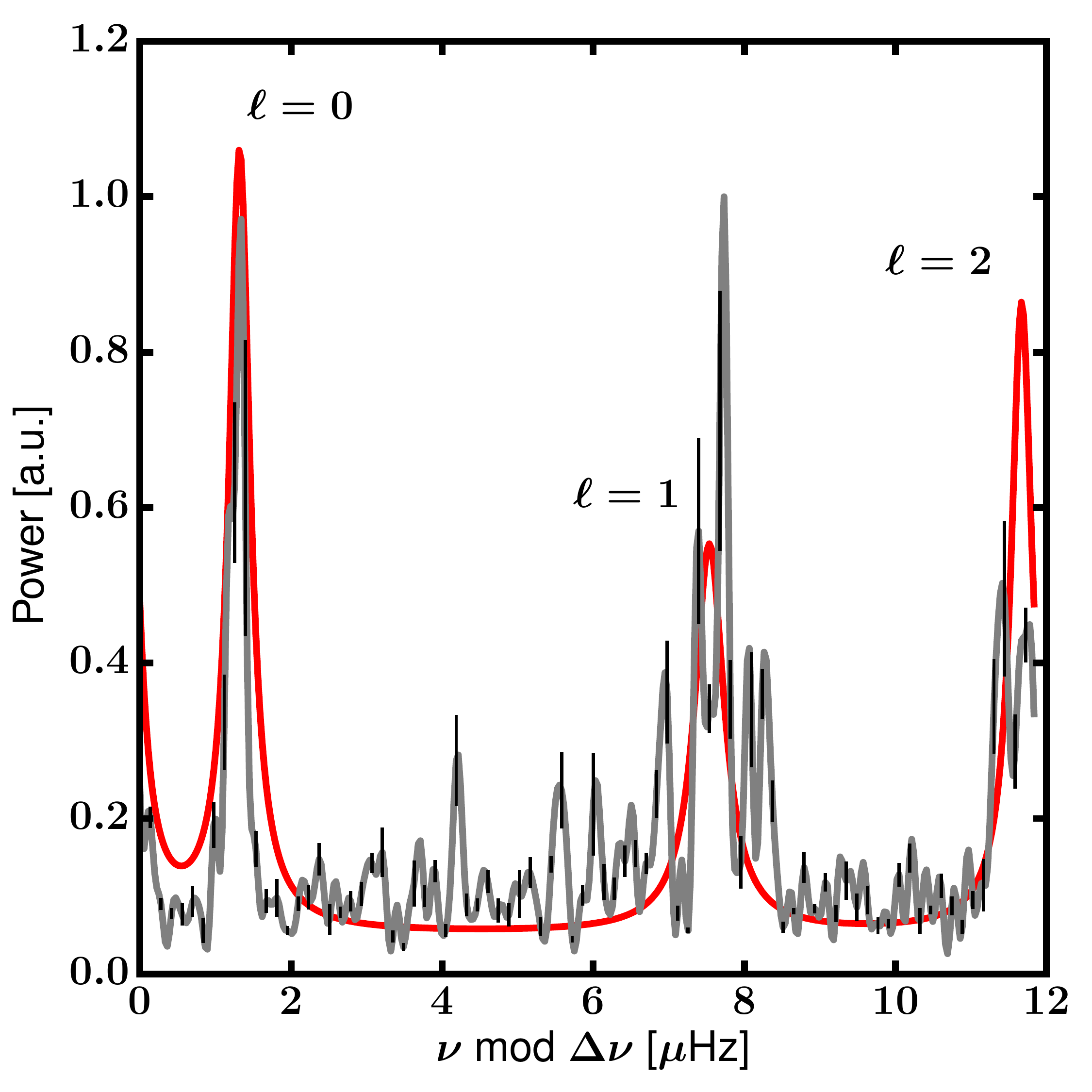}\\
\caption{An example fit by BAM to the folded central power spectrum,
  with best-fitting model (red) and data (grey); black error bars are
  calculated as described in the text, of which every
  $5^{\mathrm{th}}$ is shown, for clarity.}
\label{fig:dnu_ex}
\end{figure}

We furthermore take advantage of the correlation between $\numax$ and
$\dnu$ to place a prior on $\dnu$ in the same way we place priors on
granulation parameters described in \S\ref{sec:gran} \& \S\ref{sec:numax}. Because of the short duration
of {\it K2} light curves ($\sim 80$ days), individual modes may not be
well-resolved, and therefore the large frequency separation can be
difficult to measure. BAM measures $\dnu$ in two independent ways: one using the SYD
autocorrelation method (see \citealt{huber+2009}), and the other using the
$\dnu$-folded power spectrum centered around $\numax$ and extending on
$3\dnu$ on either side, as
shown in Figure~\ref{fig:dnu_ex}. The background contribution from the
Harvey components of the model are divided out, and the folded power
spectrum is computed by folding the spectrum on $\dnu$, where each bin of the folded spectrum contains the sum over the power by folding the spectrum $3\dnu$ on either side of $\numax$ by $\dnu$; the bins are then normalized such that the highest peak of the folded power spectrum is unity. For the majority of red giants the folded
spectrum shows three broad oscillation power excess regions
corresponding to the radial, dipole, and quadrupole modes. We do not fit an octopole mode component because
its low power usually makes it undetectable in {\it K2} data. We obtain
$\dnu$ from this diagram by modeling it using three Lorentzian profiles, appropriate for solar-like oscillation modes, corresponding to the radial ($\ell = 0$),
dipole ($\ell = 1$), and quadrupole ($\ell = 2$) modes, as follows :
\begin{equation}
\begin{split}
A_{\mathrm{folded}}(\nu_j, (\nu_{\ell}, A_{\ell},
\mathrm{FWHM}_{\ell})_{\ell=0,1,2}, \dnu, C) = \\
\sum_{\ell=0}^{\ell=2} \frac{A_\ell}{1.0 +
  \frac{[(\nu_j\ \mathrm{mod}\ \dnu) -
      \nu_{\ell}]^2}{\mathrm{FWHM}_{\ell}^2/4}} + C.
\end{split}
\label{eq:dnu}
\end{equation}
$C$ is a
constant to model the imperfections when removing the background level
in the vicinity of $\numax$. The
frequencies of the modes, $\nu_{\ell}$, in the folded central power spectrum are given by :
\begin{align*}
\nu_0 &\equiv \epsilon\\
\nu_1 &= \nu_0 - \frac{1}{2}{\dnu} + \delta\nu_{01}\\
\nu_2 &= \nu_0 - \delta \nu_{02}.
\end{align*}
The positions of the non-radial modes with respect to the
radial mode, $\epsilon$, thus follow standard
definitions \citep[e.g.,][]{bedding&kjeldsen2010}, such that a given
mode in the spectrum has a frequency, $\nu$,  given by $\nu \approx
\dnu(n+\ell/2+\epsilon)$, where $n$ is the radial order of the mode.

Placing priors on the above
parameters as detailed in Table~\ref{tab:dnu_priors} following the
procedure in \S\ref{sec:gran} \& \S\ref{sec:numax} of the form
\begin{equation*}
  \begin{split}
  P(\thetadnu&=\{(\delta_{01}, \delta_{02}), (A_0, A_1, A_2),\\
  &(\mathrm{FWHM}_0, \mathrm{FWHM}_1, \mathrm{FWHM}_2), \dnu, C\} |
  \thetaexcess)\\
  &= P((\delta_{01}, \delta_{02}),(A_0, A_1, A_2),\\
  &(\mathrm{FWHM}_0, \mathrm{FWHM}_1,
  \mathrm{FWHM}_2), \Delta \nu | b, A_{\mathrm{max}}, \numax)\\
  &= P((\delta_{01}, \delta_{02}) | \numax) P((A_0,
  A_1, A_2) | \numax)\\
  &P((\mathrm{FWHM}_0, \mathrm{FWHM}_1, \mathrm{FWHM}_2)
  | \numax) P(\Delta \nu | \numax)
  \end{split}
\end{equation*}
yields a posterior probability
\begin{equation}
  \begin{split}
P(\thetadnu | D, \thetaexcess) \propto P(\thetadnu | \thetaexcess)
\\ \prod_{j} \left[\frac{A_{\mathrm{o, folded},j}(\nu_j, \dnu)^{n_j-1}}{A_{\mathrm{folded},j}(\nu_j, \thetadnu)^{n_j}} \exp \left(-n_j\frac{A_{\mathrm{o, folded}}(\nu_j, \dnu)}{A_{\mathrm{folded}}(\nu_j, \thetadnu)}\right) \right]
\end{split}
\label{eq:logl_dnu}
\end{equation}
where we use the statistics for an averaged spectrum derived in \cite{appourchaux2003}. $A_{o, \mathrm{folded}}(\nu_j, \dnu)$ is the power at frequency bin $\nu_{j}$ in
the observed folded spectrum for a given
$\dnu$, and is a function of $\dnu$: depending on $\dnu$, the folding process will distribute the power in frequency bins, $A_{\mathrm{o, folded}}(\nu_j, \dnu)$, differently. In practice what this requires is re-computing the folded spectrum for each trial $\dnu$ in our MCMC. $A_{\mathrm{folded}}(\nu_j, \thetadnu)$ is the model for the folded spectrum (Eq.~\ref{eq:dnu}), and
$n_j$ is the number of points that went into the sum over power for that bin in the folded power spectrum.

Using the folded spectrum is particularly useful for
determining $\dnu$ from {\it K2} data because individual mode frequencies are
not very well resolved. What complicates the recovery of $\dnu$ in the
presence of degraded spectral resolution is that observed mode
amplitudes and phases (and hence frequencies) are not stable with time, and have
intrinsic scatter. This is because the oscillations are
stochastically-driven and damped \citep[e.g.,][]{woodard1984}, which
causes continuous variation in the centroid of mode frequencies and
their amplitudes. The random behavior of the stochastic mode
profile can only be mitigated by averaging spectra that are
independent in frequency or in time \citep[for a review
  of power spectrum statistics in the context of solar-like
  oscillations, see][and references
  therein]{anderson_duvall&jefferies1990}. The folded spectrum approach therefore effectively averages out the random behavior of
the modes and increases their signal-to-noise, and is what would be called an `$m$-averaged' spectrum
\citep{anderson_duvall&jefferies1990} in the context of solar
modes.

To find the optimal $\dnu$, we start with a guess value derived
from the $\dnu$--$\numax$ relation by \cite{stello+2009}
\begin{equation}
\dnu_{\mathrm{guess}} = 0.263\numax^{0.772}.
\end{equation}

We determine best-fitting values by MCMC, in which $\dnu$ is
constrained to be $0.7\dnu_{\mathrm{guess}} < \dnu <
1.3\dnu_{\mathrm{guess}}$ and apply priors as described in
Table~\ref{tab:dnu_priors}. BAM returns $\dnu$ values for stars
for which there is agreement to within 2$\sigma$
with $\dnu$ computed using the SYD autocorrelation method and for
which the
uncertainty on $\dnu$ is less than the spread in the $\dnu$ prior. The
latter requirement captures
information about how reliably the modes have been fit, and serves as a means of determining which stars have more information about
$\dnu$ than our prior choice. Note that BAM's second, separate $\dnu$ value from an autocorrelation approach acts as a sort of second opinion. This autocorrelation $\dnu$ will not in general be the same $\dnu$ that a stand-alone application of the SYD pipeline to the same star would: the autocorrelation method requires a $\numax$ to identify the region of the power spectrum that contains the power excess, and it also requires a removal of the smooth background of the power spectrum, both of which are independent of SYD in this case (for details of the autocorrelation approach to calculating $\dnu$, see \citealt{huber+2009}). We show an example of a model fit to the folded spectrum from this process in Figure~\ref{fig:dnu_ex}.

\begin{figure}
  \centering
  \includegraphics[width=0.5\textwidth]{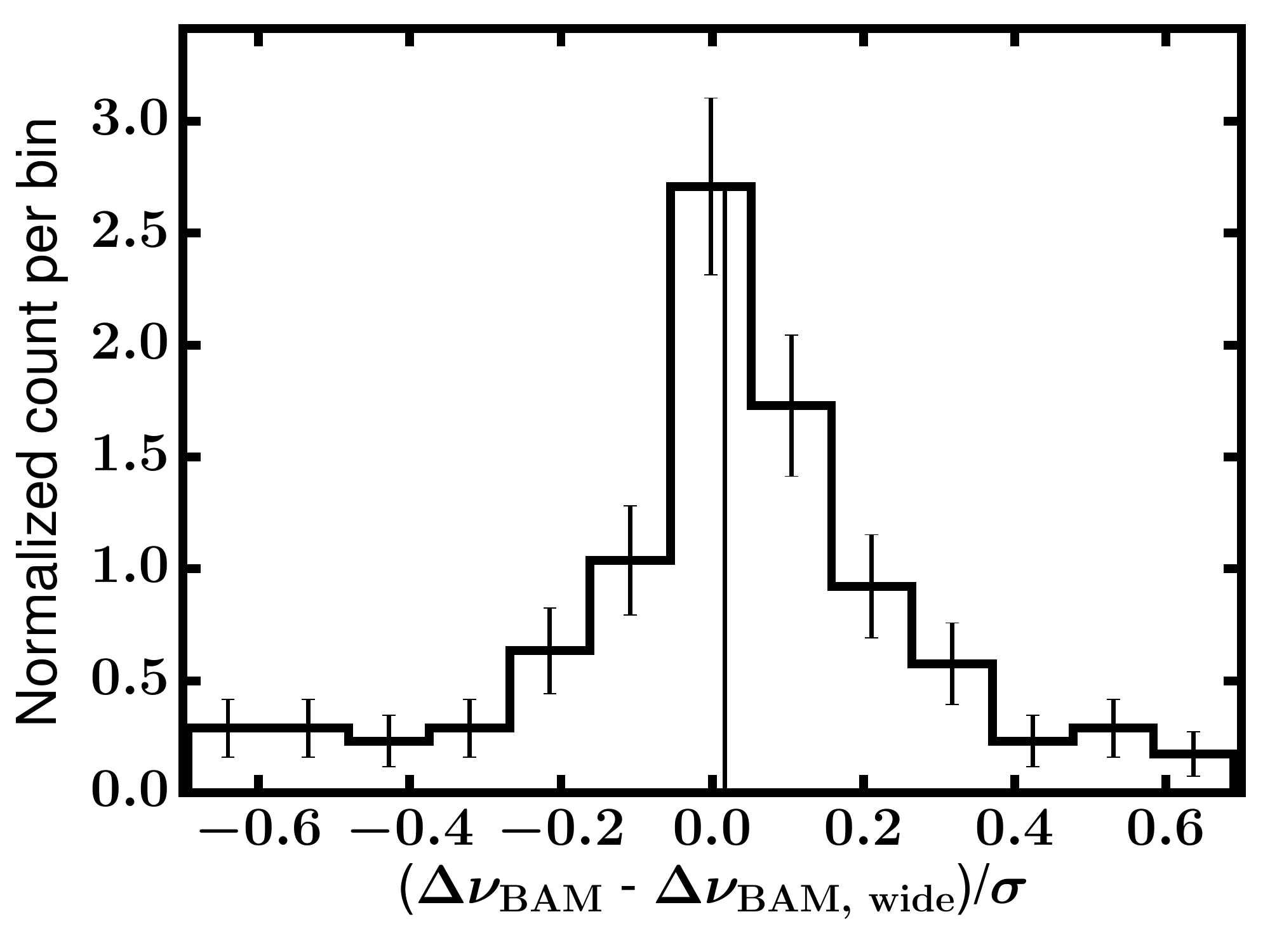}
  \caption{The difference in best-fitting $\dnu$ when using a $\dnu$ prior of
    width $0.9\dnu_{\mathrm{guess}}$ ($\Delta
    \nu_{\mathrm{BAM,\ wide}}$) versus the nominal
    $0.15\dnu_{\mathrm{guess}}$, normalized by the error in the
    difference, $\sigma$; error bars on the
  histogram bins correspond to Poisson uncertainties.The vertical line corresponds to the median of the distribution.This indicates that the
    differences between BAM runs with an expanded prior on $\dnu$
    results in insignificant differences --- $10$ times smaller than
    the error on $\dnu$ --- in the resulting $\dnu$.
  }
\label{fig:dnu_err}
\end{figure}

Importantly, the priors that are placed on $\dnu$ are not too
stringent. We tested the sensitivity of our $\dnu$ results on priors by
increasing the spread in the $\dnu$ prior to
$0.9\dnu_{\mathrm{guess}}$ from $0.15\dnu_{\mathrm{guess}}$ (see Table~\ref{tab:dnu_priors}). For confirmed
oscillators in the C1 {\it K2} GAP sample, our best-fitting
$\dnu$ values are not significantly different when using our fiducial
prior or a widened prior. We show the
difference in best-fitting $\dnu$ using these two different priors in
Figure~\ref{fig:dnu_err}. The spread is less than $0.1\sigma$ for the majority of
objects, indicating that the priors indeed do not significantly impact the
determination of $\dnu$.

\begin{figure}
\centering
\includegraphics[width=0.5\textwidth]{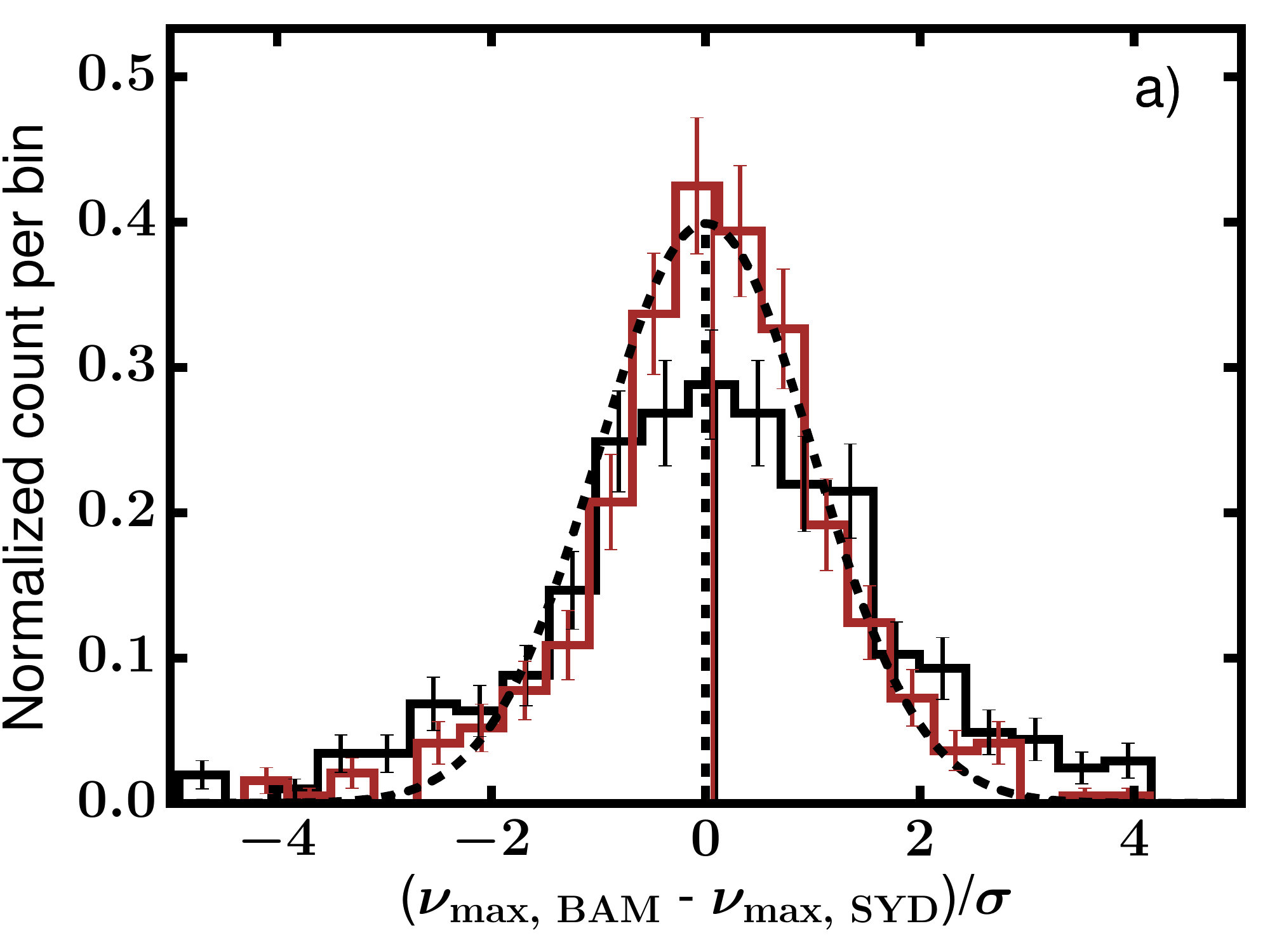}
\includegraphics[width=0.5\textwidth]{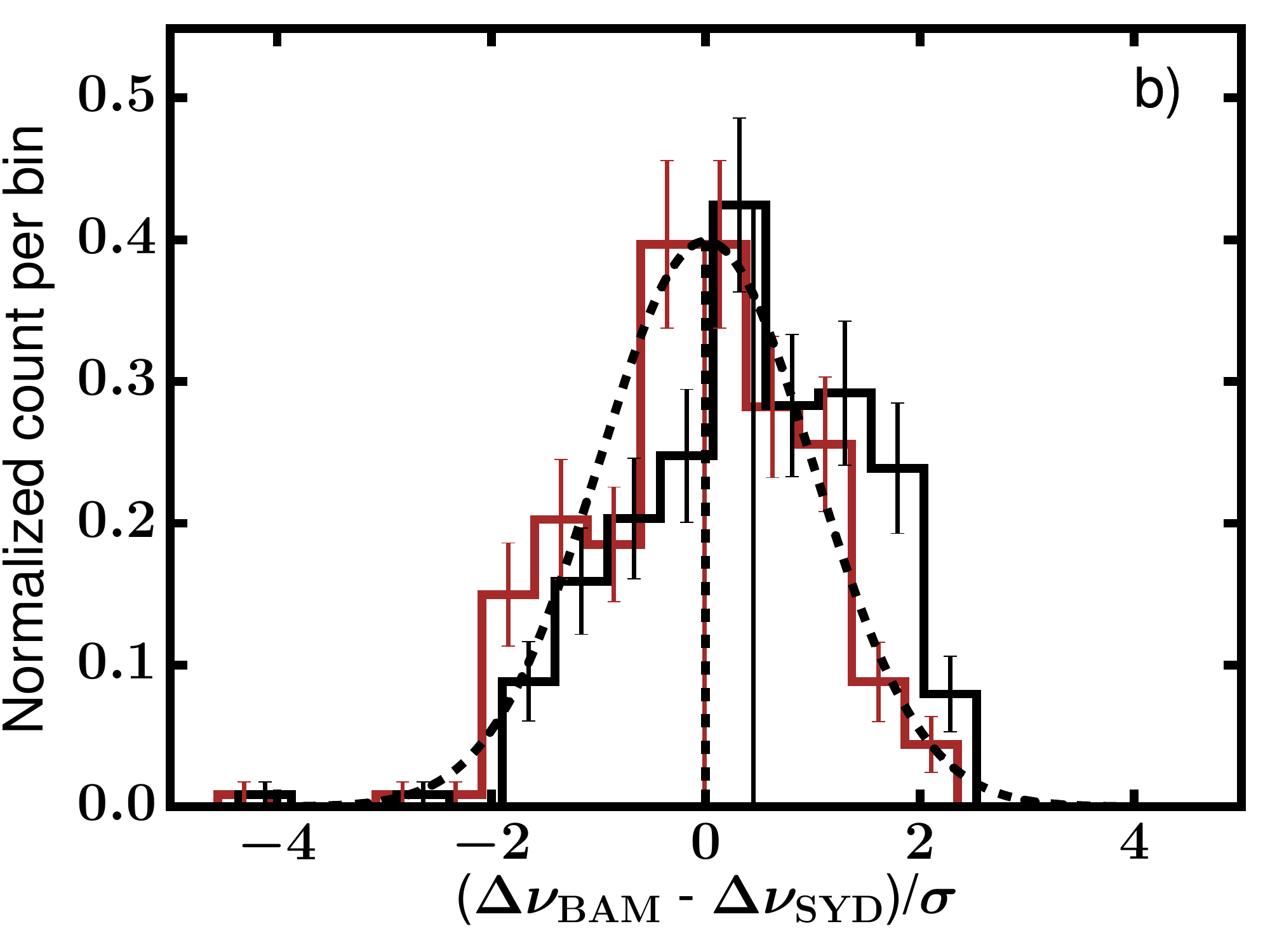}
\caption{Distributions of the differences between BAM and SYD $\numax$ (a)
  and $\dnu$ (b), normalized by the sum in quadrature of their errors,
  $\sigma \equiv \sqrt{\sigma_{\nu_{\mathrm{max,\ BAM}}}^2 +
    \sigma_{\nu_{\mathrm{max,\ SYD}}}^2}$ and $\sigma \equiv
  \sqrt{\sigma_{\dnu_{BAM}}^2 +
        \sigma_{\dnu_{SYD}}^2}$. The medians of both distributions
  are shown as vertical, solid black lines; error bars on the
  histogram bins correspond to Poisson uncertainties. The red distributions in each panel indicate the distributions of differences in BAM and SYD values after systematic differences in central value and/or uncertainties are corrected, according to the text. The dotted curve is a Gaussian, to guide the eye; the vertical dashed line is centered at zero. Stars plotted here are drawn from the C1 GAP sample deemed
  from manual inspection to be definite oscillators
  (see \protect\citealt{stello+2017}) and such that both SYD and BAM as implemented in this work
  returned $\numax$ or $\dnu$ values.}
\label{fig:bam_syd}
\end{figure}

\subsection{Comparison to SYD}
BAM parameters agree favorably with those
computed by other techniques via different pipelines, as demonstrated in
\citep{stello+2017}. As a point of comparison to a well-established asteroseismic pipeline,
Figure~\ref{fig:bam_syd} shows BAM $\numax$ and $\dnu$
values compared to those from SYD for the C1 GAP oscillator
sample. The BAM parameters for this comparison exercise have been re-derived using slightly different methodology than described in the GAP Data Release 1 (GAP DR1) release paper \citep{stello+2017} so as to be consistent with the methodology presented in this work. SYD values for $\dnu$ and $\numax$ are taken directly from GAP DR1. Only giants candidates that were verified to be such by eye in \cite{stello+2017} and that BAM selects as giants according to \S\ref{sec:selection} are considered in this comparison exercise.

The median in the normalized distribution of differences
between BAM and SYD $\dnu$ values for this GAP comparison sample (solid black
vertical line in Figure~\ref{fig:bam_syd}b)
indicates a systematic offset of $\sim 0.6\%$. The red histogram in Figure~\ref{fig:bam_syd}b shows
the $\dnu$ differences distribution if the BAM values are re-scaled downward by $0.6\%$, which brings the distribution
into better alignment with the expected Gaussian.
The median in the distribution of $\numax$ differences indicates a
marginally significant (1$\sigma$) systematic offset between the two
numax scales (solid black vertical line in Figure~\ref{fig:bam_syd}a),
and which corresponds to a difference in BAM and SYD $\numax$ scales
of $\sim 0.2\%$. There does appear to be an under-estimation of either BAM or SYD $\dnu$ values (black histogram in Figure~\ref{fig:bam_syd}a), which is ameliorated by re-scaling the error on the difference upward by $30\%$ (red histogram in Figure~\ref{fig:bam_syd}a).

Given that \cite{kallinger+2014}
found systematic differences of up to $\sim 5\%$ in $\numax$ depending
on the model used for the meso-granulation and granulation background, any small systematic difference in $\numax$ could easily be due to the different treatment of the
background between BAM and SYD. For example, the sinc term in Equation~\ref{eq:numax} is not included
in the SYD pipeline. This difference in methodology could plausibly explain the $0.6\%$ systematic
difference in $\dnu$, as well: the positions of the modes used to measure $\dnu$ will be affected by the choice of the
meso-granulation and granulation background, which are removed before calculating
the folded spectrum.

Apart from these systematic differences, we find
BAM parameters are consistent with SYD to within $\sim 1.53\%$
and $1.51\%$ for $\numax$ and $\dnu$, which correspond to the BAM GAP sample mean fractional errors on $\numax$ and $\dnu$, respectively. There is some ambiguity as to the agreement in $\numax$, where the errors on $\numax$ for either BAM or SYD may be under-estimated by up to $30\%$, given the non-Gaussianity of the $\numax$ difference distribution (black histogram in Figure~\ref{fig:bam_syd}a). Non-Gaussianity in
comparisons across pipelines was also found in \cite{stello+2017}, and
in part is caused by under- and over-estimation of errors in
\textit{K2} asteroseismic parameters (\citealt{pinsonneault+2018}, Zinn et
al., in prep.).

\subsection{Bayesian oscillator selection}
\label{sec:selection}
Because our approach for measuring the oscillation and granulation
parameters will always provide a best-fitting model, even if there is
no solar-like oscillation signal, we still need to determine if a
fit corresponds to a true detection. 
As mentioned in \S\ref{sec:introduction}, BAM's Bayesian approach
means that we can use the parameter fits to determine which stars are,
and are not,
true oscillators. 

This is essentially a problem in model comparison: does the model with a power excess term (Equation~\ref{eq:tot}) describe a star's power spectrum better or does one without power excess (Equation~\ref{eq:numax})? \cite{jeffreys1935} first formalized model comparison in a Bayesian approach using what is now called the Bayes factor, defined to be the ratio of the posterior odds in favor of a model to its prior odds. The Bayes factor derives simply from Bayes theorem, by which the posterior odds of $M_1$ can be written as

\begin{equation}
\frac{P(M_1|D)}{P(M_2|D)} = \frac{P(D|M_1)}{P(D|M_2)}\frac{P(M_1)}{P(M_2)}.
\label{eq:bayes}
\end{equation}

In our case, the probability densities, $P(D|M_1)$ and $P(D|M_2)$ correspond to integrals of Equations~\ref{eq:logl_numax}~\&Equation~\ref{eq:logl_harvey} over all of parameter space, and we assume that, a priori, a star is as likely to be a non-oscillator as an oscillator, in which case the prior odds of $M_1$, $\frac{P(M_1)}{P(M_2)} = 1$. The Bayes factor is defined as $B \equiv \frac{P(D|M_1)}{P(D|M_2)}$.

To compute the Bayes factor, one needs to integrate the conditional probability densities of Equations ~\ref{eq:logl_numax}~\&~\ref{eq:logl_harvey} over all of parameter space. Though these conditional probability densities share the same priors on granulation parameters, $P(\thetameso, \thetagran | \thetaexcess)$, they do not neatly cancel out when computing the Bayes factor because $P(D|M_1)$ and $P(D|M_2$ in Equation~\ref{eq:bayes} are each separate integrals involving these priors. Such integrals are often computationally expensive to do, and analytically intractable. Fortunately, there are various methods available to approximate the Bayes factor \citep[e.g.,][]{green95,chib01,skilling04}. We use the widely-applicable Bayesian Information Criterion \citep[WBIC;][]{watanabe2013} to compute the Bayes factor. This method generalizes the Bayesian Information Criterion \citep{schwarz1978}, such that the WBIC approximates the Bayes factor in the limit of weak priors and with the assumption that the posterior is asymptotically normal:

\begin{equation}
\ln B \approx \Delta_{\mathrm{WBIC}} \equiv 
    <\ln \mathcal{L}_{1}>_{P(\theta|D)} - <\ln \mathcal{L}_{2}>_{P(\theta|D)},
    \label{eq:B}
\end{equation}
where $<>_{P(\theta|D)} $ indicates a mean taken over the modified posteriors of Equations ~\ref{eq:logl_numax'}~\&~\ref{eq:logl_harvey'} (see below), and the likelihoods are from Equations~\ref{eq:logl_numax}~\&~\ref{eq:logl_harvey} ($\mathcal{L}_{1}\equiv \prod_{j} \left[\frac{1}{A_{\mathrm{tot}}(\nu_j)} \exp \left(-\frac{A_o(\nu_j)}{A_{\mathrm{tot}}(\nu_j)}\right) \right]$ and $\mathcal{L}_{2} \equiv \prod_{j} \left[\frac{1}{A(\nu_j)} \exp\left(-\frac{A_o(\nu_j)}{A(\nu_j)}\right)\right]$).

Crucially, the WBIC approach means that the Bayes factor can be computed trivially in a MCMC setting. We compute the means $<\ln \mathcal{L}_{1}>_{P(\theta|D)} $ and $<\ln \mathcal{L}_{2}>_{P(\theta|D)}$ using our two-step MCMC method, recalling that we perform fits to the data both with and without a power excess term (Equations~\ref{eq:tot}~\&Equation~\ref{eq:numax}). For the purposes of approximating the Bayes factor, then, we run each MCMC an additional time, except using modified conditional posteriors so that instead of Equations ~\ref{eq:logl_numax}~\&~\ref{eq:logl_harvey}, we have:

\begin{equation}
  \begin{split}
P(\thetameso, \thetagran, \thetaexcess | D) &\propto P(\thetameso, \thetagran,
\thetaexcess) \\
&\prod_{j} \left[\frac{1}{A_{\mathrm{tot}}(\nu_j)} \exp \left(-\frac{A_o(\nu_j)}{A_{\mathrm{tot}}(\nu_j)}\right) \right]^{\beta}
\end{split}
\label{eq:logl_numax'}
\end{equation}

and

\begin{equation}
  \begin{split}
    P(&\thetameso, \thetagran | D=\{(\nu_j, A_{o}(\nu_j)), j=0,1,2,...\}, \thetaexcess) \\
    &\propto P(\thetameso,
    \thetagran | \thetaexcess) \prod_{j} \left[\frac{1}{A(\nu_j)} \exp\left(-\frac{A_o(\nu_j)}{A(\nu_j)}\right)\right]^{\beta},
\end{split}
\label{eq:logl_harvey'}
\end{equation}

where $\beta \equiv 1/\ln N$, with $N$ being the number of points in the power spectrum being fit. While performing a MCMC fit using posteriors from Equations ~\ref{eq:logl_numax'}~\&~\ref{eq:logl_harvey'} in place of Equations ~\ref{eq:logl_numax}~\&~\ref{eq:logl_harvey}, we save the original likelihoods from Equations~\ref{eq:logl_numax}~\&~\ref{eq:logl_harvey} at each link in our MCMC chains. In the end, we take an average of those likelihoods, insert into Equation~\ref{eq:B}, and in this way compute the Bayes factor.

We interpret the strength of evidence for the Gaussian excess model following \cite{kass_raftery1995}, who recommend that $\ln B > 1$ would indicate positive evidence for the Gaussian excess model. We also require that the
    granulation component be resolved by imposing that the white noise
    be lower than the granulation component power (i.e., that the white noise should not dominate the power spectrum). Note that these selection criteria do not include information about $\dnu$:
identifying excess power corresponding to $\numax$ is easier than identifying $\dnu$, especially in the
presence of mixed modes exhibited in red clump stars. The sample of non-GAP red giants that we will discuss in \S\ref{sec:results} are these candidates that had evidence according to the Bayes factor of exhibiting solar-like oscillations ($\ln B > 1$): 316 giant candidates are chosen in this way from the non-GAP sample of $13016$ objects.

For every star in this sample of oscillating red giant candidates, we confirmed BAM's selections as bona fide giants or not by visual inspection of the power spectra.   We categorized each of BAM's giant candidates into one of three categories: as having 1) a spectrum with oscillation modes that are discernible
individually by eye or with excess power that is conspicuous by eye (`yes'
oscillator); 2) a spectrum with marginal evidence of excess power at a frequency consistent
with the shape of the granulation and meso-granulation components
(`maybe' oscillator); or 3) a spectrum that shows at best very weak evidence of excess power or whose model power spectrum is in clear disagreement with the observed one
(`no' oscillator). The $\numax$ inferred by eye in the `yes' and
`maybe' cases  must be within $3\muhz - 283\muhz$, such that giants that show evidence of a granulation spectrum at low frequencies
are not selected as oscillators if the power excess is not
visible above $3\muhz$. In this discernment process, the amplitude of
the power spectrum, which has a relation to $\numax$ (as formalized,
e.g., in \citealt{kallinger+2014} and in
Table~\ref{tab:numax_priors}), is allowed to be $10$-$50$ times smaller than might be
expected of a giant, to allow for cases where light from a
non-oscillator contaminates the light curve, hence reducing the
fractional brightness variation from granulation and oscillations. This effect can be
significant. For instance, if a foreground dwarf of the same
brightness as a background giant falls on the giant's aperture mask, it would dilute the
signal of the giant's power spectrum by a factor of four.

Upon this visual verification, 31 of BAM's non-GAP giant candidates were certain
oscillators; 73 possible oscillators; and 212 not oscillating giants.

\section{results and discussion}
\label{sec:results}

We apply the BAM pipeline to $13016$ C1 targets with VJ light curves
not in the GAP
sample, which have been selected for a wide range of science programs---mostly detection of planets around dwarfs. 
We identify $31$ 
red giants that have detectable oscillation excesses that satisfy the
BAM selection criteria of \S\ref{sec:selection} and that have been validated by
individual inspection --- $21$ of these are from GO
proposal target lists that did not intentionally target giants. An additional $73$ objects are potential
giants, though can not be definitely confirmed as such. $70$ of these `maybe' cases are from programs that did not intentionally target giants. Combined,
these 104 red giants and red giant candidates represent an $8\%$ increase
in the number of giants 
identified from C1 compared to those from the GAP sample
\cite{stello+2017}, which expressly targeted giants. The global oscillation
parameters and granulation parameters for the
red giants and red giant candidates are given in
Table~\ref{tab:seren}.

\subsection{Completeness and purity of observed non-GAP giants}
\label{sec:completeness}
\begin{figure}
\centering
\includegraphics[width=0.5\textwidth]{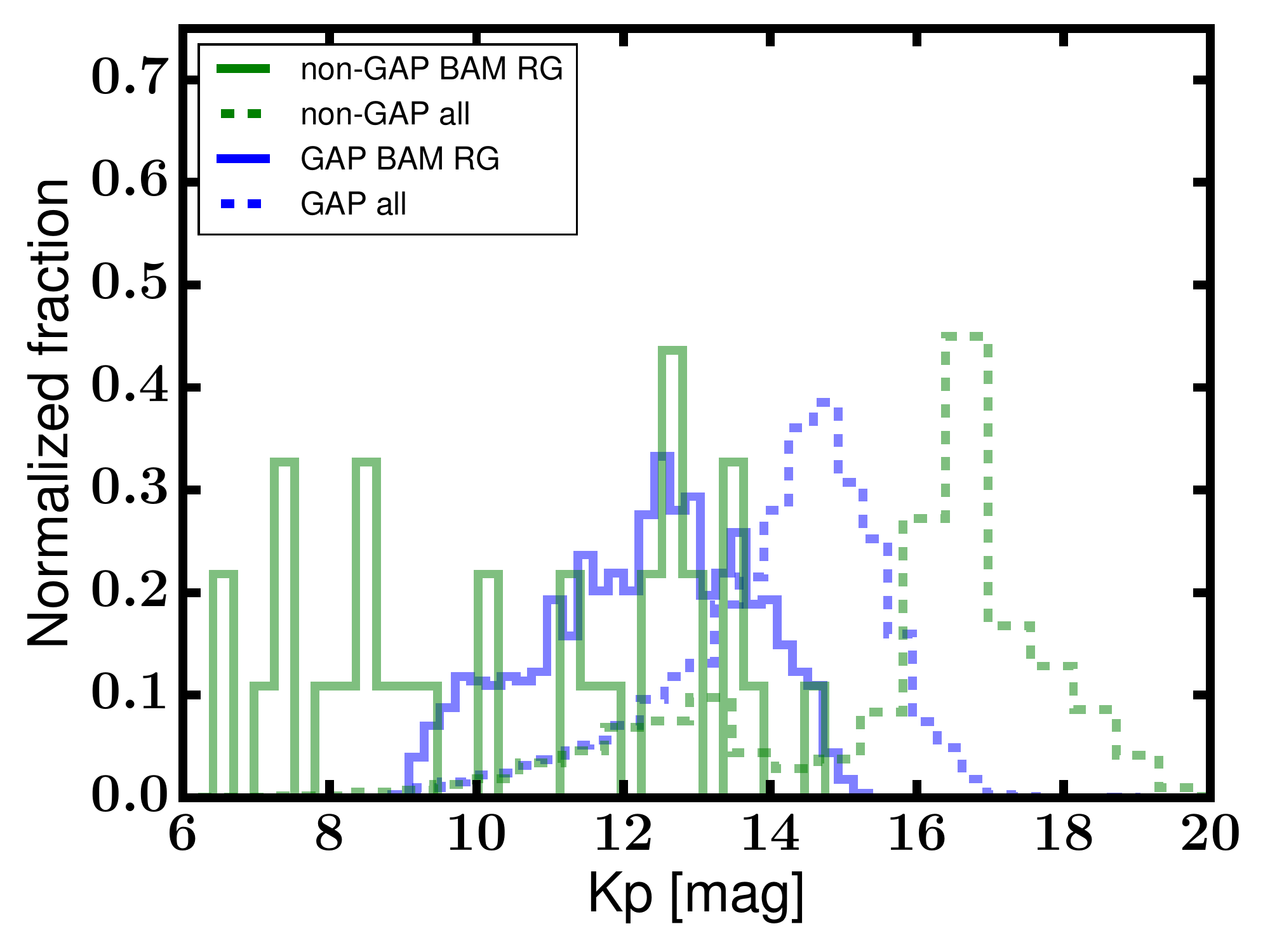}
\caption{Magnitude distribution of the BAM non-GAP giant sample of this
  work (solid green line), compared to all observed non-GAP C1 targets (dashed green line); all GAP targets (dashed blue line); and GAP oscillators from \protect\cite{stello+2017} (solid blue line).}
\label{fig:galpop}
\end{figure}

The magnitude distribution of the stars we find
in this serendipitous sample, shown in Figure~\ref{fig:galpop},
demonstrates that BAM can recover red giant oscillations in {\it K2}
down to $Kp \sim 14$ ($H \sim 12$). All the adopted magnitudes and colors we use in the following are taken from the Ecliptic Input Catalogue \citep[EPIC;][]{huber+2016}.\footnote{A few objects had photometry in the EPIC that did not correspond to the giant in question, and these mismatches were corrected by searching for the nearest, brighter neighbor in the EPIC. The EPIC IDs affected were 201269306, 201472519, and 201724514.} Note that even though the majority of the non-GAP C1 targets
have $Kp \gtrsim 15$ (dashed green), the non-GAP giant sample from this work mostly has $Kp \lesssim 15$ (solid green). This is due to white noise
dominating the spectra of giants at fainter magnitudes, and is the
reason why the number of GAP giants also drops beyond $Kp \gtrsim 13$ (solid
blue). We adopt a conservative $Kp = 13$ as our fiducial completeness limit, whose actual completeness we will test in the next section by comparing to a model of the C1 non-GAP oscillators.

\begin{figure*}
\centering
\includegraphics[width=0.3\textwidth]{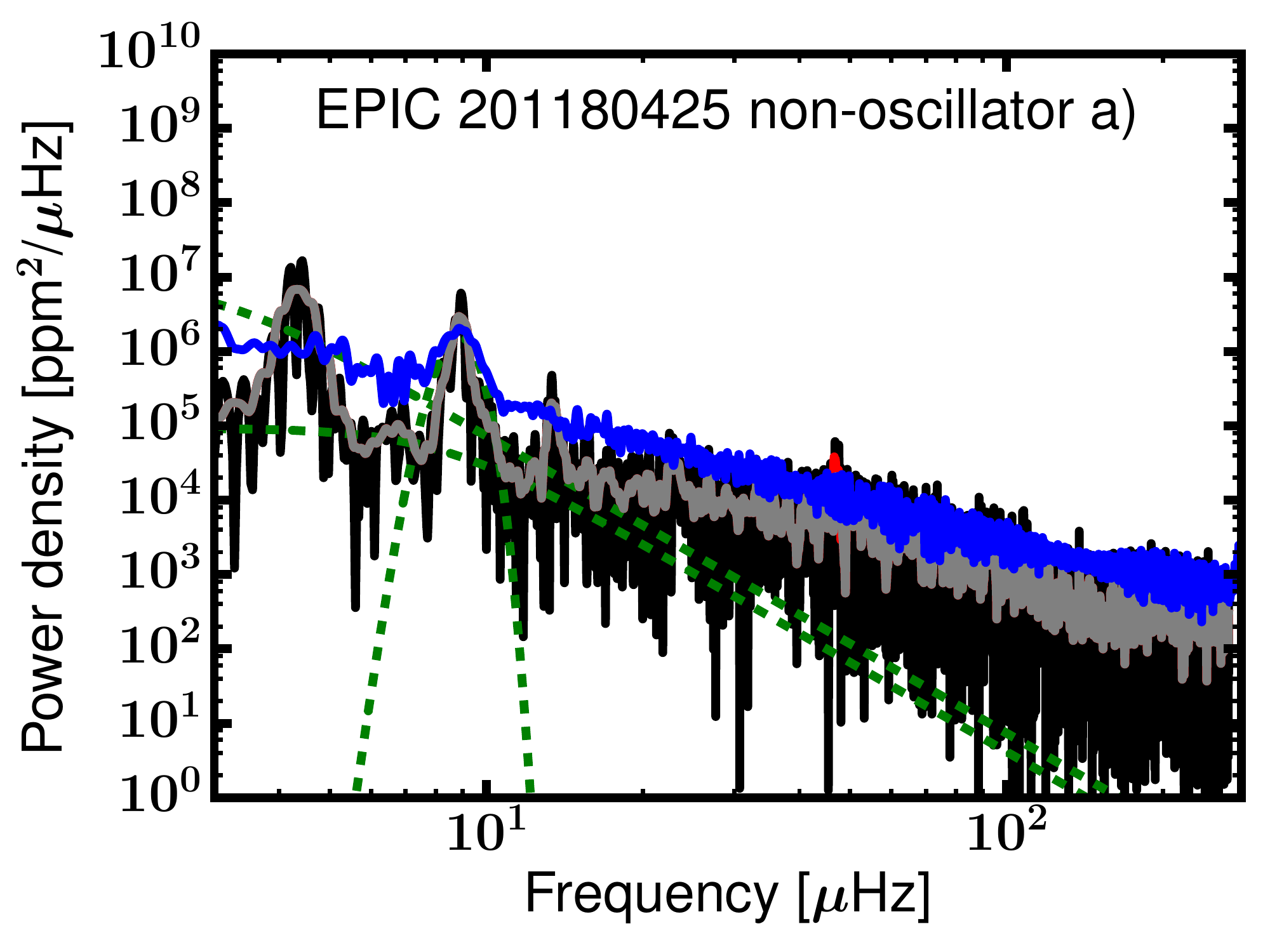}
\includegraphics[width=0.3\textwidth]{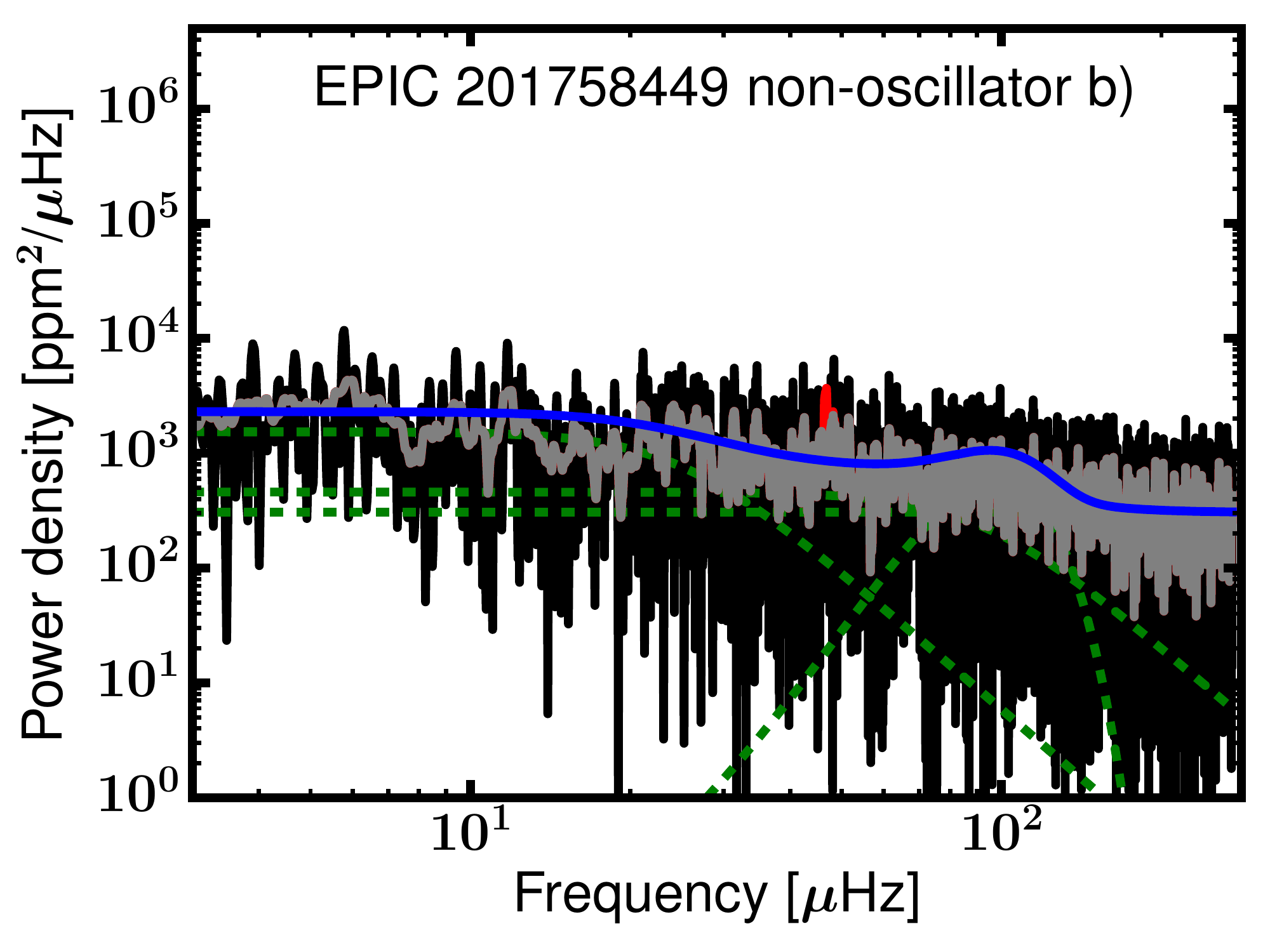}
\includegraphics[width=0.3\textwidth]{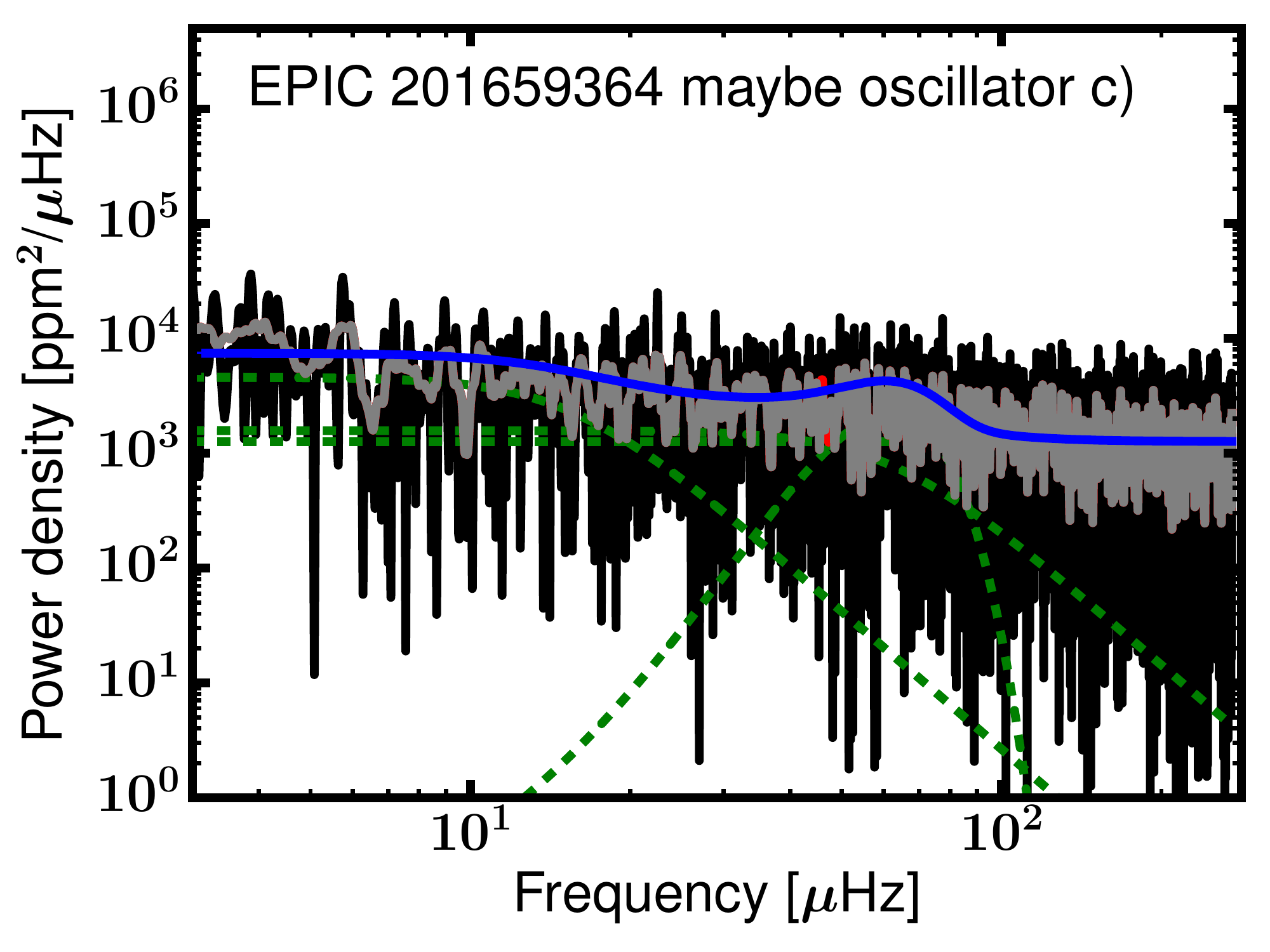}\\
\includegraphics[width=0.3\textwidth]{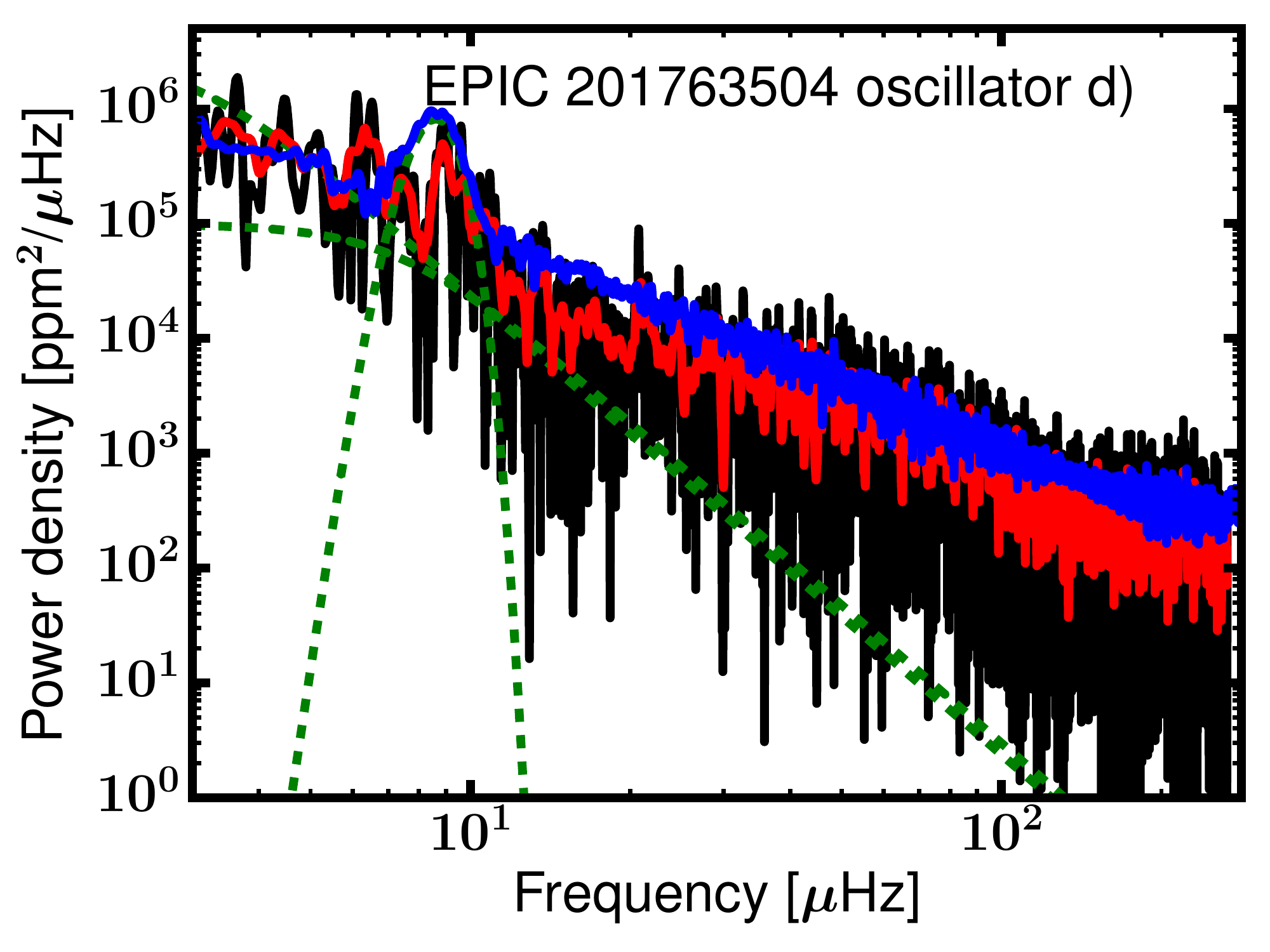}
\includegraphics[width=0.3\textwidth]{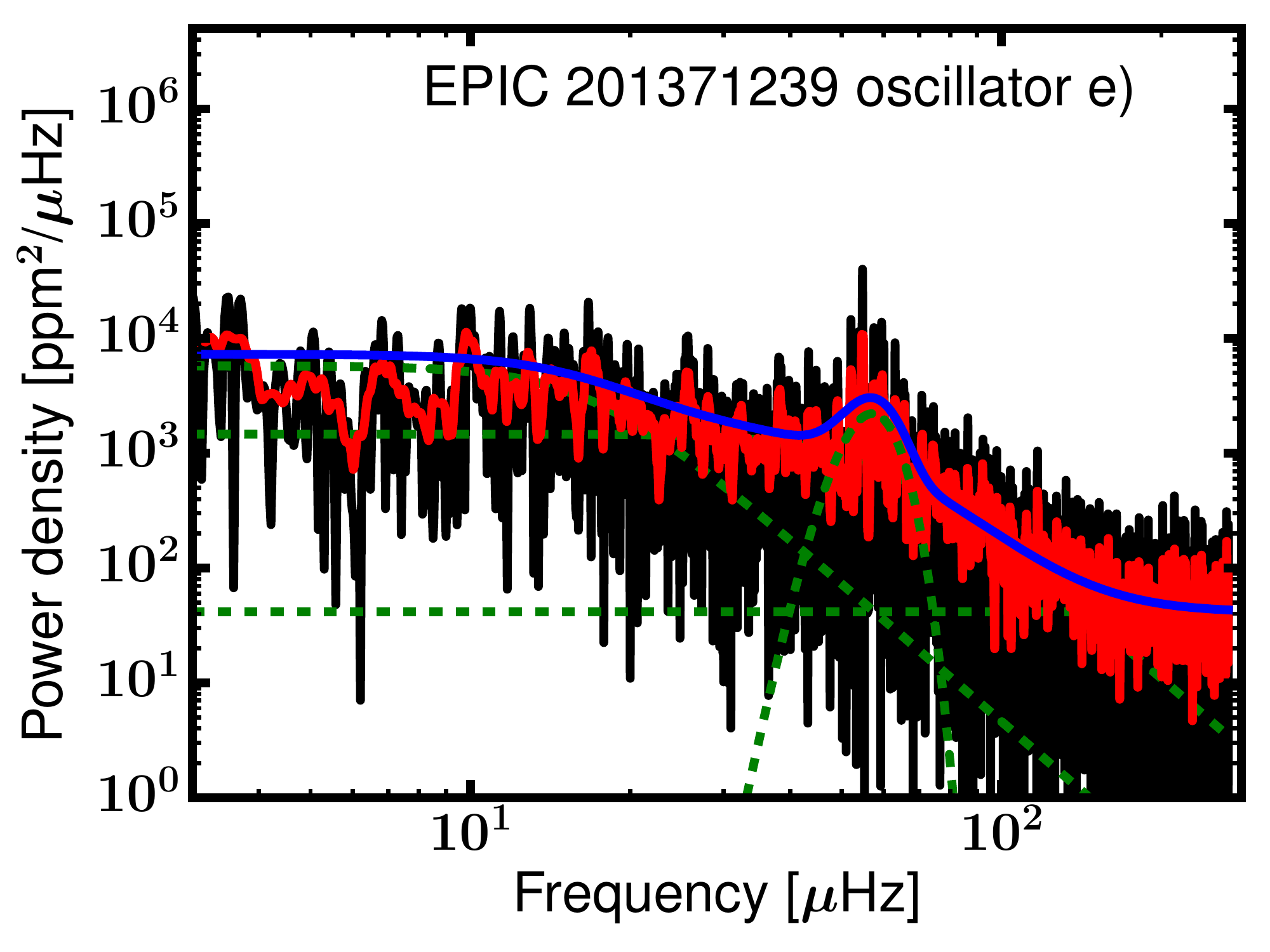}
\includegraphics[width=0.3\textwidth]{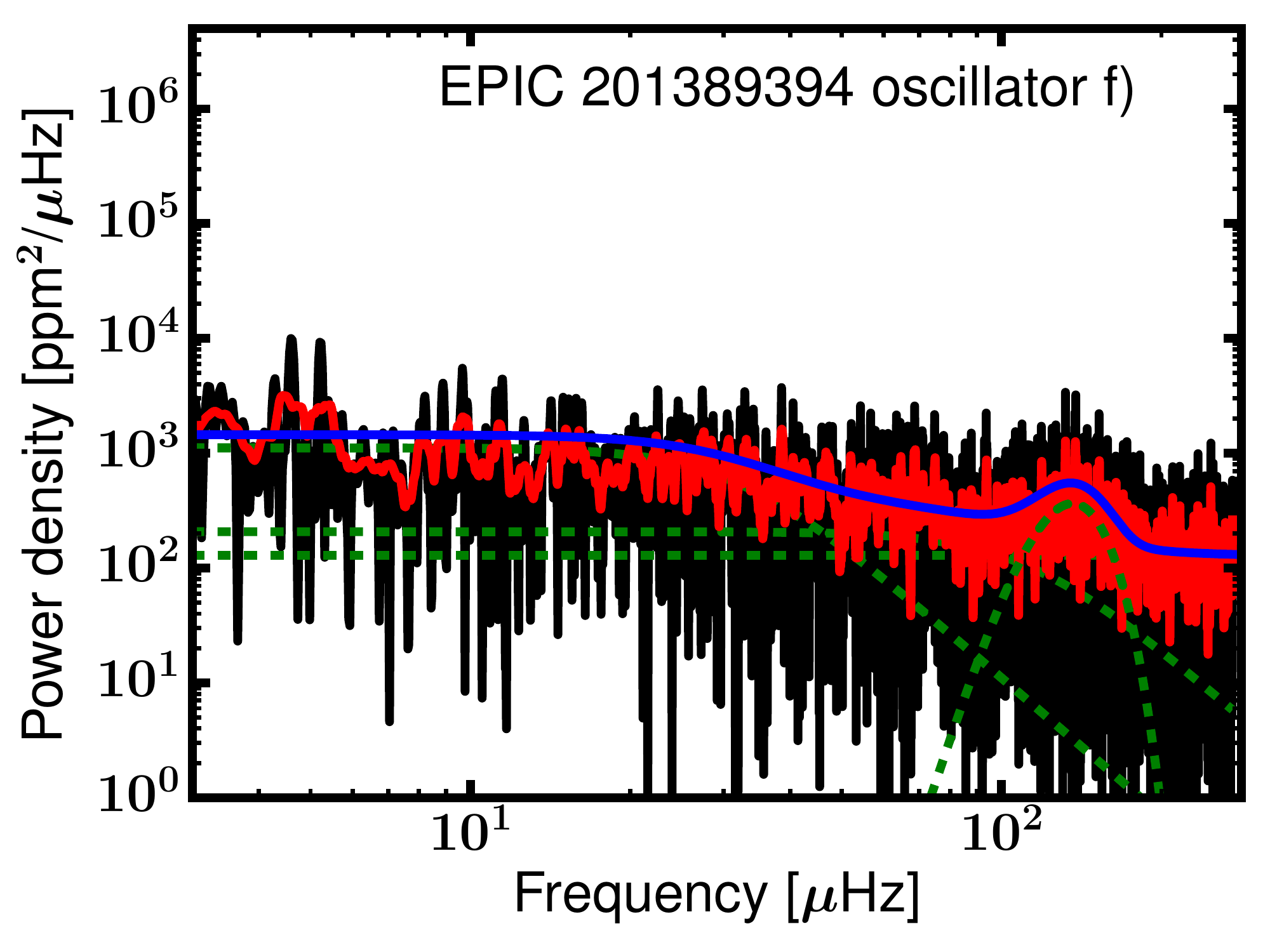}
\caption{Examples of the raw (black) and smoothed (red) power spectra
  of giant candidates selected by BAM, by requiring that the WBIC favor
  Equation~\ref{eq:tot} over Equation~\ref{eq:numax} (see \S\ref{sec:selection}). Each component of the models is shown in green
  dashed curves (white noise, Gaussian excess, and Harvey
  components), with the total model in blue. The top row shows BAM giant
  candidates determined to be false positives by visual inspection: EPIC 201180425 (a) shows a periodic signal, an alias of which BAM has mistaken for solar-like oscillations; EPIC 201758449 (b) shows a dwarf-like power spectrum that is at best a borderline no/maybe case --- BAM has converged on a suboptimal model in this case, in addition; and EPIC 201659364 (c) shows what may be a giant spectrum with no discernible oscillation modes. In all panels in this row, shown in grey is a smoothed VJ spectrum when the thruster
  firing has been removed according to the procedure described in
  \S\protect\ref{sec:data}. The bottom row shows BAM giant candidates confirmed by
  visual inspection. The model of
  EPIC 201763504 (d) has been convolved with the spectral window, which
  allows BAM to fit the correct $\numax$ at $\sim 8\muhz$ rather than
  the spectral noise at $\sim 50\muhz$ (see text).}
\label{fig:numax_ex2}
\end{figure*}

The purity of the non-GAP giant sample from BAM can be thought of as how many
giants are verified visually as giants out of all the candidates that
BAM believes are giants (i.e., $31$ out of $316$). Given that the majority of the non-GAP
targets were selected by GO programs
to be dwarfs, it is unsurprising that there are giant impostors that
BAM mistakenly selected as giant candidates. Encouragingly, we find
that BAM does not mistake the power in the frequency spectra from {\it K2}'s regular thruster
firing for genuine oscillator excess. Instead,
the objects mistakenly flagged as oscillators are due to one of a handful of failure modes. A full half of the false positives are objects exhibiting sharp, periodic signals overlaid on smooth, power-law
spectra. Unlike genuine solar-like oscillators,
however, objects falling into the latter failure mode generally exhibit multiple peaks (e.g., in
Figure~\ref{fig:numax_ex2}a). In future work, power spectra of
periodic signals could be separated from those of giants by adding a
second power excess component in Equation~\ref{eq:tot}. If the best-fitting
model preferred two regions of power excess instead of one, the spectrum would be
rejected as a possible periodic case and not a giant. The other half of the false positives are either borderline `maybe'/`no' cases where the power excess is seemingly absent, but a granulation signal is present; cases in which BAM has converged on an incorrect $\numax$ (in which case, even if the giant is oscillating, it is assigned a `no' category); or dwarfs that have enough low-frequency activity to mimic a noisy giant granulation spectrum. Examples of these false positives are shown in Figures~\ref{fig:numax_ex2}a\&b, in addition to an example of a potential giant oscillator (Figure~\ref{fig:numax_ex2}c) and examples of bona fide oscillators (Figures~\ref{fig:numax_ex2}d-f).

To get a better idea of the completeness of the sample, and to better understand
the distributions of the observed properties of the non-GAP giant
sample, we compare to a simulation that we describe in the next section.

\subsection{\texttt{Galaxia} simulation of non-GAP giants}
\label{sec:galaxia}
We model the non-GAP giant population using a \texttt{Galaxia} synthetic
population of all stars in the field of Campaign 1 (see
\citealt{sharma+2011} for a description of \texttt{Galaxia} and
\citealt{stello+2017} for a comparison of this synthetic population to
observed asteroseismic red giants from the GAP targets). Non-GAP
\texttt{Galaxia} giants are
defined to have $3 \muhz < \numax < 290 \muhz$, $Kp < 13$, and a
probability of detection greater than $95\%$ according to the
same procedure used in \cite{chaplin+2011}. However, here we assume $\sqrt{A_{\mathrm{max}}} =
2.5 \left( L/\lsun \right)^{0.9} \left(M/\msun \right)^{-1.7}
\left(\teff/\teffsun\right)^{-2.0}$ \citep{stello+2011} and noise
equal to that of {\it K2}. The use of a stellar population model of C1 like this is to make population-level statements about the concordance between the observed non-GAP giant population with a simulated one, and ideally to come to conclusions regarding the completeness and purity of the BAM non-GAP giant sample. In what follows, we will argue that there are likely inadequacies in both the recovered observed distribution due to selection effects, as well as inadequacies on the modeling side due to a difficult selection function and a probable metallicity offset in \texttt{Galaxia}'s underlying stellar models.

\begin{figure}
\centering
\includegraphics[width=0.5\textwidth]{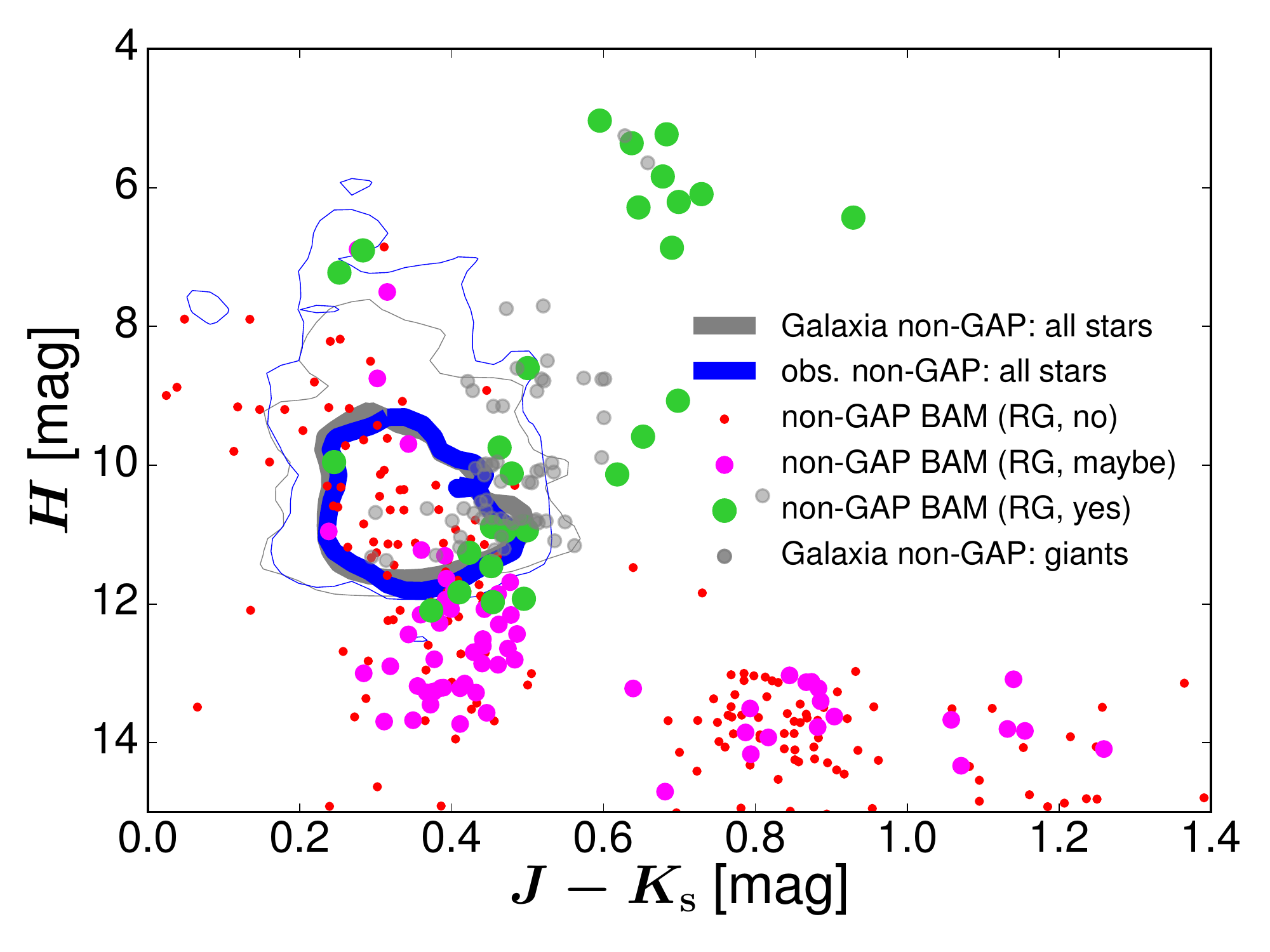}
\caption{A color--magnitude diagram for \texttt{Galaxia} stars not passing GAP selection criteria
  (grey contours); \texttt{Galaxia} giants not passing GAP selection criteria,
  with $>95\%$ probability of detection (grey dots); and observed non-GAP stars
  (blue contours). Contours enclose $68\%$ (thick lines) and
  $95\%$ (thin lines) of stars in the plotted region. Contours have
  been smoothed for illustrative purposes. Overlaid are stars from the non-GAP C1 target sample
  returned by BAM that visual inspection classified as definitely
  oscillators (green dots; 31 stars), maybe oscillators (magenta dots; 73 stars), and not
  oscillators (red dots; 212 stars).} 
\label{fig:jk}
\end{figure}

In order to make a fair comparison
between the observed non-GAP targets and the non-GAP \texttt{Galaxia}
stars, we re-sampled the \texttt{Galaxia} simulation such that it reproduced the observed
non-GAP distribution in ($J-K_{\mathrm{s}}$,$H$) space. We first binned the observed non-GAP stars in ($J-K_{\mathrm{s}}$,$H$)
space, and assigned each bin a probability of sample membership proportional
to the number of stars in that bin, and such that the
sum of each bin's probabilities summed to unity. We then binned the
\texttt{Galaxia} non-GAP stars using the same bins, and re-sampled the stars by
drawing a star one-by-one with a probability equal to the
aforementioned sample membership probability
of the bin in which it falls. The bins were chosen to optimize
agreement with the simulated and observed distributions in
($J-K_{\mathrm{s}}$,$H$) space, and were approximately ($0.05$mag,
$1$mag) in width. The re-sampling stopped when the
number of stars with $Kp < 13$ equalled the number of stars in the
observed non-GAP sample with $Kp < 13$ ($2080$ stars in
total).\footnote{$12839$ out of the $13016$ non-GAP
stars had valid $Kp$ values in the EPIC, $11579$ of those had valid $J-\K$ colors, and $2080$ of those also had $Kp <
13$.} This process results in some stars having the same
properties because there are not enough unique \texttt{Galaxia} stars to match the number of observed stars. For this
reason, we added a spread of $3\%$ on the simulated giants' $\numax$, $\dnu$,
and $2\%$ on photometry to avoid a sample with identical stars. 
The re-sampled \texttt{Galaxia} distribution is shown in the grey contours in
Figure~\ref{fig:jk}. The blue contours show the observed non-GAP
population that we wanted to simulate,
which shows the simulation is consistent with the observations. The
simulated giants within this sample, defined as mentioned above
to have $3 \muhz < \numax < 290 \muhz$, $Kp < 13$, and a
probability of detection greater than $95\%$,
are shown by the grey dots.

\subsection{Comparison to \texttt{Galaxia}}
With the \texttt{Galaxia} model for the non-GAP giants in hand, we can proceed to evaluate the agreement between simulation and observation, with implications for both the purity/completeness of the BAM sample, as well as the fidelity of the \texttt{Galaxia} simulation in its description of the data. Figure~\ref{fig:jk} shows that the recovered giants (magenta and green
dots) occupy two primary magnitude--color loci:
1) bright, red objects ($H < 7$ and $J-K_{\mathrm{s}} > 0.5$), which were not targeted in GAP because of the
the brightness cut in GAP of $H > 7$, and 2)
giants at a typical magnitude, but bluer than typical
giants ($7 < H < 13$ and
$J-K_{\mathrm{s}} < 0.5$), which were not in GAP because they have $J-K_{\mathrm{s}} <
0.5$. First, let us consider the blue
($J-K_{\mathrm{s}} < 0.5$) giants, which are the more numerous
population. That
\texttt{Galaxia} predicts the presence of this population (grey dots)
is the best indicator of agreement between our simulations and
observations. Indeed, we expect the blue population of non-GAP giants is a result of at least
two factors: 1) the GAP $J - K_{\mathrm{s}} > 0.5$ selection is arbitrary and
there are genuine oscillators with $J - K_{\mathrm{s}} < 0.5$, and 2) due to photometric errors (taken to be $\sim 0.02$ in the \texttt{Galaxia}
C1 simulation), some oscillating giants with $J - K_{\mathrm{s}} > 0.5$
will be scattered to  $J - K_{\mathrm{s}} < 0.5$. The \texttt{Galaxia}
simulation also successfully predicts the bright ($H < 7$)
giants should
exist. Note that our simulations only extend to our completeness
cut of $Kp = 13$, and so
we do not comment on \texttt{Galaxia} agreement in the regime of $H >
12$.

If the non-GAP sample were drawn from a similar distribution as our
\texttt{Galaxia} simulation, we would expect the ratio of red ($J -
K_{\mathrm{s}} > 0.5$) to blue ($J -
K_{\mathrm{s}} < 0.5$) giants in
\texttt{Galaxia} to agree with that of recovered BAM giants. We take
the ratio of the observed number of published `yes' and `maybe'
oscillators from {\it K2} GAP DR1 (\citealt{stello+2017}; with $Kp < 13$ and $J-K_{\mathrm{s}} > 0.5$ cuts applied) to
those with $J-K_{\mathrm{s}} < 0.5$ from the new, non-GAP giant sample
presented here, and compare it to the expected ratio from
\texttt{Galaxia}. For this test, the ($J-K_{\mathrm{s}}$,$H$)
distribution of the GAP population was simulated in
\texttt{Galaxia} following the sample membership probability procedure described
above, only using the GAP targets instead of the non-GAP
targets. Giants were then chosen to have $3 \muhz < \numax <
290 \muhz$, a
probability of detection greater than $95\%$, and $Kp < 13$. The resulting
ratio for \texttt{Galaxia} of $13 \pm 2$ is
significantly less than the same ratio for the BAM distribution of
`yes' and `maybe' GAP giants of $38 \pm 9.0$, accounting for Poisson errors. Either
the number of GAP giants are at odds with predictions, the number of
non-GAP giants are, or both. Looking at the
absolute numbers of giants in this ratio, 651/17 for observed BAM
giants and 821/64 for \texttt{Galaxia}, the GAP giants agree better in number with
what is expected from \texttt{Galaxia} than do the non-GAP giants. The
$70\%$ deficit in observed giants
compared to \texttt{Galaxia} for the blue, non-GAP giants indicates that \texttt{Galaxia} predicts too many
blue giants and/or BAM recovers too few blue giants. We consider both
effects, in turn.

One of the primary effects that might result in an over-prediction in
our \texttt{Galaxia} model's number of non-GAP giants is an incorrect
selection function. The \texttt{Galaxia} non-GAP sample as we have constructed it only
reproduces the color-magnitude distribution of the many GO proposal
targets that comprise the non-GAP sample. We expect this approach to
globally describe the complex selection function of the sample, given that the GO proposals select objects based on color and
magnitude cuts. Indeed, the non-GAP sample does describe well the
observed sample (Figure~\ref{fig:jk}). However, the majority of
the GO proposals that comprise the non-GAP sample also use
proper motion or reduced proper motion cuts to choose dwarfs. Although
these cuts will be functions of color and magnitude, we cannot precisely
reproduce them in color and magnitude space. Therefore, we
tested how many \texttt{Galaxia} non-GAP giants remained after
applying a rather conservative (i.e., preserving more giants than
dwarfs) reduced proper motion cut of $V + 5\log_{10} \mu > 20(V-J) -
25$. (These cuts use the kinematic information that is stored as a part of a \texttt{Galaxia} simulation.) Only 11 non-GAP stars remained after this reduced proper motion cut, which indicates that the GO
reduced proper motion cuts could explain the difference between the
observed number of non-GAP giants (17) and that otherwise predicted by
\texttt{Galaxia} (64). Another selection function could still be at 
work within the \texttt{Galaxia} model itself: an incorrect metallicity
distribution of disk stars could result in too many blue giants, whose
colors naturally depend on metallicity. A metallicity effect could
also explain the offset in red clump position with respect to the
observed red clump in \textit{K2} data, which is discussed in the next
section.

With the reduced
proper motion cut's role in mind,
we still anticipate that some of the deficit in observed numbers of non-GAP giants is
likely to reflect genuine incompleteness in the BAM giant
sample. For example, in a handful of cases in the false positive (`no's) sample, BAM performed a poor fit to the data, which will mean its Bayesian model comparison will not be valid. Also, blended light from dwarfs would also strongly select against
recovery with BAM because of a dilution of
the oscillation signal resulting in significant departures from the
amplitudes imposed by BAM's priors in Table~\ref{tab:numax_priors}.   We note also that asteroseismic giant detection with {\it K2} will miss
giants with $ \numax \lesssim 3 \muhz$ and $\numax > 283 \muhz$ --- the most
evolved giants, and those closest to the base of the red giant
branch.
 Establishing robust completeness and efficiency estimates is not the purpose of
this paper, however, and we will explore these concerns more thoroughly in the next
K2GAP data release (Zinn et al., in prep.).

\begin{figure}
\centering
\includegraphics[width=0.5\textwidth]{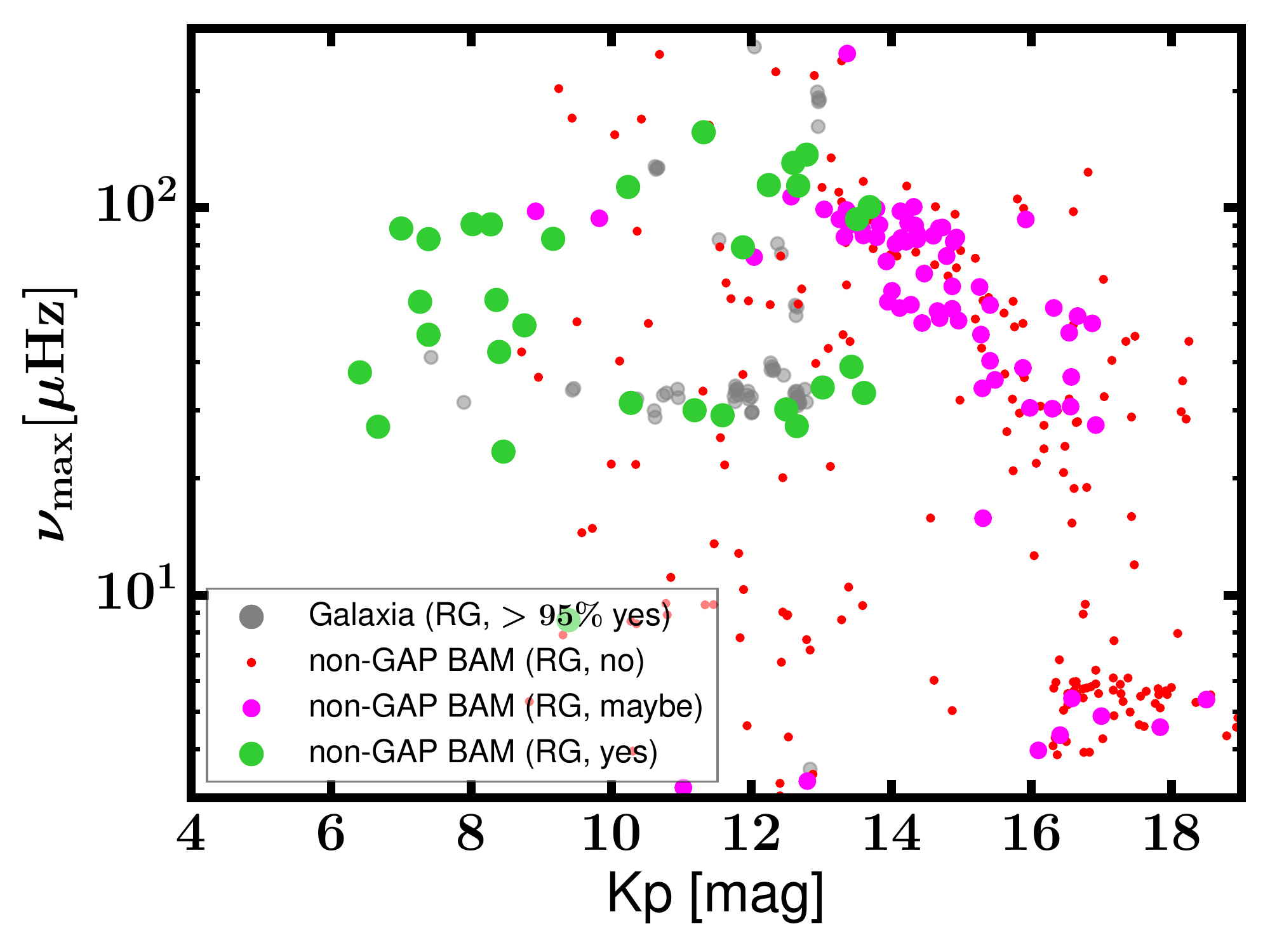}
\caption{$\numax$--$Kp$ distribution of \texttt{Galaxia} predicted
  detections of non-GAP oscillating giants
  (grey). Overlaid are stars from the non-GAP C1 target sample
  returned by BAM that visual inspection classified as definitely
  oscillators (green dots; 31 stars), maybe oscillators (magenta dots; 73 stars), and not
  oscillators (red dots; 212 stars).} 

\label{fig:kp_numax}
\end{figure}

We can also compare the \texttt{Galaxia} non-GAP red giant sample and the observed BAM non-GAP red giant sample in magnitude-$\numax$ space, as shown in Figure~\ref{fig:kp_numax}. Kolmogorov-Smirnov tests indicate that both
the $\numax$ distribution and $Kp$ distribution for the definite BAM red giants are in $\sim 3.2\sigma$ and $\sim 4.0\sigma$ tension with the \texttt{Galaxia} $\numax$ and $Kp$ distributions, assuming our adopted detection
limit of $Kp < 13$. We note at this point that the procedure to match observed and
\texttt{Galaxia} magnitude and color distributions
(\S\ref{sec:galaxia}) is stochastic because the distributions are
matched by drawing from probability distributions. This results in the
\texttt{Galaxia} giants having $\numax$ and $Kp$ distributions that
vary in their agreement with the observed non-GAP giant distributions,
fluctuating at the $0.3\sigma$ and $0.4\sigma$ level, respectively. Keeping this caveat in mind, there is still a tension in
the simulated and observed $\numax$ distributions when marginalizing
over realizations of the \texttt{Galaxia} $\numax$ distribution. That the tension in $\numax$ space decreases by $\sim 1\sigma$ with a reduced
proper motion cut (see \S\ref{sec:galaxia}), indicates this difference might
be due to the un-modelled non-GAP selection function effects of individual GO proposals. There could also certainly be a $\numax$-dependent efficiency in BAM identifying
giants. Indeed, the latter effect is seen across various pipelines when
comparing to a ground truth set of giants in \textit{K2} fields identified by eye, even while
\texttt{Galaxia} giant predictions as a function of $\numax$ agree
very well with the ground truth (K2GAP DR2; Zinn et al., in prep.).

\begin{figure}
\centering
\includegraphics[width=0.5\textwidth]{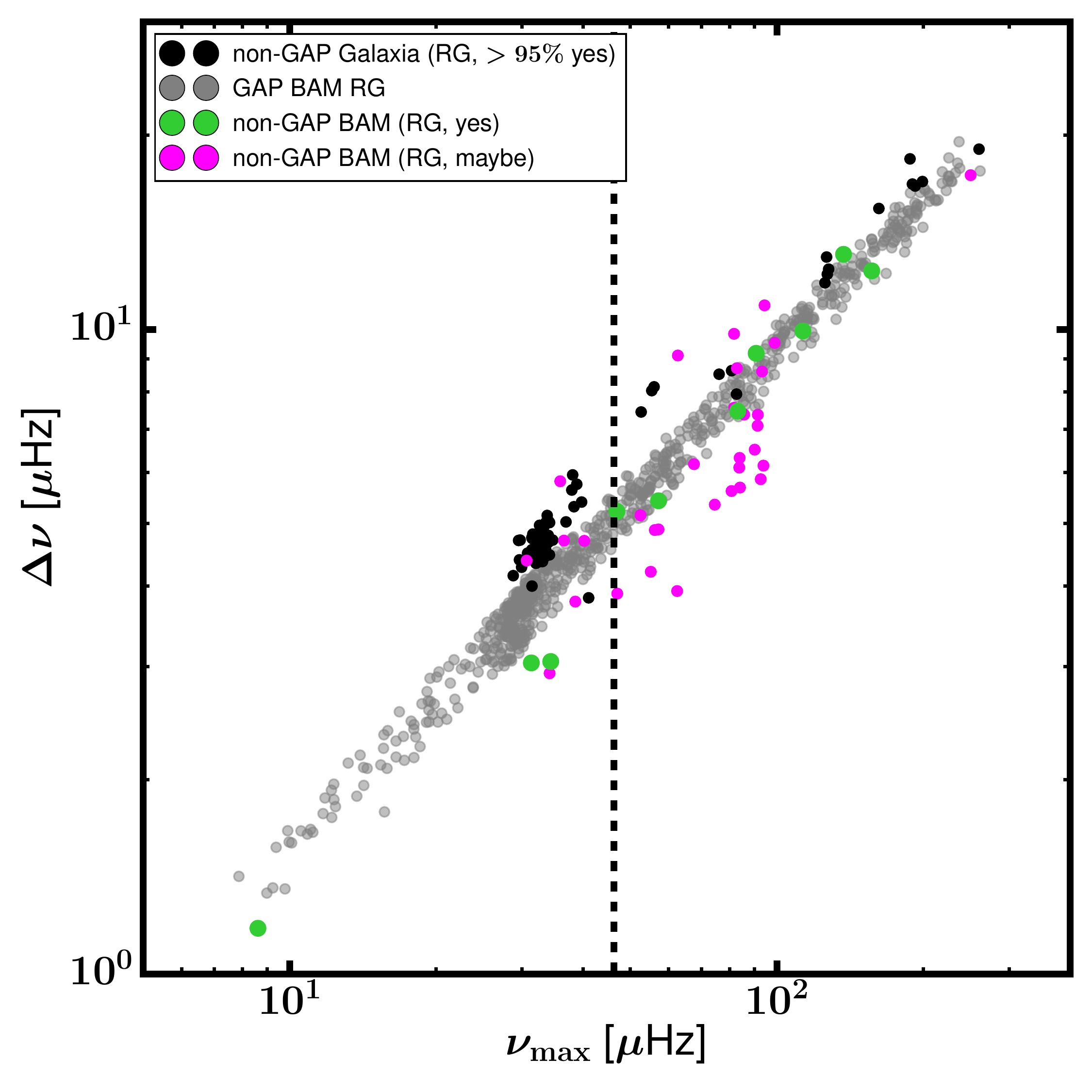}
\caption{The $\dnu$--$\numax$ relation, with the
  non-GAP giant sample shown as points with colors as in
  Figure~\protect\ref{fig:kp_numax}, comprising stars that have both $\dnu$
  and $\numax$ measured by BAM. Grey points are BAM results from K2GAP DR1
\protect\citep{stello+2017}, and black points are from our \texttt{Galaxia}
simulation of the non-GAP giant sample. The dashed line corresponds to the
nominal {\it K2} thruster firing frequency.}
\label{fig:dnu_numax_seren}
\end{figure}
\begin{figure}
\centering
\includegraphics[width=0.5\textwidth]{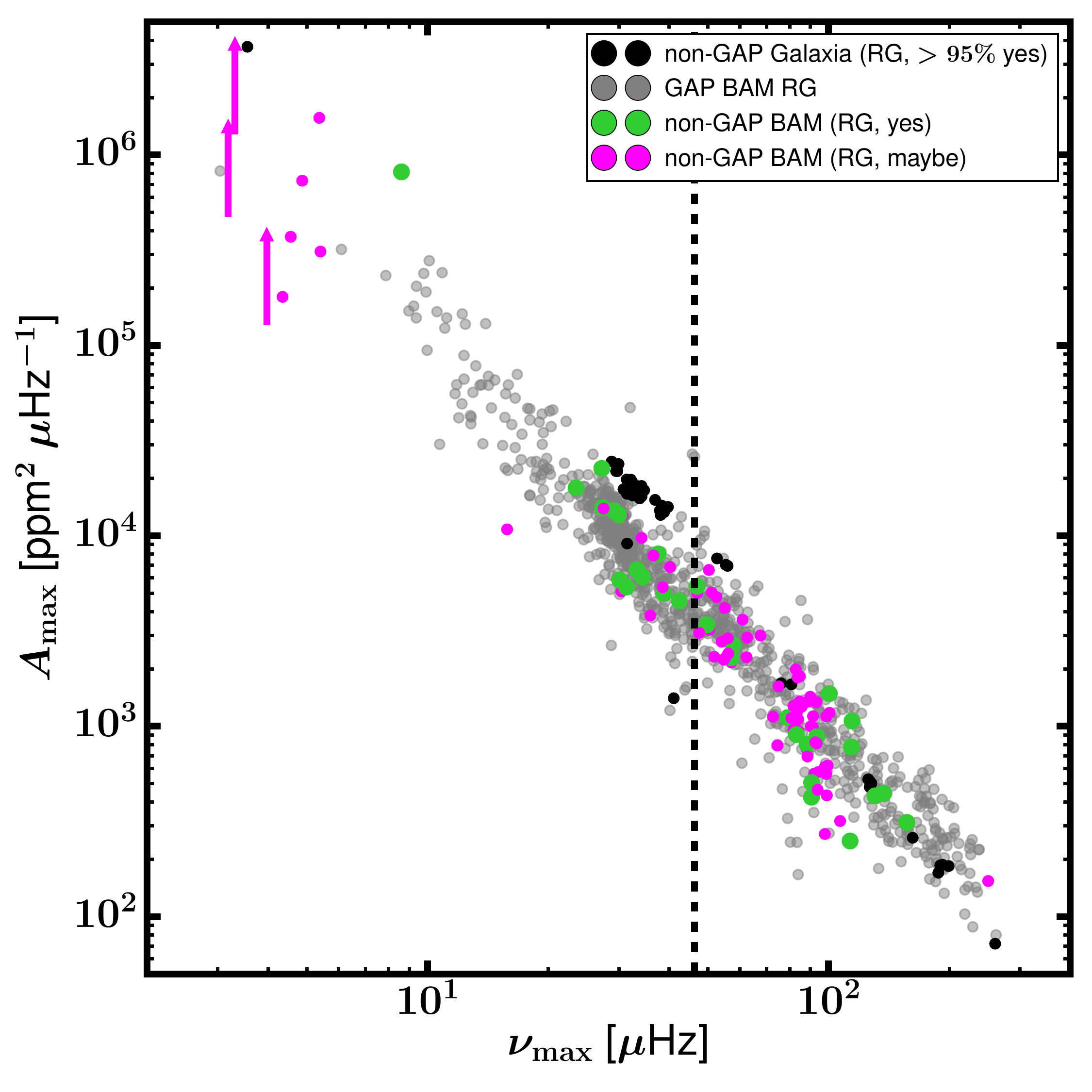}
\caption{The $A_{\mathrm{max}}$--$\numax$ relation, with the
  non-GAP giant sample shown as points with colors as in
  Figure~\protect\ref{fig:kp_numax}. Giants for which $\numax
  \lesssim 4\muhz$ are considered upper limits. Grey points are BAM results from K2GAP DR1
\protect\citep{stello+2017}, and black points are from our \texttt{Galaxia}
simulation of the non-GAP giant sample. BAM K2GAP DR1 amplitudes were not published in \protect\citep{stello+2017}, though are reproduced here. The dashed line corresponds to the
nominal \textit{K2} thruster firing frequency.}
\label{fig:dnu_amp_seren}
\end{figure}

\subsection{Properties of \texttt{Galaxia} and observed non-GAP
  giants}

\begin{figure}
\centering
\includegraphics[width=0.5\textwidth]{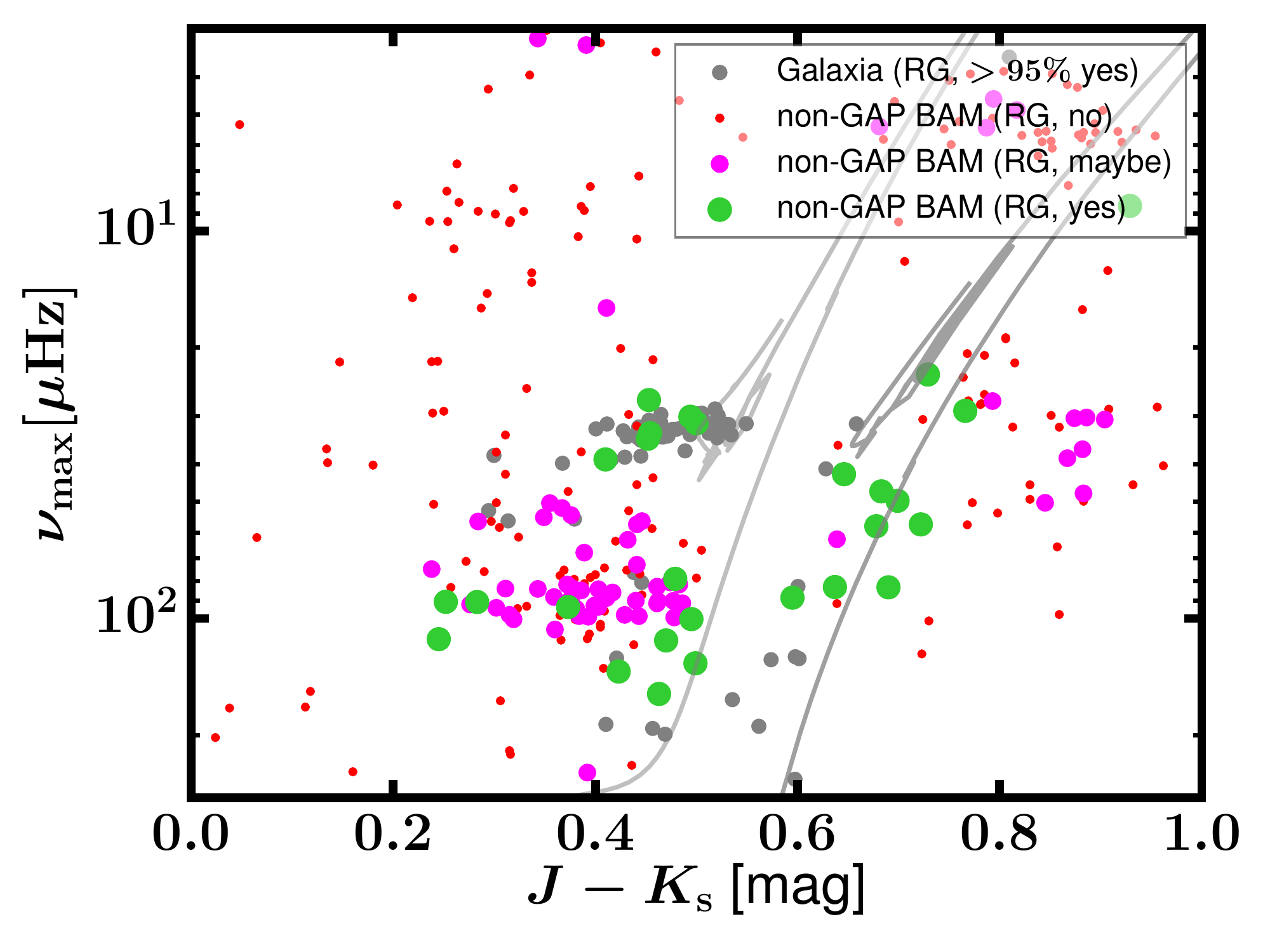}
\caption{Modified Kiel diagram, with the
  non-GAP giant sample shown as points with colors as in
  Figure~\protect\ref{fig:kp_numax}. The grey points are
  predictions from a simulation of the non-GAP stellar population in
  Campaign 1 using \texttt{Galaxia} \protect\citep{sharma+2011}. See
  text for details. Evolutionary tracks for a 1.3$\msun$ star with [Fe/H] $=0$ (dark grey) and [Fe/H] $=-1$ (light grey) from MIST \protect\citep{dotter+2016a,choi+2016a}
are shown for visualization purposes.} 
\label{fig:kiel}
\end{figure}

We show in Figures~\ref{fig:dnu_numax_seren}
and~\ref{fig:dnu_amp_seren} the $\dnu$--$\numax$ and
$A_{\mathrm{max}}$--$\numax$ relations for this sample (colored points), as well as for the \texttt{Galaxia} model (black points). We have also included BAM GAP giants published in \cite{stello+2017}, for reference (grey points). The agreement
between model and observed properties in these spaces is good, except
for the clump, for which \texttt{Galaxia} predicts a too-high $\dnu$ and
$A_{\mathrm{max}}$. We can see that \texttt{Galaxia} over-predicts $\dnu$
and $A_{\mathrm{max}}$ (and does not under-predict $\numax$) because
the $\numax$ location of the
over-density in GAP BAM stars at $\numax \sim 30 \muhz$
agrees with the location of the over-density in the non-GAP
\texttt{Galaxia} stars. Figure~\ref{fig:kiel} shows a modified Kiel
diagram, in which  $J-K_{\mathrm{s}}$ color is used instead of temperature and $\numax$
instead of gravity\footnote{Note the reversed y-axis: a smaller $\numax$ means
a smaller gravity, and so is in the sense of a normal Kiel
diagram.}. In this space, we can see that nearly all of the observed
non-GAP sample is found at or below the clump (at $\numax \sim 30\muhz$),
and that the location of the \texttt{Galaxia} clump overlaps with
several of the presumable observed red clump stars, confirming that the
\texttt{Galaxia} clump $\numax$ locus is not discrepant with the observed locus. That the modeled clump $\dnu$ locus is offset from the observed clump $\dnu$ locus is another indication that the \texttt{Galaxia} models could be relying on a Galactic metallicity distribution at odds with the actual one --- a conclusion that one arrives at when comparing Galaxia stellar parameters to those from asteroseismology in other {\it K2} campaigns \citep{sharma+2019}.

\subsection{Implications for dwarf selection purity}
A summary of the number of
`yes' and `maybe' giants broken down by the GO target list
from which they arise is shown in Table~\ref{tab:purity}. $21$ of the sample of non-GAP giants are
serendipitous: they are only targets from GO proposals
that do not intentionally select giants. This, in turn, allows us to say that the
purity of giant exclusion across {\it K2} C1 GO proposals is $\sim
99\%$, based on the observed confirmed number of serendipitous giants
found among the GO target lists that do not purport
to select giants (those that intentionally target giants are not included in our calculation of
dwarf purity, and are noted in Table~\ref{tab:purity}). The purity decreases a negligible amount if also including the BAM non-GAP `maybe' giants. This estimated dwarf selection
purity is an upper bound because we have certainly not
recovered all the giants due to reasons discussed in
\S\ref{sec:galaxia}. In this estimate, we have only counted targets
that are within our completeness limit of $Kp < 13$. In this sense, we
confirm that the \textit{K2} dwarf samples chosen with color and proper
motion cuts are generally free from giants for $Kp < 13$.

\section{conclusion}
\label{sec:conclusion}
In this paper, we have presented the BAM pipeline, which calculates
global oscillation parameters in a Bayesian framework. A major advantage of the
Bayesian fitting method we have employed is its natural basis for
probabilistic selection of likely true oscillators among a collection
of light curves. In the process of developing this pipeline and
applying it to {\it K2} Campaign 1 (C1) stars, including both Galactic Archaeology Program \citep[GAP;][]{stello+2015,stello+2017} giant targets and non-GAP dwarf targets, we have found the following:

\begin{enumerate}
  \item We have identified as-of-yet an
unidentified noise pattern present in \cite{vanderburg&johnson2014} light curves of C1 stars that causes a splitting of
the nominal thruster firing frequency artefact at $47.22\muhz$ in a
time-dependent manner.
\item We have additionally shown that it is necessary
to account for the spectral window in fitting the spectra of solar-like
oscillators in order to model the unphysical spectral leakage in the power spectrum of oscillators with $\numax \lesssim 15$
$\muhz$. In this work, we have
done so by convolving models of the granulation with the observed window function.
\item We have benchmarked our asteroseismic parameters against the existing SYD
asteroseismic pipeline, and quantified statistical and systematic
errors for BAM parameters accordingly. We find typical errors for
\textit{K2} BAM giants in $\numax$ and $\dnu$ of $\sim
1.53\% (\mathrm{random}) \pm 0.2\% (\mathrm{systematic})$ and $1.51\% (\mathrm{random}) \pm
0.6\% (\mathrm{systematic})$.
\item As an example application of BAM, we have also presented a sample of $104$ non-GAP BAM red giants and red giant candidates from C1 identified by their solar-like oscillations, $91$ of
which were not selected by Guest Observer proposals to be giants, and hence are serendipitous
discoveries.
\item The size of the non-GAP BAM red giant sample suggests that \textit{K2} C1 dwarf samples chosen with color and proper motion cuts are generally free from giants for $Kp < 13$ to a high degree (upper bound of $\sim 99\%$ pure).
\item Simulated \texttt{Galaxia} C1 non-GAP giant populations are in tension with the $Kp$ and $\numax$
distributions of observed non-GAP giants with $Kp < 13$ found by BAM. When considering also the higher-than-observed number of blue ($J - K_{\mathrm{s}} <
0.5$) giants in the \texttt{Galaxia} model, the disagreement between model and observation can be explained by the proper motion cuts used
to select the non-GAP targets. There is also likely incompleteness in
the BAM giant detection process, which
will be addressed in future work. Finally, the \texttt{Galaxia} metallicity
distribution is likely different than the distribution of the non-GAP stars \citep{sharma+2019}.

\end{enumerate}

BAM promises to robustly identify and
characterize solar-like oscillators in {\it K2} and the TESS
mission \citep{ricker+2014}, which is observing hundreds of thousands
of red giants with at least 30 minute cadence. Though it will perform
at least as well as {\it K2} in resolving oscillations on the lower giant
branch, the majority of TESS's red giant data will have roughly half
the temporal baseline of
{\it K2}, and therefore will be a factor of two worse in spectral
resolution. Spectral resolution is particularly important in
identifying the low frequency oscillators like those presented
here. In this sense, BAM's Bayesian fitting techniques will take
advantage of the information in (`global') features of the power spectrum
that are less sensitive to degraded frequency resolution, in
order to robustly identify $\numax$ for TESS giants.

\acknowledgments
We thank the referee for their comments, which improved both BAM and
the contents of this paper. J.~Z. acknowledges support from NASA grants 80NSSC18K0391 and
NNX17AJ40G. D.~H. acknowledges support by the National Science Foundation
(AST-1717000) and the National Aeronautics and Space Administration
under Grants NNX14AB92G and NNX16AH45G issued through the \textit{Kepler}
Participating Scientist Program and the \textit{K2} Guest Observer Program.
D.~S. is the recipient of an Australian Research Council Future Fellowship (project number
FT1400147). Parts of this research were conducted by the Australian
Research Council Centre of Excellence for All Sky Astrophysics in 3
Dimensions (ASTRO 3D), through project number CE170100013. This research was supported in part by the National Science Foundation under Grant No. NSF PHY-1748958 and by the Heising-Simons Foundation.

This paper includes data collected by the {\it K2} mission. We thank
all NASA employees that have contributed to the remarkable success of
the {\it K2} mission.  Funding
for the {\it K2} mission is provided by the NASA Science Mission
directorate. Some of the data presented
in this paper were obtained from the Mikulski Archive
for Space Telescopes (MAST).

\begin{longrotatetable}
  \vspace{1cm}
  \begin{deluxetable*}{ccccccccccccccccccccc}
    \tabletypesize{\tiny}
    \setlength{\tabcolsep}{0.02in}
          \tablecaption{Campaign 1 non-GAP BAM asteroseismic
            parameters for giants and giant candidates    \label{tab:seren}}
    \tablehead{\colhead{EPIC} & \colhead{2MASS} & \colhead{RA} &   \colhead{Dec} & \colhead{$\numax$} & \colhead{$\sigma_{\numax}$} &
    \colhead{$\dnu$} & \colhead{$\sigma_{\dnu}$} &
    \colhead{$A_{\mathrm{max}}$} &
    \colhead{$\sigma_{A_{\mathrm{max}}}$} &
    \colhead{$\sigma_{\mathrm{meso}}$} &
    \colhead{$\sigma_{\sigma_{\mathrm{meso}}}$} &
    \colhead{$\tau_{\mathrm{meso}}$} &
    \colhead{$\sigma_{\tau_{\mathrm{meso}}}$}  &
    \colhead{$\sigma_{\mathrm{gran}}$} &
    \colhead{$\sigma_{\sigma_{\mathrm{gran}}}$} &
    \colhead{$\tau_{\mathrm{gran}}$} & \colhead{
      $\sigma_{\tau_{\mathrm{gran}}}$} & \colhead{WN} &
    \colhead{$\sigma_{\mathrm{WN}}$} &
    \colhead{giant?\tablenotemark{a}}\\
      \colhead{} & \colhead{}& \colhead{deg} & \colhead{deg} &
      \colhead{$\muhz$} & \colhead{$\muhz$} & \colhead{$\muhz$} &
      \colhead{$\muhz$} & \colhead{ppm$^2$ $\muhz^{-1}$} &
      \colhead{ppm$^2$ $\muhz^{-1}$} & \colhead{ppm} & \colhead{ppm} &
      \colhead{$\muhz^{-1}$} & \colhead{$\muhz^{-1}$} & \colhead{ppm}
      & \colhead{ppm} & \colhead{$\muhz^{-1}$} &
      \colhead{$\muhz^{-1}$} & \colhead{ppm$^2$ $\muhz^{-1}$} &
      \colhead{ppm$^2$ $\muhz^{-1}$} & \colhead{}
    }
    \startdata
201147434 & 11452628-0521209 & 176.359512 &  -5.355842 & 67.604 & 2.676 & 6.179 & 0.696 & 2.994e+03 & 9.890e+02 & 4.081e+02 & 4.279e+01 & 1.977e-02 & 1.790e-03 & 4.936e+02 & 3.934e+01 & 4.681e-03 & 4.733e-04 & 1.125e+03 & 1.211e+01 & 1 \\
201151637 & 11360854-0515250 & 174.035575 &  -5.256981 & 57.160 & 4.415 & 4.894 & 0.021 & 2.166e+03 & 1.287e+03 & 4.491e+02 & 3.038e+01 & 2.230e-02 & 3.120e-03 & 5.273e+02 & 2.435e+01 & 5.841e-03 & 3.477e-04 & 5.719e+02 & 1.328e+01 & 1 \\
201156121 & 11315343-0508579 & 172.972654 &  -5.149433 & 51.849 & 3.264 &   \nodata &   \nodata & 2.314e+03 & 1.125e+03 & 3.754e+02 & 3.378e+01 & 2.725e-02 & 3.432e-03 & 4.534e+02 & 2.530e+01 & 6.410e-03 & 5.875e-04 & 1.159e+03 & 2.382e+01 & 1 \\
201163464 & 11353033-0458260 & 173.876408 &  -4.973911 & 52.509 & 3.362 & 5.149 & 0.018 & 4.768e+03 & 3.751e+03 & 5.603e+02 & 4.441e+01 & 2.319e-02 & 2.310e-03 & 6.859e+02 & 3.957e+01 & 5.976e-03 & 4.948e-04 & 2.425e+03 & 5.196e+01 & 1 \\
201198517 & 11391931-0408141 & 174.830492 &  -4.137267 & 42.417 & 0.645 &   \nodata &   \nodata & 4.554e+03 & 6.148e+02 & 1.235e+03 & 5.464e+01 & 3.896e-02 & 2.187e-03 & 5.012e+02 & 4.072e+01 & 5.776e-03 & 3.189e-04 & 6.436e+01 & 5.346e+00 & 2 \\
201214537 & 11461882-0345352 & 176.578525 &  -3.759797 & 51.098 & 4.421 &   \nodata &   \nodata & 5.040e+03 & 3.956e+03 & 5.377e+02 & 3.964e+01 & 2.843e-02 & 3.819e-03 & 7.327e+02 & 3.243e+01 & 5.561e-03 & 3.705e-04 & 2.272e+03 & 4.871e+01 & 1 \\
201228269 & 11245189-0332138 & 171.216150 &  -3.537211 & 50.382 & 2.234 &   \nodata &   \nodata & 3.211e+03 & 2.136e+03 & 4.665e+02 & 3.551e+01 & 2.524e-02 & 3.378e-03 & 5.765e+02 & 2.944e+01 & 5.763e-03 & 4.952e-04 & 1.290e+03 & 3.113e+01 & 1 \\
201244712 & 11233476-0317217 & 170.894883 &  -3.289417 & 49.686 & 0.704 &   \nodata &   \nodata & 3.406e+03 & 6.118e+02 & 8.720e+02 & 3.486e+01 & 2.863e-02 & 1.788e-03 & 6.132e+02 & 2.231e+01 & 6.004e-03 & 1.920e-04 & 8.177e+01 & 2.313e+00 & 2 \\
201251246 & 11291517-0311199 & 172.313354 &  -3.188886 & 3.309 &   \nodata &   \nodata &   \nodata & 2.566e+06 & 8.814e+05 & 4.222e+03 & 5.350e+02 & 3.516e-01 & 3.131e-02 & 3.295e+03 & 2.028e+02 & 8.672e-02 & 6.170e-03 & 6.972e+01 & 4.575e+00 & 1 \\
201253257 & \nodata & 179.712362 &  -3.158483 & 55.077 & 4.786 &   \nodata &   \nodata & 4.172e+03 & 3.050e+03 & 4.662e+02 & 4.151e+01 & 2.424e-02 & 3.321e-03 & 7.106e+02 & 3.180e+01 & 5.625e-03 & 3.720e-04 & 1.646e+03 & 3.615e+01 & 1 \\
201262747 & 11542756-0300552 & 178.614854 &  -3.015358 & 98.614 & 4.590 &   \nodata &   \nodata & 5.610e+02 & 2.737e+02 & 3.355e+02 & 1.901e+01 & 1.305e-02 & 1.457e-03 & 3.358e+02 & 1.925e+01 & 3.774e-03 & 3.295e-04 & 3.586e+02 & 8.709e+00 & 1 \\
201269306 & 11235529-0254559 & 170.980417 &  -2.915544 & 114.376 & 0.673 &   \nodata &   \nodata & 1.065e+03 & 9.483e+01 & 4.929e+02 & 1.667e+01 & 1.176e-02 & 6.325e-04 & 3.587e+02 & 1.476e+01 & 3.408e-03 & 1.805e-04 & 8.608e+01 & 2.869e+00 & 2 \\
201272934 & 11190372-0251388 & 169.765562 &  -2.860719 & 89.965 & 7.486 &   \nodata &   \nodata & 1.427e+03 & 9.458e+02 & 4.507e+02 & 3.229e+01 & 1.387e-02 & 1.517e-03 & 5.431e+02 & 3.221e+01 & 3.682e-03 & 3.046e-04 & 1.025e+03 & 2.614e+01 & 1 \\
201305005 & 11374947-0222333 & 174.456163 &  -2.375928 & 30.647 & 2.360 & 4.379 & 0.058 & 1.281e+04 & 9.055e+03 & 1.019e+03 & 7.423e+01 & 4.373e-02 & 3.892e-03 & 8.133e+02 & 6.819e+01 & 9.908e-03 & 9.029e-04 & 6.525e+03 & 1.110e+02 & 1 \\
201310650 & 11351161-0217353 & 173.798379 &  -2.293164 & 47.040 & 4.038 & 3.891 & 0.022 & 5.000e+03 & 3.903e+03 & 5.739e+02 & 4.526e+01 & 2.915e-02 & 3.419e-03 & 7.150e+02 & 3.875e+01 & 6.180e-03 & 5.059e-04 & 2.635e+03 & 5.559e+01 & 1 \\
201348966 & 11305698-0143174 & 172.737462 &  -1.721542 & 38.852 & 1.057 &   \nodata &   \nodata & 4.974e+03 & 1.102e+03 & 5.837e+02 & 3.973e+01 & 3.047e-02 & 2.635e-03 & 7.666e+02 & 2.502e+01 & 7.802e-03 & 2.314e-04 & 1.159e+02 & 2.557e+00 & 2 \\
201361246 & 12005920-0132144 & 180.246675 &  -1.537347 & 47.547 & 5.332 &   \nodata &   \nodata & 3.093e+03 & 2.187e+03 & 5.869e+02 & 4.346e+01 & 2.613e-02 & 4.348e-03 & 6.068e+02 & 3.927e+01 & 7.203e-03 & 7.921e-04 & 1.866e+03 & 4.179e+01 & 1 \\
201361404 & 11112565-0132069 & 167.856942 &  -1.535272 & 100.372 & 0.903 &   \nodata &   \nodata & 1.482e+03 & 3.635e+02 & 4.316e+02 & 2.491e+01 & 1.144e-02 & 1.093e-03 & 4.287e+02 & 2.480e+01 & 3.316e-03 & 2.431e-04 & 3.794e+02 & 1.135e+01 & 2 \\
201370145 & 11130872-0124345 & 168.286287 &  -1.409614 & 82.959 & 1.019 &   \nodata &   \nodata & 1.266e+03 & 1.318e+02 & 5.517e+02 & 2.180e+01 & 1.510e-02 & 8.499e-04 & 4.355e+02 & 1.811e+01 & 4.215e-03 & 1.461e-04 & 4.157e+01 & 1.464e+00 & 2 \\
201371239 & 11111931-0123405 & 167.830442 &  -1.394597 & 57.161 & 0.813 & 5.421 & 0.018 & 2.287e+03 & 3.498e+02 & 5.523e+02 & 2.636e+01 & 1.894e-02 & 1.235e-03 & 5.211e+02 & 2.139e+01 & 5.413e-03 & 1.759e-04 & 4.157e+01 & 1.183e+00 & 2 \\
201374034 & 11062714-0121136 & 166.613092 &  -1.353819 & 98.965 & 6.816 & 9.528 & 0.028 & 4.333e+02 & 2.267e+02 & 2.891e+02 & 1.616e+01 & 1.252e-02 & 1.068e-03 & 2.729e+02 & 1.612e+01 & 3.694e-03 & 3.402e-04 & 2.290e+02 & 5.629e+00 & 1 \\
201375598 & 11194625-0119486 & 169.942750 &  -1.330181 & 4.559 & 0.941 & 0.731 & 0.016 & 3.717e+05 & 3.404e+05 & 2.476e+03 & 5.092e+02 & 2.255e-01 & 5.046e-02 & 2.637e+03 & 4.187e+02 & 7.317e-02 & 1.506e-02 & 8.340e+04 & 1.384e+03 & 1 \\
201379262 & 11234562-0116277 & 170.940087 &  -1.274369 & 40.242 & 5.022 & 4.699 & 0.051 & 6.862e+03 & 5.470e+03 & 6.450e+02 & 5.438e+01 & 3.529e-02 & 5.425e-03 & 8.063e+02 & 4.876e+01 & 7.269e-03 & 8.057e-04 & 4.611e+03 & 9.516e+01 & 1 \\
201379481 & \nodata & 168.657175 &  -1.270828 & 37.570 & 0.758 &   \nodata &   \nodata & 8.026e+03 & 1.018e+03 & 1.091e+03 & 4.372e+01 & 3.813e-02 & 2.345e-03 & 5.955e+02 & 3.462e+01 & 5.662e-03 & 2.493e-04 & 3.954e+01 & 3.093e+00 & 2 \\
201389394 & 11082147-0107156 & 167.089500 &  -1.121086 & 137.060 & 1.990 & 13.075 & 0.049 & 4.438e+02 & 5.241e+01 & 3.329e+02 & 1.265e+01 & 1.006e-02 & 5.729e-04 & 2.704e+02 & 1.439e+01 & 2.838e-03 & 2.119e-04 & 1.289e+02 & 4.161e+00 & 2 \\
201400095 & 11210314-0057399 & 170.263029 &  -0.961072 & 50.276 & 4.850 &   \nodata &   \nodata & 6.613e+03 & 5.340e+03 & 6.171e+02 & 4.821e+01 & 2.535e-02 & 2.918e-03 & 7.880e+02 & 4.464e+01 & 6.060e-03 & 5.457e-04 & 3.689e+03 & 7.561e+01 & 1 \\
201411387 & 11232145-0047048 & 170.839412 &  -0.784719 & 38.573 & 3.427 & 3.782 & 0.032 & 5.361e+03 & 3.500e+03 & 7.352e+02 & 5.184e+01 & 3.525e-02 & 3.285e-03 & 7.006e+02 & 4.463e+01 & 7.732e-03 & 6.768e-04 & 3.371e+03 & 6.706e+01 & 1 \\
201414774 & 11252575-0044016 & 171.357304 &  -0.733783 & 4.869 & 0.363 &   \nodata &   \nodata & 7.325e+05 & 3.186e+05 & 2.949e+03 & 5.548e+02 & 2.350e-01 & 2.549e-02 & 3.273e+03 & 2.135e+02 & 6.460e-02 & 4.725e-03 & 2.207e+04 & 4.225e+02 & 1 \\
201415775 & 11333341-0043060 & 173.389225 &  -0.718339 & 27.298 & 0.294 &   \nodata &   \nodata & 1.405e+04 & 2.691e+03 & 9.926e+02 & 7.152e+01 & 4.837e-02 & 2.932e-03 & 5.915e+02 & 3.877e+01 & 9.928e-03 & 4.971e-04 & 3.234e+01 & 2.713e+00 & 2 \\
201420175 & 11334091-0039085 & 173.420575 &  -0.652361 & 33.244 & 0.805 &   \nodata &   \nodata & 6.643e+03 & 1.246e+03 & 6.136e+02 & 4.458e+01 & 3.751e-02 & 3.055e-03 & 7.234e+02 & 2.309e+01 & 7.648e-03 & 2.576e-04 & 1.427e+02 & 3.323e+00 & 2 \\
201438887 & 11361104-0022409 & 174.046021 &  -0.378047 & 30.272 & 2.411 &   \nodata &   \nodata & 1.274e+04 & 9.403e+03 & 9.159e+02 & 7.223e+01 & 4.535e-02 & 3.834e-03 & 8.394e+02 & 5.464e+01 & 9.200e-03 & 8.658e-04 & 5.250e+03 & 1.055e+02 & 1 \\
201454791 & 11453665-0009037 & 176.402708 &  -0.151056 & 130.534 & 1.069 &   \nodata &   \nodata & 4.305e+02 & 3.249e+01 & 3.025e+02 & 1.218e+01 & 9.235e-03 & 5.605e-04 & 2.150e+02 & 1.463e+01 & 2.935e-03 & 2.704e-04 & 3.790e+01 & 1.434e+00 & 2 \\
201459357 & 11255698-0005024 & 171.487429 &  -0.084022 & 3.972 &   \nodata &   \nodata &   \nodata & 2.560e+05 & 2.209e+05 & 2.544e+03 & 4.206e+02 & 2.651e-01 & 3.377e-02 & 2.398e+03 & 1.666e+02 & 8.400e-02 & 8.039e-03 & 1.014e+04 & 1.748e+02 & 1 \\
201467358 & 11422234+0002007 & 175.593208 &   0.033297 & 113.094 & 0.658 & 9.942 & 0.032 & 2.495e+02 & 2.104e+01 & 2.694e+02 & 8.544e+00 & 1.167e-02 & 4.811e-04 & 1.895e+02 & 7.096e+00 & 3.102e-03 & 1.254e-04 & 8.390e+00 & 4.052e-01 & 2 \\
201472519 & 11441126+0006347 & 176.046942 &   0.109667 & 29.995 & 0.735 &   \nodata &   \nodata & 1.290e+04 & 2.561e+03 & 9.782e+02 & 7.404e+01 & 3.403e-02 & 2.911e-03 & 9.752e+02 & 5.070e+01 & 1.020e-02 & 4.305e-04 & 3.790e+01 & 1.323e+00 & 2 \\
201503634 & 11302841+0034387 & 172.618392 &   0.577436 & 5.408 & 0.378 &   \nodata &   \nodata & 3.108e+05 & 1.452e+05 & 2.163e+03 & 3.068e+02 & 2.004e-01 & 2.316e-02 & 2.238e+03 & 1.779e+02 & 6.368e-02 & 6.244e-03 & 1.052e+04 & 1.785e+02 & 1 \\
201508025 & 11110893+0038392 & 167.787258 &   0.644258 & 88.027 & 7.028 &   \nodata &   \nodata & 7.613e+02 & 5.025e+02 & 3.710e+02 & 2.034e+01 & 1.372e-02 & 1.205e-03 & 3.613e+02 & 1.763e+01 & 4.171e-03 & 3.231e-04 & 2.669e+02 & 6.714e+00 & 1 \\
201508594 & 11260798+0039130 & 171.533283 &   0.653625 & 34.168 & 4.443 & 2.927 & 0.029 & 9.747e+03 & 9.512e+03 & 7.774e+02 & 5.589e+01 & 3.514e-02 & 3.750e-03 & 7.343e+02 & 4.661e+01 & 8.281e-03 & 8.927e-04 & 3.624e+03 & 5.932e+01 & 1 \\
201512825 & 11432842+0043107 & 175.868425 &   0.719658 & 79.011 & 0.804 &   \nodata &   \nodata & 1.109e+03 & 9.890e+01 & 3.580e+02 & 1.692e+01 & 1.495e-02 & 1.033e-03 & 3.385e+02 & 1.569e+01 & 4.117e-03 & 1.831e-04 & 3.418e+01 & 1.180e+00 & 2 \\
201515047 & 11504834+0045163 & 177.701492 &   0.754572 & 249.915 & 5.820 & 17.348 & 0.006 & 1.538e+02 & 1.885e+01 & 2.526e+02 & 8.249e+00 & 7.260e-03 & 4.611e-04 & 2.237e+02 & 8.993e+00 & 1.578e-03 & 1.442e-04 & 6.159e+01 & 4.456e-01 & 1 \\
201525065 & 11450722+0054340 & 176.280079 &   0.909461 & 23.455 & 0.507 &   \nodata &   \nodata & 1.784e+04 & 2.212e+03 & 1.727e+03 & 4.306e+01 & 6.146e-02 & 2.692e-03 & 5.076e+02 & 5.357e+01 & 1.706e-02 & 1.278e-03 & 1.775e+01 & 2.482e+00 & 2 \\
201526688 & 11461610+0056057 & 176.567104 &   0.934933 & 36.565 & 4.112 & 4.703 & 0.047 & 7.871e+03 & 6.151e+03 & 7.356e+02 & 6.506e+01 & 3.705e-02 & 4.139e-03 & 8.143e+02 & 5.311e+01 & 8.109e-03 & 6.968e-04 & 4.712e+03 & 8.413e+01 & 1 \\
201555742 & 11272268+0122258 & 171.844575 &   1.373844 & 46.977 & 0.559 & 5.220 & 0.010 & 5.428e+03 & 6.818e+02 & 7.309e+02 & 3.682e+01 & 2.409e-02 & 2.847e-03 & 6.774e+02 & 3.570e+01 & 6.013e-03 & 2.915e-04 & 7.148e+01 & 2.407e+00 & 2 \\
201584221 & 11102278+0147570 & 167.594996 &   1.799169 & 54.805 & 2.963 &   \nodata &   \nodata & 2.233e+03 & 1.476e+03 & 4.817e+02 & 2.958e+01 & 2.349e-02 & 2.103e-03 & 5.422e+02 & 2.718e+01 & 5.404e-03 & 3.761e-04 & 1.134e+03 & 2.482e+01 & 1 \\
201601287 & 11151790+0203483 & 168.824604 &   2.063439 & 57.788 & 0.667 &   \nodata &   \nodata & 2.676e+03 & 4.368e+02 & 7.412e+02 & 3.328e+01 & 2.258e-02 & 1.598e-03 & 6.237e+02 & 2.412e+01 & 5.807e-03 & 1.875e-04 & 7.126e+01 & 1.356e+00 & 2 \\
201614936 & 11352447+0216273 & 173.851979 &   2.274258 & 4.350 & 0.743 &   \nodata &   \nodata & 1.796e+05 & 1.486e+05 & 2.381e+03 & 4.011e+02 & 2.420e-01 & 4.810e-02 & 2.115e+03 & 2.292e+02 & 7.717e-02 & 1.132e-02 & 8.692e+03 & 6.729e+01 & 1 \\
201615435 & 11082866+0216551 & 167.119350 &   2.281997 & 90.062 & 6.909 & 6.507 & 0.006 & 5.094e+02 & 3.152e+02 & 3.104e+02 & 2.022e+01 & 1.320e-02 & 1.357e-03 & 3.715e+02 & 1.554e+01 & 4.037e-03 & 2.998e-04 & 2.541e+02 & 7.043e+00 & 1 \\
201619206 & 11204791+0220248 & 170.199588 &   2.340211 & 30.413 & 2.190 &   \nodata &   \nodata & 5.101e+03 & 3.665e+03 & 7.108e+02 & 5.437e+01 & 4.403e-02 & 3.538e-03 & 5.629e+02 & 4.172e+01 & 9.954e-03 & 8.876e-04 & 2.377e+03 & 4.599e+01 & 1 \\
201620616 & 12012949+0221443 & 180.372892 &   2.362361 & 97.836 & 1.976 &   \nodata &   \nodata & 2.716e+02 & 4.170e+01 & 2.263e+02 & 1.223e+01 & 1.206e-02 & 9.010e-04 & 2.105e+02 & 1.176e+01 & 3.834e-03 & 2.662e-04 & 7.108e+01 & 1.758e+00 & 1 \\
201629034 & 12000308+0229322 & 180.012879 &   2.492258 & 91.308 & 4.808 & 7.088 & 0.014 & 9.854e+02 & 3.294e+02 & 3.062e+02 & 2.563e+01 & 1.374e-02 & 1.447e-03 & 3.960e+02 & 2.100e+01 & 4.076e-03 & 3.098e-04 & 4.810e+02 & 1.133e+01 & 1 \\
201635902 & 11164224+0235580 & 169.176033 &   2.599492 & 56.061 & 3.460 &   \nodata &   \nodata & 2.892e+03 & 1.311e+03 & 4.498e+02 & 3.385e+01 & 2.604e-02 & 3.015e-03 & 4.963e+02 & 3.266e+01 & 5.784e-03 & 4.880e-04 & 1.286e+03 & 2.702e+01 & 1 \\
201636027 & 11335625+0236055 & 173.484437 &   2.601469 & 3.183 &   \nodata &   \nodata &   \nodata & 9.458e+05 & 2.102e+05 & 2.693e+03 & 2.718e+02 & 3.542e-01 & 3.987e-02 & 2.063e+03 & 1.608e+02 & 1.128e-01 & 8.169e-03 & 8.455e+00 & 1.157e+00 & 1 \\
201643723 & 11215719+0243098 & 170.488400 &   2.719483 & 35.924 & 3.088 & 5.813 & 0.460 & 3.812e+03 & 3.107e+03 & 6.480e+02 & 4.616e+01 & 3.839e-02 & 3.996e-03 & 6.961e+02 & 3.003e+01 & 7.896e-03 & 5.978e-04 & 1.973e+03 & 3.963e+01 & 1 \\
201659364 & 11162508+0257536 & 169.104496 &   2.964933 & 62.417 & 4.133 & 3.927 & 0.022 & 2.301e+03 & 1.301e+03 & 4.730e+02 & 3.324e+01 & 2.055e-02 & 2.340e-03 & 5.534e+02 & 3.206e+01 & 5.147e-03 & 3.930e-04 & 1.266e+03 & 2.928e+01 & 1 \\
201667626 & 11241443+0306133 & 171.060271 &   3.103350 & 30.137 & 1.307 &   \nodata &   \nodata & 5.854e+03 & 1.984e+03 & 6.626e+02 & 5.740e+01 & 3.759e-02 & 3.099e-03 & 8.035e+02 & 3.007e+01 & 9.093e-03 & 2.954e-04 & 5.443e+01 & 1.397e+00 & 2 \\
201680913 & 12014373+0319161 & 180.432225 &   3.321144 & 55.107 & 2.315 & 4.207 & 0.013 & 2.271e+03 & 1.026e+03 & 3.333e+02 & 2.600e+01 & 2.490e-02 & 3.454e-03 & 4.304e+02 & 2.612e+01 & 5.496e-03 & 4.999e-04 & 1.001e+03 & 2.259e+01 & 1 \\
201696051 & 12044299+0334230 & 181.179146 &   3.573111 & 88.357 & 3.333 &   \nodata &   \nodata & 8.036e+02 & 1.932e+02 & 4.894e+02 & 2.670e+01 & 1.251e-02 & 8.702e-04 & 4.415e+02 & 2.593e+01 & 4.323e-03 & 2.850e-04 & 1.950e+02 & 5.450e+00 & 2 \\
201697539 & 11482814+0335534 & 177.117221 &   3.598100 & 15.789 & 0.769 &   \nodata &   \nodata & 1.080e+04 & 5.597e+03 & 1.201e+03 & 1.332e+02 & 8.025e-02 & 6.970e-03 & 6.971e+02 & 6.035e+01 & 1.973e-02 & 1.773e-03 & 2.106e+03 & 3.720e+01 & 1 \\
201700607 & 11282997+0338594 & 172.124917 &   3.649897 & 81.719 & 5.318 & 7.558 & 0.052 & 9.614e+02 & 4.903e+02 & 3.906e+02 & 2.278e+01 & 1.605e-02 & 1.690e-03 & 3.830e+02 & 2.185e+01 & 4.127e-03 & 3.552e-04 & 5.028e+02 & 1.114e+01 & 1 \\
201702907 & \nodata & 178.026996 &   3.688792 & 74.988 & 7.398 &   \nodata &   \nodata & 1.618e+03 & 1.135e+03 & 4.119e+02 & 3.300e+01 & 1.480e-02 & 1.662e-03 & 5.439e+02 & 2.490e+01 & 4.621e-03 & 2.693e-04 & 5.542e+02 & 1.268e+01 & 1 \\
201704568 & 11261591+0342583 & 171.566321 &   3.716222 & 97.876 & 2.453 &   \nodata &   \nodata & 6.111e+02 & 1.180e+02 & 2.721e+02 & 1.385e+01 & 1.457e-02 & 1.240e-03 & 2.750e+02 & 1.497e+01 & 3.569e-03 & 2.790e-04 & 2.393e+02 & 5.845e+00 & 1 \\
201713224 & 11264919+0351517 & 171.704958 &   3.864342 & 31.338 & 0.406 & 3.038 & 0.002 & 5.379e+03 & 9.801e+02 & 4.816e+02 & 3.372e+01 & 3.413e-02 & 3.187e-03 & 5.307e+02 & 2.088e+01 & 9.077e-03 & 3.144e-04 & 1.661e+01 & 4.746e-01 & 2 \\
201720476 & 11143920+0359154 & 168.663354 &   3.987589 & 83.725 & 7.192 & 6.103 & 0.722 & 1.801e+03 & 1.140e+03 & 5.019e+02 & 4.172e+01 & 1.395e-02 & 1.486e-03 & 5.545e+02 & 3.616e+01 & 4.061e-03 & 3.630e-04 & 1.023e+03 & 2.439e+01 & 1 \\
201722766 & 11240739+0401380 & 171.030762 &   4.027258 & 88.538 & 4.371 &   \nodata &   \nodata & 6.930e+02 & 2.708e+02 & 3.104e+02 & 1.907e+01 & 1.441e-02 & 1.370e-03 & 3.437e+02 & 1.855e+01 & 3.847e-03 & 3.201e-04 & 3.768e+02 & 9.177e+00 & 1 \\
201723568 & 11360352+0402289 & 174.014671 &   4.041367 & 5.372 & 0.363 & 0.743 & 0.024 & 1.565e+06 & 9.312e+05 & 2.169e+03 & 3.964e+02 & 1.931e-01 & 2.499e-02 & 2.648e+03 & 3.212e+02 & 6.169e-02 & 6.429e-03 & 1.907e+05 & 2.827e+03 & 1 \\
201724514 & 11582220+0403262 & 179.592525 &   4.057289 & 29.111 & 0.611 &   \nodata &   \nodata & 1.355e+04 & 2.703e+03 & 1.070e+03 & 7.104e+01 & 3.660e-02 & 2.477e-03 & 9.358e+02 & 4.733e+01 & 9.489e-03 & 4.001e-04 & 4.201e+01 & 1.246e+00 & 2 \\
201724852 & 11122791+0403471 & 168.116350 &   4.063117 & 94.344 & 6.219 & 10.897 & 0.046 & 5.750e+02 & 2.997e+02 & 3.066e+02 & 1.903e+01 & 1.342e-02 & 1.369e-03 & 3.039e+02 & 1.823e+01 & 3.961e-03 & 3.720e-04 & 3.552e+02 & 8.529e+00 & 1 \\
201726163 & 11150267+0405078 & 168.761167 &   4.085508 & 93.244 & 2.575 & 8.602 & 0.038 & 1.338e+03 & 3.435e+02 & 3.942e+02 & 3.385e+01 & 1.153e-02 & 1.159e-03 & 5.179e+02 & 2.848e+01 & 4.104e-03 & 2.688e-04 & 5.624e+02 & 1.442e+01 & 1 \\
201729267 & 11230257+0408173 & 170.760737 &   4.138147 & 92.008 & 2.000 &   \nodata &   \nodata & 5.634e+02 & 1.083e+02 & 2.784e+02 & 1.703e+01 & 1.185e-02 & 8.909e-04 & 3.272e+02 & 1.350e+01 & 3.366e-03 & 1.919e-04 & 1.166e+02 & 3.677e+00 & 1 \\
201733406 & 11213386+0412299 & 170.391112 &   4.208306 & 90.679 & 1.445 & 9.181 & 1.372 & 5.055e+02 & 7.600e+01 & 2.350e+02 & 1.280e+01 & 1.409e-02 & 1.177e-03 & 2.298e+02 & 1.339e+01 & 3.770e-03 & 3.028e-04 & 1.147e+02 & 3.052e+00 & 2 \\
201743103 & 11170064+0421565 & 169.252658 &   4.365717 & 84.644 & 5.676 &   \nodata &   \nodata & 1.348e+03 & 6.447e+02 & 4.056e+02 & 2.541e+01 & 1.562e-02 & 1.488e-03 & 4.229e+02 & 2.519e+01 & 4.144e-03 & 3.356e-04 & 6.171e+02 & 1.551e+01 & 1 \\
201747404 & 11390558+0426188 & 174.773221 &   4.438600 & 156.608 & 1.034 & 12.324 & 0.674 & 3.123e+02 & 2.315e+01 & 2.616e+02 & 8.140e+00 & 9.993e-03 & 6.149e-04 & 2.087e+02 & 8.171e+00 & 2.669e-03 & 1.708e-04 & 3.179e+01 & 1.310e+00 & 2 \\
201749662 & 11153895+0428400 & 168.912304 &   4.477769 & 90.575 & 1.455 &   \nodata &   \nodata & 4.238e+02 & 5.398e+01 & 2.296e+02 & 1.100e+01 & 1.732e-02 & 1.492e-03 & 1.801e+02 & 1.138e+01 & 3.903e-03 & 3.090e-04 & 9.589e+01 & 2.272e+00 & 2 \\
201750985 & 11292465+0429584 & 172.352729 &   4.499581 & 27.465 & 0.976 &   \nodata &   \nodata & 1.390e+04 & 4.681e+03 & 9.214e+02 & 7.938e+01 & 4.705e-02 & 4.105e-03 & 7.151e+02 & 6.105e+01 & 1.146e-02 & 1.011e-03 & 5.044e+03 & 9.507e+01 & 1 \\
201751998 & 11160874+0431029 & 169.036404 &   4.517472 & 54.093 & 1.772 &   \nodata &   \nodata & 2.776e+03 & 1.131e+03 & 3.731e+02 & 2.817e+01 & 2.365e-02 & 2.788e-03 & 5.354e+02 & 2.422e+01 & 5.059e-03 & 3.211e-04 & 8.587e+02 & 1.943e+01 & 1 \\
201752633 & 11121723+0431442 & 168.071796 &   4.528894 & 106.785 & 6.842 &   \nodata &   \nodata & 3.175e+02 & 1.511e+02 & 3.065e+02 & 1.540e+01 & 1.126e-02 & 9.700e-04 & 3.141e+02 & 1.355e+01 & 3.527e-03 & 2.658e-04 & 1.077e+02 & 3.486e+00 & 1 \\
201758449 & 11175773+0437487 & 169.490575 &   4.630186 & 99.409 & 5.084 &   \nodata &   \nodata & 6.210e+02 & 2.615e+02 & 3.469e+02 & 1.888e+01 & 1.279e-02 & 1.011e-03 & 3.648e+02 & 1.952e+01 & 3.452e-03 & 2.673e-04 & 3.053e+02 & 8.457e+00 & 1 \\
201761560 & 11152022+0441023 & 168.834287 &   4.683992 & 88.898 & 3.367 &   \nodata &   \nodata & 1.347e+03 & 5.248e+02 & 4.625e+02 & 2.959e+01 & 1.286e-02 & 1.096e-03 & 4.554e+02 & 2.774e+01 & 4.162e-03 & 3.229e-04 & 7.687e+02 & 6.284e+00 & 1 \\
201763504 & 11483335+0443022 & 177.138908 &   4.717314 & 8.609 & 0.139 & 1.177 & 0.158 & 8.146e+05 & 1.573e+05 & 5.359e+03 & 3.332e+02 & 1.075e-01 & 6.684e-03 & 1.519e+03 & 2.472e+02 & 4.300e-02 & 3.608e-03 & 3.880e-01 & 9.678e-01 & 2 \\
201765667 & 11145908+0445177 & 168.746204 &   4.754925 & 85.684 & 4.221 & 7.375 & 0.050 & 1.266e+03 & 4.148e+02 & 3.232e+02 & 2.353e+01 & 1.742e-02 & 1.930e-03 & 4.094e+02 & 2.394e+01 & 4.006e-03 & 3.893e-04 & 6.068e+02 & 1.375e+01 & 1 \\
201766812 & 11150311+0446301 & 168.762996 &   4.775025 & 100.462 & 4.872 &   \nodata &   \nodata & 1.175e+03 & 4.130e+02 & 4.083e+02 & 2.828e+01 & 1.300e-02 & 1.230e-03 & 4.734e+02 & 2.518e+01 & 3.561e-03 & 3.001e-04 & 5.436e+02 & 1.521e+01 & 1 \\
201772439 & 11112479+0452234 & 167.853371 &   4.873231 & 98.648 & 2.035 &   \nodata &   \nodata & 1.127e+03 & 2.578e+02 & 4.209e+02 & 2.540e+01 & 1.434e-02 & 1.340e-03 & 4.923e+02 & 1.984e+01 & 3.732e-03 & 2.733e-04 & 3.643e+02 & 1.059e+01 & 1 \\
201774359 & 11162408+0454236 & 169.100333 &   4.906556 & 72.644 & 5.287 &   \nodata &   \nodata & 1.120e+03 & 5.598e+02 & 3.210e+02 & 1.958e+01 & 1.865e-02 & 1.844e-03 & 3.651e+02 & 1.782e+01 & 4.662e-03 & 3.498e-04 & 4.071e+02 & 9.163e+00 & 1 \\
201774883 & 11163452+0454529 & 169.143833 &   4.914697 & 83.505 & 5.950 &   \nodata &   \nodata & 1.197e+03 & 6.494e+02 & 3.223e+02 & 2.548e+01 & 1.453e-02 & 1.803e-03 & 3.983e+02 & 2.150e+01 & 4.235e-03 & 3.786e-04 & 6.295e+02 & 1.350e+01 & 1 \\
201781960 & 11103146+0502268 & 167.631062 &   5.040825 & 81.685 & 6.281 & 9.842 & 0.731 & 1.279e+03 & 8.737e+02 & 4.916e+02 & 3.903e+01 & 1.377e-02 & 1.557e-03 & 4.964e+02 & 3.245e+01 & 4.500e-03 & 3.862e-04 & 8.033e+02 & 1.852e+01 & 1 \\
201786083 & 11125055+0506500 & 168.210683 &   5.113931 & 80.724 & 5.662 & 5.613 & 0.018 & 1.103e+03 & 6.570e+02 & 3.848e+02 & 2.605e+01 & 1.420e-02 & 1.407e-03 & 4.465e+02 & 2.663e+01 & 4.407e-03 & 3.542e-04 & 7.038e+02 & 1.520e+01 & 1 \\
201788284 & 11101971+0509085 & 167.582175 &   5.152367 & 90.274 & 4.934 &   \nodata &   \nodata & 9.987e+02 & 4.206e+02 & 2.776e+02 & 2.451e+01 & 1.322e-02 & 1.486e-03 & 3.942e+02 & 2.029e+01 & 3.674e-03 & 2.790e-04 & 5.087e+02 & 4.786e+00 & 1 \\
201797512 & 11235883+0519178 & 170.995146 &   5.321614 & 113.952 & 0.674 &   \nodata &   \nodata & 7.773e+02 & 7.031e+01 & 3.354e+02 & 1.110e+01 & 1.249e-02 & 6.951e-04 & 2.692e+02 & 1.141e+01 & 3.120e-03 & 1.892e-04 & 5.433e+01 & 1.841e+00 & 2 \\
201797810 & 11110894+0519382 & 167.787300 &   5.327325 & 82.795 & 4.936 & 8.705 & 0.027 & 1.988e+03 & 8.873e+02 & 3.949e+02 & 3.293e+01 & 1.766e-02 & 2.020e-03 & 5.200e+02 & 3.276e+01 & 3.878e-03 & 3.651e-04 & 9.518e+02 & 2.400e+01 & 1 \\
201825690 & \nodata & 167.827858 &   5.836272 & 61.013 & 2.837 &   \nodata &   \nodata & 3.625e+03 & 1.530e+03 & 5.475e+02 & 3.983e+01 & 2.117e-02 & 2.681e-03 & 6.541e+02 & 4.011e+01 & 5.117e-03 & 4.068e-04 & 1.739e+03 & 4.163e+01 & 1 \\
201830769 & 11134232+0555598 & 168.426371 &   5.933281 & 83.880 & 4.531 & 6.318 & 0.702 & 1.087e+03 & 4.195e+02 & 4.978e+02 & 2.741e+01 & 1.443e-02 & 1.171e-03 & 4.417e+02 & 2.863e+01 & 4.251e-03 & 3.460e-04 & 6.077e+02 & 1.508e+01 & 1 \\
201839927 & \nodata & 174.142337 &   6.106439 & 27.190 & 0.344 &   \nodata &   \nodata & 2.258e+04 & 4.173e+03 & 1.576e+03 & 2.150e+02 & 5.011e-02 & 4.196e-03 & 8.107e+02 & 6.529e+01 & 9.456e-03 & 8.908e-04 & 1.334e+02 & 1.323e+01 & 2 \\
201843056 & 11392070+0609576 & 174.836300 &   6.166011 & 93.862 & 2.045 & 6.147 & 0.013 & 4.633e+02 & 9.604e+01 & 3.635e+02 & 1.424e+01 & 1.622e-02 & 1.213e-03 & 2.968e+02 & 1.352e+01 & 3.880e-03 & 2.572e-04 & 1.565e+02 & 4.012e+00 & 1 \\
201843394 & 11105997+0610194 & 167.749917 &   6.172050 & 84.029 & 5.168 & 5.683 & 0.012 & 9.430e+02 & 4.859e+02 & 4.234e+02 & 2.726e+01 & 1.553e-02 & 1.771e-03 & 5.008e+02 & 2.251e+01 & 4.261e-03 & 3.092e-04 & 4.732e+02 & 1.274e+01 & 1 \\
201843809 & 11113052+0610473 & 167.877196 &   6.179844 & 93.392 & 3.923 &   \nodata &   \nodata & 8.797e+02 & 3.696e+02 & 4.266e+02 & 2.785e+01 & 1.140e-02 & 1.027e-03 & 4.573e+02 & 2.556e+01 & 3.769e-03 & 2.820e-04 & 3.918e+02 & 1.103e+01 & 2 \\
201846331 & 11535849+0613495 & 178.493754 &   6.230472 & 74.569 & 4.161 & 5.347 & 0.005 & 7.916e+02 & 3.287e+02 & 4.056e+02 & 1.744e+01 & 1.933e-02 & 1.399e-03 & 3.334e+02 & 1.620e+01 & 4.535e-03 & 2.940e-04 & 1.820e+02 & 4.964e+00 & 1 \\
201852681 & 11423531+0621280 & 175.647167 &   6.357800 & 34.347 & 0.918 & 3.053 & 0.016 & 6.068e+03 & 1.227e+03 & 6.915e+02 & 4.496e+01 & 3.609e-02 & 2.687e-03 & 7.504e+02 & 2.751e+01 & 8.389e-03 & 2.798e-04 & 9.884e+01 & 2.325e+00 & 2 \\
201877455 & 11554329+0651459 & 178.930404 &   6.862772 & 83.084 & 1.515 & 7.465 & 0.041 & 9.033e+02 & 1.019e+02 & 6.456e+02 & 2.115e+01 & 1.486e-02 & 8.137e-04 & 4.026e+02 & 2.368e+01 & 4.202e-03 & 2.230e-04 & 4.734e+01 & 1.724e+00 & 2 \\
201881721 & 11505049+0656500 & 177.710400 &   6.947208 & 56.167 & 3.041 & 4.883 & 0.053 & 2.408e+03 & 1.224e+03 & 4.115e+02 & 2.739e+01 & 2.580e-02 & 2.923e-03 & 5.347e+02 & 2.637e+01 & 4.729e-03 & 2.945e-04 & 9.605e+02 & 2.122e+01 & 1 \\
201887247 & 11552041+0703364 & 178.835058 &   7.060106 & 93.323 & 5.441 &   \nodata &   \nodata & 8.127e+02 & 3.055e+02 & 3.199e+02 & 1.927e+01 & 1.462e-02 & 1.235e-03 & 3.370e+02 & 1.832e+01 & 3.863e-03 & 3.019e-04 & 3.730e+02 & 9.010e+00 & 1 \\
201907942 & 11252290+0729247 & 171.345400 &   7.490258 & 62.643 & 3.740 & 9.109 & 1.316 & 2.919e+03 & 1.089e+03 & 4.923e+02 & 3.618e+01 & 2.036e-02 & 2.139e-03 & 4.964e+02 & 3.419e+01 & 5.737e-03 & 4.723e-04 & 1.382e+03 & 2.843e+01 & 1 \\
201908986 & 11433272+0730459 & 175.886379 &   7.512714 & 92.651 & 2.459 & 5.856 & 0.578 & 8.239e+02 & 3.185e+02 & 3.036e+02 & 2.007e+01 & 1.413e-02 & 1.444e-03 & 3.869e+02 & 1.698e+01 & 3.735e-03 & 2.881e-04 & 3.751e+02 & 1.027e+01 & 1 \\
201913188 & 11515330+0736060 & 177.972125 &   7.601786 & 91.410 & 3.206 & 7.372 & 0.009 & 1.128e+03 & 2.642e+02 & 2.886e+02 & 2.170e+01 & 1.397e-02 & 1.732e-03 & 4.050e+02 & 1.973e+01 & 4.000e-03 & 2.965e-04 & 4.892e+02 & 1.142e+01 & 1 \\
201944519 & 11360006+0818157 & 174.000267 &   8.304389 & 84.794 & 5.014 &   \nodata &   \nodata & 1.827e+03 & 8.741e+02 & 4.207e+02 & 3.387e+01 & 1.372e-02 & 1.680e-03 & 6.126e+02 & 2.536e+01 & 4.357e-03 & 3.054e-04 & 8.930e+02 & 2.046e+01 & 1 \\
  \enddata
  \vspace{-0.2cm}
 \tablenotetext{a}{Certain giants (2) and giant candidates (1). See
   text for details on the classification.}
    \vspace{-0.2cm}
   \tablecomments{Properties of the C1 solar-like oscillator sample returned by running all non-GAP C1 stars through BAM, and which visual inspection indicated were certain giants (`yes' in the text) or giant candidates (`maybe' in the text). Stars with $\numax \lesssim 4 \muhz$ should be considered upper limits,
     as mentioned in the text, and are not assigned errors.}

  \end{deluxetable*}
\end{longrotatetable}

\begin{table*}
  \begin{tabular*}{\textwidth}{cccp{7.0cm}}
    \hline \hline
    Guest Observer ID & Giant fraction (yes) & Giant fraction (maybe) & Notes \\ \hline
GO1001 & $     0/     3$ & $     0/     3$ &  \\
GO1002 & $     1/    30$ & $     0/    30$ &  \\
GO1003 & $     0/     2$ & $     0/     2$ & Targeted extremely red stars, many likely to be AGB and long-period variables, which would not have been selected by BAM because their frequencies would be below our cutoff of $3\muhz$. \\
GO1005 & $     0/    16$ & $     0/    16$ &  \\
GO1006 & $     0/    20$ & $     0/    20$ &  \\
GO1014 & $     0/     1$ & $     0/     1$ &  \\
GO1021 & $     0/     1$ & $     0/     1$ &  \\
GO1023 & $     0/     4$ & $     0/     4$ &  \\
GO1026 & $     0/     2$ & $     0/     2$ & Targeted eclipsing binaries, some of which may be giants. \\
GO1027 & $     2/    50$ & $     1/    50$ & Targeted AF-type stars, the coolest of which might be oscillating giants. \\
GO1029 & $     0/     1$ & $     0/     1$ &  \\
GO1030 & $     0/     1$ & $     0/     1$ &  \\
GO1036 & $     0/    38$ & $     0/    38$ &  \\
GO1038 & $     0/    12$ & $     1/    12$ & Targeted potential oscillators. \\
GO1040 & $     5/     6$ & $     0/     6$ & Targeted bright giants. \\
GO1043 & $     0/    25$ & $     0/    25$ &  \\
GO1046 & $     0/     3$ & $     0/     3$ & Targeted bright stars, among them three subgiants, which likely will not oscillate below the long cadence Nyquist frequency of $\sim 283\muhz$. \\
GO1052 & $     0/     1$ & $     0/     1$ &  \\
GO1053 & $     0/     1$ & $     0/     1$ &  \\
GO1054 & $     9/  2092$ & $     5/  2092$ &  \\
GO1055 & $     0/    39$ & $     0/    39$ &  \\
GO1057 & $     0/     1$ & $     0/     1$ & Targeted giant oscillators, and this object was missed by BAM. \\
GO1061 & $     2/     7$ & $     0/     7$ &  \\
GO1062 & $     3/     4$ & $     0/     4$ &  \\
GO1066 & $     3/     3$ & $     0/     3$ & Targeted subgiants. \\
GO1068 & $     0/     3$ & $     0/     3$ & Targeted eclipsing binaries, some of which may be giants. \\
GO1069 & $     0/     6$ & $     0/     6$ &  \\
GO1072 & $     0/     4$ & $     0/     4$ &  \\
GO1073 & $     0/    10$ & $     0/    10$ &  \\
GO1074 & $     1/     3$ & $     0/     3$ & Targeted extra-galactic objects. \\
  \end{tabular*}
    \caption{The number of confirmed and marginal giants discussed in this paper
    found in the observed targets of various Guest Observer proposals
    gives an indication of the success at rejecting giants using color
    and proper motion cuts. Note that the tabulated numbers only
    include targets that had long cadence data. Unless otherwise noted above, the Guest Observer proposals did
    not, to our knowledge, target giants. We have not listed GO1059,
    because that is the GAP.}
    \label{tab:purity}
  \end{table*}

\vspace{-10cm}

\clearpage

\bibliography{bib}

\begin{thebibliography}{}
\expandafter\ifx\csname natexlab\endcsname\relax\def\natexlab#1{#1}\fi
\providecommand{\url}[1]{\href{#1}{#1}}

\bibitem[{{Aigrain} {et~al.}(2016){Aigrain}, {Parviainen}, \&
  {Pope}}]{aigrain_parviainen&pope2016}
{Aigrain}, S., {Parviainen}, H., \& {Pope}, B.~J.~S. 2016, \mnras, 459, 2408

\bibitem[{{Anderson} {et~al.}(1990){Anderson}, {Duvall}, \&
  {Jefferies}}]{anderson_duvall&jefferies1990}
{Anderson}, E.~R., {Duvall}, Jr., T.~L., \& {Jefferies}, S.~M. 1990, \apj, 364,
  699

\bibitem[{{Angus} {et~al.}(2016){Angus}, {Foreman-Mackey}, \&
  {Johnson}}]{angus_foreman-mackey_johnson2016}
{Angus}, R., {Foreman-Mackey}, D., \& {Johnson}, J.~A. 2016, \apj, 818, 109

\bibitem[{{Appourchaux}(2003)}]{appourchaux2003}
{Appourchaux}, T. 2003, \aap, 412, 903

\bibitem[{{Armstrong} {et~al.}(2015){Armstrong}, {Kirk}, {Lam}, {McCormac},
  {Walker}, {Brown}, {Osborn}, {Pollacco}, \& {Spake}}]{armstrong+2015}
{Armstrong}, D.~J., {Kirk}, J., {Lam}, K.~W.~F., {et~al.} 2015, \aap, 579, A19

\bibitem[{{Baglin} {et~al.}(2006){Baglin}, {Michel}, {Auvergne}, \& {COROT
  Team}}]{baglin+2006}
{Baglin}, A., {Michel}, E., {Auvergne}, M., \& {COROT Team}. 2006, in ESA
  Special Publication, Vol. 624, Proceedings of SOHO 18/GONG 2006/HELAS I,
  Beyond the spherical Sun, 34.1

\bibitem[{{Bedding} \& {Kjeldsen}(2010)}]{bedding&kjeldsen2010}
{Bedding}, T.~R., \& {Kjeldsen}, H. 2010, Communications in Asteroseismology,
  161, 3

\bibitem[{{Bedding} {et~al.}(2010){Bedding}, {Huber}, {Stello}, {Elsworth},
  {Hekker}, {Kallinger}, {Mathur}, {Mosser}, {Preston}, {Ballot}, {Barban},
  {Broomhall}, {Buzasi}, {Chaplin}, {Garc{\'{\i}}a}, {Gruberbauer}, {Hale}, {De
  Ridder}, {Frandsen}, {Borucki}, {Brown}, {Christensen-Dalsgaard},
  {Gilliland}, {Jenkins}, {Kjeldsen}, {Koch}, {Belkacem}, {Bildsten}, {Bruntt},
  {Campante}, {Deheuvels}, {Derekas}, {Dupret}, {Goupil}, {Hatzes}, {Houdek},
  {Ireland}, {Jiang}, {Karoff}, {Kiss}, {Lebreton}, {Miglio}, {Montalb{\'a}n},
  {Noels}, {Roxburgh}, {Sangaralingam}, {Stevens}, {Suran}, {Tarrant}, \&
  {Weiss}}]{bedding+2010}
{Bedding}, T.~R., {Huber}, D., {Stello}, D., {et~al.} 2010, \apjl, 713, L176

\bibitem[{{Borucki} {et~al.}(2008){Borucki}, {Koch}, {Basri}, {Batalha},
  {Brown}, {Caldwell}, {Christensen-Dalsgaard}, {Cochran}, {Dunham}, {Gautier},
  {Geary}, {Gilliland}, {Jenkins}, {Kondo}, {Latham}, {Lissauer}, \&
  {Monet}}]{borucki+2008}
{Borucki}, W., {Koch}, D., {Basri}, G., {et~al.} 2008, in IAU Symposium, Vol.
  249, Exoplanets: Detection, Formation and Dynamics, ed. Y.-S. {Sun},
  S.~{Ferraz-Mello}, \& J.-L. {Zhou}, 17--24

\bibitem[{{Brown} {et~al.}(1991){Brown}, {Gilliland}, {Noyes}, \&
  {Ramsey}}]{brown+1991}
{Brown}, T.~M., {Gilliland}, R.~L., {Noyes}, R.~W., \& {Ramsey}, L.~W. 1991,
  \apj, 368, 599

\bibitem[{{Chaplin} {et~al.}(2011){Chaplin}, {Kjeldsen}, {Bedding},
  {Christensen-Dalsgaard}, {Gilliland}, {Kawaler}, {Appourchaux}, {Elsworth},
  {Garc{\'{\i}}a}, {Houdek}, {Karoff}, {Metcalfe}, {Molenda-{\.Z}akowicz},
  {Monteiro}, {Thompson}, {Verner}, {Batalha}, {Borucki}, {Brown}, {Bryson},
  {Christiansen}, {Clarke}, {Jenkins}, {Klaus}, {Koch}, {An}, {Ballot}, {Basu},
  {Benomar}, {Bonanno}, {Broomhall}, {Campante}, {Corsaro}, {Creevey}, {Esch},
  {Gai}, {Gaulme}, {Hale}, {Handberg}, {Hekker}, {Huber}, {Mathur}, {Mosser},
  {New}, {Pinsonneault}, {Pricopi}, {Quirion}, {R{\'e}gulo}, {Roxburgh},
  {Salabert}, {Stello}, \& {Suran}}]{chaplin+2011}
{Chaplin}, W.~J., {Kjeldsen}, H., {Bedding}, T.~R., {et~al.} 2011, \apj, 732,
  54

\bibitem[{Chib \& Jeliazkov(2001)}]{chib01}
Chib, S., \& Jeliazkov, I. 2001, Journal of the American Statistical
  Association, 96, 270.
\newblock \url{https://doi.org/10.1198/016214501750332848}

\bibitem[{{Choi} {et~al.}(2016){Choi}, {Dotter}, {Conroy}, {Cantiello},
  {Paxton}, \& {Johns on}}]{choi+2016a}
{Choi}, J., {Dotter}, A., {Conroy}, C., {et~al.} 2016, \apj, 823, 102

\bibitem[{{De Ridder} {et~al.}(2009){De Ridder}, {Barban}, {Baudin}, {Carrier},
  {Hatzes}, {Hekker}, {Kallinger}, {Weiss}, {Baglin}, {Auvergne}, {Samadi},
  {Barge}, \& {Deleuil}}]{de_ridder+2009}
{De Ridder}, J., {Barban}, C., {Baudin}, F., {et~al.} 2009, \nat, 459, 398

\bibitem[{{Dotter}(2016)}]{dotter+2016a}
{Dotter}, A. 2016, \apjs, 222, 8

\bibitem[{{Foreman-Mackey} {et~al.}(2013){Foreman-Mackey}, {Hogg}, {Lang}, \&
  {Goodman}}]{foreman-mackey+2013}
{Foreman-Mackey}, D., {Hogg}, D.~W., {Lang}, D., \& {Goodman}, J. 2013, \pasp,
  125, 306

\bibitem[{Green(1995)}]{green95}
Green, P.~J. 1995, Biometrika, 82, 711.
\newblock \url{https://dx.doi.org/10.1093/biomet/82.4.711}

\bibitem[{{Hekker} {et~al.}(2009){Hekker}, {Kallinger}, {Baudin}, {De Ridder},
  {Barban}, {Carrier}, {Hatzes}, {Weiss}, \& {Baglin}}]{hekker+2009}
{Hekker}, S., {Kallinger}, T., {Baudin}, F., {et~al.} 2009, \aap, 506, 465

\bibitem[{{Hekker} {et~al.}(2010){Hekker}, {Broomhall}, {Chaplin}, {Elsworth},
  {Fletcher}, {New}, {Arentoft}, {Quirion}, \& {Kjeldsen}}]{hekker+2010}
{Hekker}, S., {Broomhall}, A.-M., {Chaplin}, W.~J., {et~al.} 2010, \mnras, 402,
  2049

\bibitem[{{Huber} {et~al.}(2009){Huber}, {Stello}, {Bedding}, {Chaplin},
  {Arentoft}, {Quirion}, \& {Kjeldsen}}]{huber+2009}
{Huber}, D., {Stello}, D., {Bedding}, T.~R., {et~al.} 2009, Communications in
  Asteroseismology, 160, 74

\bibitem[{{Huber} {et~al.}(2010){Huber}, {Bedding}, {Stello}, {Mosser},
  {Mathur}, {Kallinger}, {Hekker}, {Elsworth}, {Buzasi}, {De Ridder},
  {Gilliland}, {Kjeldsen}, {Chaplin}, {Garc{\'{\i}}a}, {Hale}, {Preston},
  {White}, {Borucki}, {Christensen-Dalsgaard}, {Clarke}, {Jenkins}, \&
  {Koch}}]{huber+2010}
{Huber}, D., {Bedding}, T.~R., {Stello}, D., {et~al.} 2010, \apj, 723, 1607

\bibitem[{{Huber} {et~al.}(2016){Huber}, {Bryson}, {Haas}, {Barclay},
  {Barentsen}, {Howell}, {Sharma}, {Stello}, \& {Thompson}}]{huber+2016}
{Huber}, D., {Bryson}, S.~T., {Haas}, M.~R., {et~al.} 2016, \apjs, 224, 2

\bibitem[{{Jeffreys}(1935)}]{jeffreys1935}
{Jeffreys}, H. 1935, Proceedings of the Cambridge Philosophy Society, 31, 203

\bibitem[{{Kallinger} {et~al.}(2016){Kallinger}, {Hekker}, {Garcia}, {Huber},
  \& {Matthews}}]{kallinger+2016}
{Kallinger}, T., {Hekker}, S., {Garcia}, R.~A., {Huber}, D., \& {Matthews},
  J.~M. 2016, Science Advances, 2, 1500654

\bibitem[{{Kallinger} {et~al.}(2010){Kallinger}, {Mosser}, {Hekker}, {Huber},
  {Stello}, {Mathur}, {Basu}, {Bedding}, {Chaplin}, {De Ridder}, {Elsworth},
  {Frandsen}, {Garc{\'{\i}}a}, {Gruberbauer}, {Matthews}, {Borucki}, {Bruntt},
  {Christensen-Dalsgaard}, {Gilliland}, {Kjeldsen}, \& {Koch}}]{kallinger+2010}
{Kallinger}, T., {Mosser}, B., {Hekker}, S., {et~al.} 2010, \aap, 522, A1

\bibitem[{{Kallinger} {et~al.}(2014){Kallinger}, {De Ridder}, {Hekker},
  {Mathur}, {Mosser}, {Gruberbauer}, {Garc{\'{\i}}a}, {Karoff}, \&
  {Ballot}}]{kallinger+2014}
{Kallinger}, T., {De Ridder}, J., {Hekker}, S., {et~al.} 2014, \aap, 570, A41

\bibitem[{{Kass}(1995)}]{kass_raftery1995}
{Kass}, R.~E., R.~A.~E. 1995, Journal of the American Statistical Association,
  90, 773

\bibitem[{{Kjeldsen} \& {Bedding}(1995)}]{kjeldsen&bedding1995}
{Kjeldsen}, H., \& {Bedding}, T.~R. 1995, \aap, 293, 87

\bibitem[{{Kjeldsen} \& {Bedding}(2011)}]{kjeldsen&bedding2011}
---. 2011, \aap, 529, L8

\bibitem[{{Luger} {et~al.}(2016){Luger}, {Agol}, {Kruse}, {Barnes}, {Becker},
  {Foreman-Mackey}, \& {Deming}}]{luger+2016}
{Luger}, R., {Agol}, E., {Kruse}, E., {et~al.} 2016, \aj, 152, 100

\bibitem[{{Lund} {et~al.}(2015){Lund}, {Handberg}, {Davies}, {Chaplin}, \&
  {Jones}}]{lund+2015}
{Lund}, M.~N., {Handberg}, R., {Davies}, G.~R., {Chaplin}, W.~J., \& {Jones},
  C.~D. 2015, \apj, 806, 30

\bibitem[{{Mathur} {et~al.}(2010){Mathur}, {Garc{\'{\i}}a}, {R{\'e}gulo},
  {Creevey}, {Ballot}, {Salabert}, {Arentoft}, {Quirion}, {Chaplin}, \&
  {Kjeldsen}}]{mathur+2010}
{Mathur}, S., {Garc{\'{\i}}a}, R.~A., {R{\'e}gulo}, C., {et~al.} 2010, \aap,
  511, A46

\bibitem[{{Mosser} \& {Appourchaux}(2009)}]{mosser&appourchaux2009}
{Mosser}, B., \& {Appourchaux}, T. 2009, \aap, 508, 877

\bibitem[{{Mosser} {et~al.}(2010){Mosser}, {Belkacem}, {Goupil}, {Miglio},
  {Morel}, {Barban}, {Baudin}, {Hekker}, {Samadi}, {De Ridder}, {Weiss},
  {Auvergne}, \& {Baglin}}]{mosser+2010}
{Mosser}, B., {Belkacem}, K., {Goupil}, M.-J., {et~al.} 2010, \aap, 517, A22

\bibitem[{{Murphy} {et~al.}(2013){Murphy}, {Shibahashi}, \&
  {Kurtz}}]{murphy+2013a}
{Murphy}, S.~J., {Shibahashi}, H., \& {Kurtz}, D.~W. 2013, \mnras, 430, 2986

\bibitem[{Nordlund {et~al.}(2009)Nordlund, Stein, \&
  Asplund}]{nordlund_stein_asplund2009}
Nordlund, {\AA}., Stein, R.~F., \& Asplund, M. 2009, Living Reviews in Solar
  Physics, 6, 2.
\newblock \url{https://doi.org/10.12942/lrsp-2009-2}

\bibitem[{{Pinsonneault} {et~al.}(2018){Pinsonneault}, {Elsworth}, {Tayar},
  {Serenelli}, {Stello}, {Zinn}, {Mathur}, {Garc{\'{\i}}a}, {Johnson},
  {Hekker}, {Huber}, {Kallinger}, {M{\'e}sz{\'a}ros}, {Mosser}, {Stassun},
  {Girardi}, {Rodrigues}, {Silva Aguirre}, {An}, {Basu}, {Chaplin}, {Corsaro},
  {Cunha}, {Garc{\'{\i}}a-Hern{\'a}ndez}, {Holtzman}, {J{\"o}nsson},
  {Shetrone}, {Smith}, {Sobeck}, {Stringfellow}, {Zamora}, {Beers},
  {Fern{\'a}ndez-Trincado}, {Frinchaboy}, {Hearty}, \&
  {Nitschelm}}]{pinsonneault+2018}
{Pinsonneault}, M.~H., {Elsworth}, Y.~P., {Tayar}, J., {et~al.} 2018, \apjs,
  239, 32

\bibitem[{{Ricker} {et~al.}(2014){Ricker}, {Winn}, {Vanderspek}, {Latham},
  {Bakos}, {Bean}, {Berta-Thompson}, {Brown}, {Buchhave}, {Butler}, {Butler},
  {Chaplin}, {Charbonneau}, {Christensen-Dalsgaard}, {Clampin}, {Deming},
  {Doty}, {De Lee}, {Dressing}, {Dunham}, {Endl}, {Fressin}, {Ge}, {Henning},
  {Holman}, {Howard}, {Ida}, {Jenkins}, {Jernigan}, {Johnson}, {Kaltenegger},
  {Kawai}, {Kjeldsen}, {Laughlin}, {Levine}, {Lin}, {Lissauer}, {MacQueen},
  {Marcy}, {McCullough}, {Morton}, {Narita}, {Paegert}, {Palle}, {Pepe},
  {Pepper}, {Quirrenbach}, {Rinehart}, {Sasselov}, {Sato}, {Seager},
  {Sozzetti}, {Stassun}, {Sullivan}, {Szentgyorgyi}, {Torres}, {Udry}, \&
  {Villasenor}}]{ricker+2014}
{Ricker}, G.~R., {Winn}, J.~N., {Vanderspek}, R., {et~al.} 2014, in \procspie,
  Vol. 9143, Space Telescopes and Instrumentation 2014: Optical, Infrared, and
  Millimeter Wave, 914320

\bibitem[{{Scargle}(1982)}]{scargle1982}
{Scargle}, J.~D. 1982, \apj, 263, 835

\bibitem[{{Schwarz}(1978)}]{schwarz1978}
{Schwarz}, G. 1978, Annals of Statistics, 6, 461

\bibitem[{{Sharma} {et~al.}(2011){Sharma}, {Bland-Hawthorn}, {Johnston}, \&
  {Binney}}]{sharma+2011}
{Sharma}, S., {Bland-Hawthorn}, J., {Johnston}, K.~V., \& {Binney}, J. 2011,
  \apj, 730, 3

\bibitem[{{Sharma} {et~al.}(2019){Sharma}, {Stello}, {Bland-Hawthorn},
  {Hayden}, {Zinn}, {Kallinger}\, {Hon}, {Asplund}, {Buder}, {De Silva},
  {D'Orazi}, {Freeman}, {Kos}, {Lin}, {Lind}, {Martell}, {Simpson},
  {Wittenmyer}, {Zucker}, {Zwitter}, {Bedding}, {Chen}, {Cotar}, {Esdaile},
  {Horner}, {Huber}, {Khanna}, {Li}, {Ting}, {Nataf}, {Nordlander}, {Saddon},
  {Wright}, \& {Wyse}}]{sharma+2019}
{Sharma}, S., {Stello}, D., {Bland-Hawthorn}, J., {et~al.} 2019, arXiv e-prints

\bibitem[{Skilling(2004)}]{skilling04}
Skilling, J. 2004, AIP Conference Proceedings, 735, 395.
\newblock \url{https://aip.scitation.org/doi/abs/10.1063/1.1835238}

\bibitem[{{Stello} {et~al.}(2016){Stello}, {Cantiello}, {Fuller}, {Garcia}, \&
  {Huber}}]{stello+2016}
{Stello}, D., {Cantiello}, M., {Fuller}, J., {Garcia}, R.~A., \& {Huber}, D.
  2016, \pasa, 33, e011

\bibitem[{{Stello} {et~al.}(2009){Stello}, {Chaplin}, {Basu}, {Elsworth}, \&
  {Bedding}}]{stello+2009}
{Stello}, D., {Chaplin}, W.~J., {Basu}, S., {Elsworth}, Y., \& {Bedding}, T.~R.
  2009, \mnras, 400, L80

\bibitem[{{Stello} {et~al.}(2011){Stello}, {Huber}, {Kallinger}, {Basu},
  {Mosser}, {Hekker}, {Mathur}, {Garc{\'{\i}}a}, {Bedding}, {Kjeldsen},
  {Gilliland}, {Verner}, {Chaplin}, {Benomar}, {Meibom}, {Grundahl},
  {Elsworth}, {Molenda-{\.Z}akowicz}, {Szab{\'o}}, {Christensen-Dalsgaard},
  {Tenenbaum}, {Twicken}, \& {Uddin}}]{stello+2011}
{Stello}, D., {Huber}, D., {Kallinger}, T., {et~al.} 2011, \apjl, 737, L10

\bibitem[{{Stello} {et~al.}(2013){Stello}, {Huber}, {Bedding}, {Benomar},
  {Bildsten}, {Elsworth}, {Gilliland}, {Mosser}, {Paxton}, \&
  {White}}]{stello+2013}
{Stello}, D., {Huber}, D., {Bedding}, T.~R., {et~al.} 2013, \apjl, 765, L41

\bibitem[{{Stello} {et~al.}(2014){Stello}, {Compton}, {Bedding},
  {Christensen-Dalsgaard}, {Kiss}, {Kjeldsen}, {Bellamy}, {Garc{\'{\i}}a}, \&
  {Mathur}}]{stello+2014}
{Stello}, D., {Compton}, D.~L., {Bedding}, T.~R., {et~al.} 2014, \apjl, 788,
  L10

\bibitem[{{Stello} {et~al.}(2015){Stello}, {Huber}, {Sharma}, {Johnson},
  {Lund}, {Handberg}, {Buzasi}, {Silva Aguirre}, {Chaplin}, {Miglio},
  {Pinsonneault}, {Basu}, {Bedding}, {Bland-Hawthorn}, {Casagrande}, {Davies},
  {Elsworth}, {Garcia}, {Mathur}, {Di Mauro}, {Mosser}, {Schneider},
  {Serenelli}, \& {Valentini}}]{stello+2015}
{Stello}, D., {Huber}, D., {Sharma}, S., {et~al.} 2015, \apjl, 809, L3

\bibitem[{{Stello} {et~al.}(2017){Stello}, {Zinn}, {Elsworth}, {Garcia},
  {Kallinger}, {Mathur}, {Mosser}, {Sharma}, {Chaplin}, {Davies}, {Huber},
  {Jones}, {Miglio}, \& {Silva Aguirre}}]{stello+2017}
{Stello}, D., {Zinn}, J., {Elsworth}, Y., {et~al.} 2017, \apj, 835, 83

\bibitem[{{Ulrich}(1986)}]{ulrich1986}
{Ulrich}, R.~K. 1986, \apjl, 306, L37

\bibitem[{{Vanderburg} \& {Johnson}(2014)}]{vanderburg&johnson2014}
{Vanderburg}, A., \& {Johnson}, J.~A. 2014, \pasp, 126, 948

\bibitem[{{Vanderplas} {et~al.}(2012){Vanderplas}, {Connolly}, {Ivezi{\'c}}, \&
  {Gray}}]{astroML}
{Vanderplas}, J., {Connolly}, A., {Ivezi{\'c}}, {\v Z}., \& {Gray}, A. 2012, in
  Conference on Intelligent Data Understanding (CIDU), 47 --54

\bibitem[{{Watanabe}(2013)}]{watanabe2013}
{Watanabe}, S. 2013, Journal of Machine Learning Research, 14, 867

\bibitem[{{Woodard}(1984)}]{woodard1984}
{Woodard}, M.~F. 1984, PhD thesis, {University of California, San Diego}

\bibitem[{{Yu} {et~al.}(2018){Yu}, {Huber}, {Bedding}, {Stello}, {Hon},
  {Murphy}, \& {Khanna}}]{yu+2018}
{Yu}, J., {Huber}, D., {Bedding}, T.~R., {et~al.} 2018, \apjs, 236, 42

\end{thebibliography}
\label{lastpage}
\end{document}